\documentclass[11pt]{article}

\usepackage{booktabs}
\usepackage{slashed}
\usepackage{subfigure}
\usepackage{amsmath,amssymb,amsfonts,graphicx,cite,color,feynarts,soul,axodraw}
\usepackage{epsfig}
\usepackage{mciteplus}
\usepackage{wasysym}
\usepackage{multirow}
\usepackage{upgreek}
\usepackage{psfrag}
\usepackage{url}
\usepackage{rotate}
\usepackage[normalem]{ulem}
\usepackage{tikz}
\usepackage{multirow}
\usepackage{booktabs}
\usepackage{bbold}
\usepackage{feynmf}
\usepackage{float}
\usepackage{afterpage}

\include{paperdef}
\graphicspath{{figs/}}

\newcommand{\ra}[1]{\renewcommand{\arraystretch}{#1}}

\begin{document}
\hfill {\tt DESY 15-073}

\def\thefootnote{\fnsymbol{footnote}}

\begin{center}
\Large\bf\boldmath
\vspace*{2cm} 
Improved prediction for the mass of the W boson in the NMSSM
\unboldmath
\end{center}
\vspace{0.4cm}
\begin{center}
O.~St{\aa}l$^{1,}$\footnote{Electronic address: oscar.stal@fysik.su.se},
G.~Weiglein$^{2,}$\footnote{Electronic address: Georg.Weiglein@desy.de},
L.~Zeune$^{3}$\footnote{Electronic address: l.k.zeune@uva.nl} \\[0.4cm] 
\vspace{0.4cm}
 {\sl $^1$ The Oskar Klein Centre, Department of Physics\\
  Stockholm University, SE-106 91 Stockholm, Sweden}\\[0.2cm]
{\sl $^2$ Deutsches Elektronen-Synchrotron DESY\\
 Notkestra{\ss}e 85, D-22607 Hamburg, Germany}\\[0.2cm]
 {\sl $^3$ ITFA, University of Amsterdam\\
 Science Park 904, 1018 XE, Amsterdam, The Netherlands}\\[0.2cm]
\end{center}
\vspace{0.2cm}

\renewcommand{\thefootnote}{\arabic{footnote}}
\setcounter{footnote}{0}

\begin{abstract}
Electroweak precision observables, being highly sensitive to loop contributions of new physics, 
provide a powerful tool to test the theory and to discriminate between different models of the underlying physics. 
In that context, the $W$ boson mass, $\MW$, plays a crucial role. The
accuracy of the $\MW$ measurement has been significantly improved over the
last years, and further improvement of the experimental accuracy is expected
from future LHC measurements. In order to fully exploit the precise
experimental determination, an accurate theoretical prediction for $\MW$ in
the Standard Model (SM) and extensions of it
is of central importance. We present the currently most
accurate prediction for the $W$ boson mass in the Next-to-Minimal
Supersymmetric extension of the
Standard Model (NMSSM), including the full one-loop result
and all available higher-order corrections of SM and SUSY type. The
evaluation of $\MW$ is performed in a flexible framework, which facilitates
the extension to other models beyond the SM. 
We show numerical results for the $W$ boson mass in the NMSSM,
focussing on phenomenologically interesting scenarios, in which the Higgs signal can be interpreted as the lightest or second lightest $\cp$-even Higgs boson of the NMSSM.
We find that, for both Higgs signal interpretations, the NMSSM $\MW$ prediction is well compatible with the measurement.
We study the SUSY contributions to $\MW$ in detail and investigate in particular the genuine NMSSM effects from the Higgs and neutralino sectors. 
\end{abstract}
\newpage

\section{Introduction}
\label{sect:Introduction}

Supersymmetry (SUSY) is regarded to be the most appealing extension of the Standard Model (SM), as it provides a natural mechanism to explain a light Higgs boson as observed by ATLAS \cite{atlas:2012gk} and CMS \cite{cms:2012gu}. 
Supersymmetry realised around the TeV-scale also comes with further
desirable features, such as a possible dark matter candidate and the unification of gauge couplings.

The superpotential of the Minimal Supersymmetric extension of the 
Standard Model (MSSM)
contains a term bilinear in the two Higgs doublets, $W_{(2)}\sim \mu H_2
H_1$. In this term a dimensionful parameter, $\mu$, is present, which in the
MSSM has no natural connection to the SUSY breaking scale. The difficulty to
motivate a phenomenologically acceptable value in this context is called the
$\mu$-problem of the MSSM~\cite{Kim:1983dt}. This problem is addressed in
the Next-to-Minimal Supersymmetric extension of the Standard Model (NMSSM), where the Higgs sector of the MSSM gets enlarged by an additional singlet. 
The corresponding term in the superpotential is replaced by
a coupling $W_{(3)}\sim \lambda S H_2 H_1$,
and the $\mu$ parameter arises
dynamically from the vev of the singlet, $S$, and 
may therefore be related to the SUSY breaking scale.

Besides the solution of the $\mu$-problem, there are additional motivations to study the NMSSM. The physical spectrum contains seven Higgs bosons, which leads to a rich and interesting phenomenology. Compared to the MSSM, the singlet field modifies the Higgs mass relations such that the tree-level mass of the lightest neutral Higgs boson can be increased.
Consequently, the radiative corrections needed to shift the mass of the lightest Higgs mass up to $125 \gev$ can be smaller.
This relaxes the requirement of heavy stops, or a large splitting in the
stop sector, in NMSSM parameter regions where the tree-level Higgs mass is
larger than the maximal MSSM value (see e.g.~\citere{Ellwanger:2009dp}).
The NMSSM singlet-doublet mixing could also modify the couplings of the $125
\gev$ boson to explain a potentially enhanced rate in the diphoton signal
(see e.g.\ 
\citeres{Ellwanger:2010nf, Benbrik:2012rm,King:2012tr,Choi:2012he}).

Extensive direct searches for supersymmetric particles are carried out by
the LHC experiments. These searches have so far not resulted in a signal,
which leads to limits on the particle masses, see e.g.\
\citeres{lhcbook,AtlasSusy,CMSSusy} for a compilation of the results.
Indirect methods are complementary to direct searches for physics beyond the SM at the LHC and future collider experiments. Whereas direct methods attempt to observe traces in the detectors arising from of the direct production of particles of new physics models, indirect methods look for the quantum effects induced by virtual exchange of new states.
Even if not yet seen directly, signs of new physics may show up as small deviations between precise measurements and SM predictions. 
Electroweak precision observables (EWPOs), such as the $W$ boson mass,
$M_W$, the sine of the effective leptonic weak mixing angle at the $Z$
boson resonance, $\sin^2 \theta_{\rm w}^{\rm eff}$, or the anomalous magnetic moment of the muon, $(g-2)_{\mu}$, (among others)
are all highly sensitive to loop contributions involving in principle all
the particles of the considered model. They can both be theoretically 
predicted and experimentally measured with such a 
high precision that they can be utilised to
 test the SM, to distinguish between different extensions,
and to derive indirect constraints on the parameters of a model. 
Input from indirect methods can be of great interest to direct searches for new particles. This was demonstrated, for instance, by the discovery of the top quark with a measured mass in remarkable agreement with the indirect prediction \cite{Abe:1995hr,Abachi:1995iq}.

In this paper we focus on the $W$ boson mass.
The accuracy of the measurement of
$\MW$ has been significantly improved in the last years with the results
presented by the Tevatron experiments CDF~\cite{Aaltonen:2012bp} and D\O~\cite{Abazov:2012bv}. The current world average is~\cite{Group:2012gb,ALEPH:2005ab}
\begin{equation}
\MW^{\rm exp} = 80.385 \pm 0.015 \gev.
\label{eq:MW}
\end{equation}
This precise measurement makes $M_W$ particularly suitable for electroweak
precision tests, even more since the precision is expected to be improved
further when including the full dataset from the Tevatron and upcoming
results from the LHC.
Of central importance for the theoretical precision that can be achieved on $\MW$ is the top quark mass measurement, since the experimental error on the input parameter $\mt$ constitutes a dominant source of (parametric)
uncertainty, see e.g.\
\citere{Heinemeyer:2003ud}.
The Tevatron~\cite{Lancaster:2011wr} and LHC~\cite{atlascmstop,atlastop1,atlastop2,Aad:2015nba,Chatrchyan:2012cz,Chatrchyan:2012ea,Chatrchyan:2013xza} measurements of $m_t$ have been combined \cite{ATLAS:2014wva} to yield
\begin{equation}
 \mt^{\rm exp} = 173.34\pm0.27\pm0.71\gev.
 \label{eq:mt}
\end{equation}
In contrast to the 
$\MW$ measurement, a considerable improvement 
of the precision on $\mt$ beyond~\refeq{eq:mt} seems less likely at the LHC.
Furthermore, it is not straightforward to relate  $\mt^{\rm exp}$
measured at a hadron collider (using kinematic information about the top decay products) to a theoretically well-defined mass parameter. 
The quantity measured with high precision 
at the Tevatron and the LHC is expected to be close to the top pole
mass with an uncertainty at the level of about 
$1 \gev$~\cite{Skands:2007zg,Hoang:2008xm,Hoang:2014oea}. 
For the calculation of $\MW$ presented in this paper
we adopt the interpretation of the measured value $\mt^{\rm exp}$ as the pole mass, but
the results could easily be re-expressed in terms of a properly
defined short distance mass (such as the \msbar\ or \drbar\ mass). 
At an $e^+e^-$ linear collider, the situation would improve significantly.
Estimates for the ILC show an expected precision $\Delta M_W^{\text{ILC}}
\sim 2.5 - 5 \mev$ and $\Delta \mt^{\text{ILC}} = 0.1
\gev$~\cite{Moortgat-Picka:2015yla,Baak:2013fwa}, where the stated
precision for $\mt$ accounts both for the uncertainty in the determination
of the actually measured mass parameter and the uncertainty related to the
conversion into a suitable and theoretically well-defined parameter
such as the $\overline{\rm{MS}}$ mass.

For exploiting the precise current and (possible) future $\MW$
measurements, theoretical predictions for $\MW$ with comparable accuracy
are desired both in the SM and extensions of it. In order to
be able to
discriminate between different models it is necessary that the precision is
comparable. In the SM, the most advanced evaluation of $\MW$ includes the
full one-loop~\cite{Sirlin:1980nh,Marciano:1980pb} and
two-loop~\cite{qcd2SMa,qcd2SMb,qcd2SMc,qcd2SMd,qcd2SMe,qcd2SMf,StremplatDiplom,
Freitas:2000gg, Freitas:2002ja, 2lfermc,2lbos,2Lbosa,2Lbosb}
result, as well as the leading
three- and four-loop corrections~\cite{Avdeev:1994db,qcd3SMa,
qcd3SMb,Chetyrkin:1996cf,Faisst:2003px,vanderBij:2000cg,Boughezal:2004ef,Boughezal:2006xk,Chetyrkin:2006bj,Schroder:2005db}.
A simple parametrization for $\MW^{\rm SM}$ has also been
developed~\cite{Awramik:2003rn}, see also \citere{Awramik:2006uz}.
Within the SM the LHC Higgs signal at $125.09 \text{ GeV}$ \cite{Aad:2015zhl} is interpreted as the SM Higgs boson. Setting 
$M_H^{\text{SM}}\simeq \MHexp \gev$, the value of $\MW$ can be predicted in the SM without any free parameters. The result (for $m_t= 173.34 \text{ GeV}$, $M_H^{\text{SM}} = 125.09 \text{ GeV}$) is\footnote{
We updated the SM $\MW$ prediction as discussed below in \refse{subsec:SMho}. This leads to a small difference compared to the SM value given in \citere{Heinemeyer:2013dia}.
}
\begin{equation}
M_W^{\rm SM} = 80.358 \text{ GeV},
\end{equation}
which differs by $\sim 1.8\, \sigma$ from the experimental value given in Eq.~\eqref{eq:MW}.
The theoretical uncertainty from missing higher-order corrections has
been estimated to be around $4$~MeV in the SM for a light Higgs
boson~\cite{Awramik:2003rn}.

For supersymmetric theories, the one-loop result for $\MW$~\cite{Barbieri:1983wy,Lim:1983re,Eliasson:1984yu,Hioki:1985wz,
Grifols:1984xs,Barbieri:1989dc,Drees:1990dx,Drees:1991zk,
Chankowski:1994tn,Garcia:1993sb,Pierce:1996zz,Heinemeyer:2006px} 
and leading two-loop corrections \cite{Djouadi:1996pa,Djouadi:1998sq,Heinemeyer:2002jq,Haestier:2005ja} have been 
obtained for the MSSM.
A precise prediction for $M_W^{\rm MSSM}$, taking into account all relevant
higher-order corrections of both SM- and SUSY-type, was presented
in~\citere{Heinemeyer:2013dia}. A first prediction for $M_W$ in the
NMSSM has also been presented in~\citere{Domingo:2011uf}. For the study of
other EWPOs (mainly focusing on $Z$ decays) in the NMSSM, see~\citeres{Cao:2008rc,Cao:2010na,Domingo:2011uf}.

In this work we follow the procedure employed in the MSSM to present a new prediction for $\MW$ in the NMSSM with the same level 
of accuracy as the current best MSSM prediction~\cite{Heinemeyer:2013dia}. 
We combine the complete NMSSM one-loop result with the 
state-of-the-art SM result and leading SUSY higher-order corrections.
Our framework allows to output, besides $M_W$, also the quantity 
$\Delta r$ directly (see \refse{sect:mwdet}), 
which summarises all (non-QED) quantum correction to the muon decay amplitude. Besides its importance for electroweak precision tests, $\Delta r$ is needed whenever a theoretical prediction is parametrized in terms of the Fermi constant $G_{\mu}$ (instead of $\MW$ or $\alpha(M_Z)$). Our NMSSM prediction for $\MW$  provides the flexibility 
to analyse SUSY loop contributions 
analytically and
to treat possible threshold effects or numerical instabilities.
We perform a detailed numerical analysis of $\MW$ in the NMSSM with the latest experimental results taken into account.  
We focus on the effects induced by the extended Higgs and neutralino sectors, and in particular on benchmark scenarios where the LHC Higgs boson is interpreted as either the lightest or the second lightest $\cp$-even Higgs boson of the NMSSM.

This paper is organised as follows: in \refse{sec:NMSSM} we give a short introduction to the NMSSM, focussing on the Higgs and neutralino sectors. In \refse{sect:mwdet} we describe the determination of the $W$ boson mass in the NMSSM.
We outline the calculation of the one-loop contributions and the incorporation of higher-order contributions.
In \refse{sec:mwsetup} we give the numerical results, analysing the NMSSM contributions to the $W$ boson mass, before we conclude in \refse{sect:conclusions}.

\section{The Next-to-Minimal Supersymmetric Standard Model}
\label{sec:NMSSM}

In this section we introduce the NMSSM and specify our notation. We
focus on the particle sectors which differ from the MSSM. Since the SM
fermions and their superpartners appear in the same way in both models,
the sfermion sector of the NMSSM is unchanged with respect to the MSSM. Also the chargino sector is identical to that in the MSSM since no new charged degrees of freedom are introduced. 
For these sectors we use the same notation as employed in~\citere{Heinemeyer:2013dia}.

In addition to the two Higgs doublets of the MSSM, the NMSSM also contains a 
Higgs singlet, $S$, which couples only to the Higgs sector.\footnote{For the Higgs doublets and the Higgs singlet we use the same notation for the  supermultiplets and for its scalar component.}
Considering the $Z_3$-symmetric version of the NMSSM, the superpotential takes the form
\BE
W^{\text{NMSSM}}= \bar{u} {\bf y}_u Q H_2 - \bar{d} {\bf y}_d Q H_1 - \bar{e} {\bf y}_l L H_1 + \lambda S H_2 H_1 + \frac{1}{3} \kappa S^3\,.
\label{eq:mssmsuperpo}
\EE 
The new contributions of the Higgs singlet to the soft breaking terms are
\BE
\begin{aligned}
{\cal L}^{\text{NMSSM}}_{\text{soft}} ={\cal L}^{\text{MSSM,mod}}_{\text{soft}} 
- m_S^2 |S|^2 - (\lambda A_{\lambda} S H_2 H_1 + \frac{1}{3} \kappa A_{\kappa} S^3 + \mathrm{h.c.}),
\label{eq:softnmssm}
\end{aligned}
\EE
where ${\cal L}^{\text{MSSM,mod}}_{\text{soft}}$ is the soft-breaking
Lagrangian ${\cal L}^{\text{MSSM}}_{\text{soft}}$ of the MSSM 
(see e.g.\ Eq.~(6.3.1) of \citere{Martin:1997ns}), but without the term 
$b H_2 H_1$. The singlet couplings make it possible to dynamically
generate an effective $\mu$ parameter as
\begin{equation}
\mueff= \la \langle S \rangle.
\label{def-mueff}
\end{equation}

The additional contributions (and the modified effective $\mu$ term) in the superpotential and in the soft breaking terms
lead to a Higgs potential which contains the additional soft breaking parameters $m_S^2$, $\Ala$, $\Aka$,
as well as the superpotential trilinear couplings $\la$ and $\ka$. 
Like in the MSSM, there is no $\cp$-violation at tree-level in the
couplings of the Higgs doublets. The new doublet-singlet couplings allow in principle for $\cp$-violation at tree level, but we will not consider this possibility here. 
We choose all parameters to be real.

The minimum of the NMSSM Higgs potential triggers electroweak symmetry breaking, after which 
 the Higgs doublets can be expanded around their minima according to
\begin{equation}
H_1 = \left(\begin{array}{c}
v_1+\frac{1}{\sqrt{2}}\left(\phi_1 - \mathrm{i}\chi_1\right)\\
- \phi_1^-
\end{array}\right),
\qquad
H_2 = \left(\begin{array}{c}
\phi_2^+\\
v_2+\frac{1}{\sqrt{2}}\left(\phi_2+\mathrm{i}\chi_2\right)
\end{array}\right).
\label{eq:higgsdoublets1}
\end{equation}
Similarly, the singlet scalar component can be expanded as
\begin{equation}
S=v_s+\frac{1}{\sqrt{2}}\left(\phi_s+\mathrm{i}\chi_s\right),
\end{equation} 
where $v_s$ is the (non-zero) vacuum expectation value of the singlet.

 The bilinear part of the Higgs potential can be written as
 \BE
 \begin{aligned}
 \VHiggs = \tedz \begin{pmatrix} \phi_1,\phi_2, \phi_S \end{pmatrix} 
 &\matr{M}_{\phi\phi\phi}
 \begin{pmatrix} \phi_1 \\ \phi_2 \\ \phi_S \end{pmatrix} +
 \tedz \begin{pmatrix} \chi_1, \chi_2, \chi_S \end{pmatrix}
 \matr{M}_{\chi\chi\chi}
 \begin{pmatrix} \chi_1 \\ \chi_2 \\ \chi_S \end{pmatrix} \\&+
  \begin{pmatrix} \phi^-_1,\phi^-_2  \end{pmatrix}
 \matr{M}_{\phi^\pm\phi^\pm}
 \begin{pmatrix} \phi^+_1 \\ \phi^+_2  \end{pmatrix} + \cdots, 
 \end{aligned}
 \EE
 with the mass matrices
 $\matr{M}_{\phi\phi\phi}$, $\matr{M}_{\chi\chi\chi}$ 
 and $\matr{M}_{\phi^\pm\phi^\pm}$.

The mixing of the $\cp$-even, $\cp$-odd and charged Higgs fields
occurring in the mass eigenstates
is described by three unitary matrices $\Ueven$, $\Uodd$, and
$\Uchar$, where
\begin{align}
\label{eq:RotateToMassES}
\begin{pmatrix} \He \\ \Hz \\ \Hd \end{pmatrix} 
= \Ueven 
\begin{pmatrix} \phi_1 \\ \phi_2 \\ \phi_S \end{pmatrix}, \quad
\begin{pmatrix}  \Ae \\ \Az \\ G \end{pmatrix} 
= \Uodd 
\begin{pmatrix} \chi_1 \\ \chi_2 \\ \chi_S \end{pmatrix}, \quad
\begin{pmatrix}  H^\pm \\ G^\pm  \end{pmatrix} = \Uchar 
\begin{pmatrix} \phi_1^\pm \\ \phi_2^\pm \end{pmatrix} .
\end{align}
These transform the Higgs fields such that the resulting (diagonal) mass matrices become
\begin{equation}
\matr{M}_{hhh}^{\rm diag} = \Ueven
\matr{M}_{\phi\phi\phi} \left(\Ueven\right)^{\dagger}, \;
\matr{M}_{aaG}^{\rm diag} = \Uodd
\matr{M}_{\chi\chi\chi} \left(\Uodd\right)^{\dagger} 
 \; {\rm and } \;
\matr{M}_{H^\pm G^\pm}^{\rm diag} = 
\Uchar \matr{M}_{\phi^\pm\phi^\pm}
\left(\Uchar\right)^{\dagger} 
\label{eq:RotMat}
\end{equation}
The $\cp$-even mass eigenstates, $\He$, $\Hz$ and $\Hd$, are ordered such that $\mHe \le \mHz \le \mHd$, and similarly for the two  $\cp$-odd Higgs bosons, $\Ae$ and $\Az$. Unchanged from the SM there are also the Goldstone bosons, $G$ and $G^\pm$. Finally, there is the charged Higgs pair, $H^\pm$ with mass given by
\BE
M^2_{H^{\pm}} = \hat{m}_A^2 + M_W^2 -\lambda^2 v^2.
\label{eq:mhpnmssm}
\EE
Here $\hat{m}_A$ is the effective $\cp$-odd doublet mass given by
\BE
\hat{m}_A^2 = \frac{\lambda v_s}{\sin \beta \,\cos \beta}\left(A_{\lambda}+\kappa v_s\right)\,.
\label{eq:mahatnmssm}
\EE

The superpartner of the singlet scalar enlarging the NMSSM Higgs sector is called the singlino, $\tilde{S}$. It
extends the neutralino sector with a fifth mass eigenstate. In the basis $(\tilde{B}, \tilde{W}^0,
\tilde{H}_1^0, \tilde{H}_2^0, \tilde{S})$ the neutralino mass matrix at tree level is given by 
\begin{equation}
\begin{aligned}
&\matr{M}_{\neu{}}=\left(\begin{array}{ccccc} 
M_1 & 0 & -M_Z s_W \cos \beta& M_Z s_W \sin \beta& 0\\ 
0 & M_2 & M_Z c_W \cos \beta& -M_Z c_W \sin \beta& 0\\ 
-M_Z s_W \cos \beta & M_Z c_W \cos \beta & 0 & -\mueff& -\lambda v_2\\
M_Z s_W \sin \beta & -M_Z c_W \sin \beta & -\mueff & 0& -\lambda v_1 \\
0 & 0 & -\lambda v_2 & -\lambda v_1 & 2K\mueff
\end{array}\right),\\
\end{aligned}
\end{equation}
where $K\equiv\ka/\la$.
This mass matrix can be diagonalised by a single unitary matrix $N$ such that 
\begin{equation}
\mathrm{diag}(\mneu{1}, \mneu{2},\mneu{3},\mneu{4},\mneu{5})=N^* \matr{M}_{\neu{}} N^\dagger ,
\label{eq:neuNNMSSM}
\end{equation}
which gives the mass eigenvalues ordered as $\mneu{i} \le \mneu{j}$ for $i<j$. 

\section{Predicting the W~boson mass}
\label{sect:mwdet}

\subsection{Determination of $\MW$}
\begin{figure}
\centering
\includegraphics[height=0.2\columnwidth]{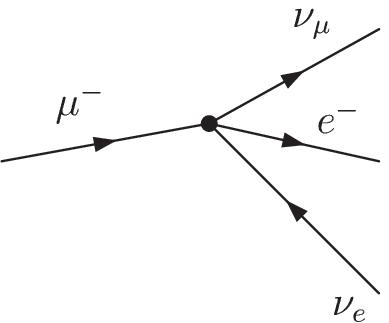}\hspace{1.5cm}
\includegraphics[height=0.2\columnwidth]{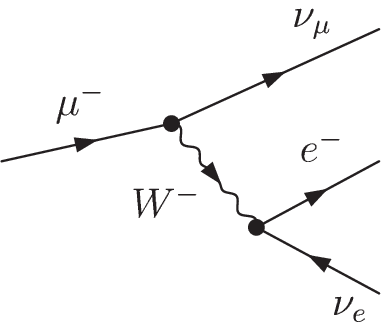}
\caption{Left: tree-level diagram with a four-fermion vertex describing muon
decay in the Fermi model. Right: $W$ boson exchange mediating muon decay in
the electroweak SM (in unitary gauge).} 
\label{fig:muondec}
\end{figure}
The $W$ boson mass can theoretically be predicted 
from the muon decay rate.
Muons decay to almost $100\%$ via 
$\mu \rightarrow e \bar{\nu_{e}} \nu_{\mu}$~\cite{Agashe:2014kda}.  
This decay was historically described first within the Fermi model
(left diagram in~\reffi{fig:muondec}). 
Comparing the muon decay amplitude calculated in the Fermi model to the same quantity calculated in the full SM or 
extensions thereof (the leading-order contribution in unitary gauge 
is depicted in~\reffi{fig:muondec}) yields the relation
\begin{equation}
\frac{G_{\mu}}{\sqrt{2}}=\frac{M_Z^2\,e^2}{8 \,M_W^2 \left(M_Z^2-M_W^2\right)}\left(1+\Delta r (M_W,M_Z,m_t, \, ... \, ,X)\right).
\label{eq:mwdef}
\end{equation}
This relates the $W$ boson mass to the Fermi constant, $G_{\mu}$, which by
definition contains the QED corrections to the four-fermion contact vertex
up to
$O(\alpha^2)~$\cite{Behrends:1955mb,Kinoshita:1958ru,vanRitbergen:1999fi,Steinhauser:1999bx,Pak:2008qt},
and to the other parameters $M_Z$ and $e$, which are
known experimentally with very high precision. The Fermi constant itself is
determined with high accuracy from precise measurements of the muon life 
time~\cite{Webber:2010zf}.

The factor $\Delta r$ in \refeq{eq:mwdef} summarises all higher-order contributions to the 
muon decay amplitude after subtracting the Fermi-model type virtual QED
corrections, which are already included in the definition of $G_\mu$.
Working in the on-shell renormalization scheme, \refeq{eq:mwdef} corresponds
to a relation between the physical masses of the $W$ and $Z$~bosons.

Neglecting the masses and momenta of the external fermions all loop diagrams can be expressed as a term proportional 
to the Born matrix element \cite{Sirlin:1980nh,Freitas:2002ja}
\begin{equation}
{\cal M}_{\text{Loop,i}}=\Delta r_i \text{ }\mathcal{M}_{\text{Born}}\, , \;\;\; \Dr = \sum_i \Dr_i\,.
\label{eq:drborn}
\end{equation}
In different models, different particles can contribute as virtual
particles in the loop diagrams to the muon-decay amplitude. Therefore,
the quantity $\Delta r$ depends on the specific model parameters (indicated by the $X$ in \refeq{eq:mwdef}), and
\refeq{eq:mwdef} provides a  
model-dependent prediction for the $W$~boson mass. 
The quantity $\Delta r$ itself does depend on $\MW$ as well; 
hence, the value of $\MW$ as the solution of \refeq{eq:mwdef} has to
be determined numerically. In practice this is done by iteration. 

In order to exploit the $W$ boson mass for electroweak precision tests 
a precise theoretical
prediction for $\Dr$ within and beyond the SM is needed.
In the next two subsections we describe our calculation of $\Delta r$ in the NMSSM. 
A new one-loop calculation has been performed which is combined with all available higher order corrections of SM- and SUSY-type.

\subsection{One-loop calculation of $\Delta r$ in the NMSSM}\label{sec:oneloopdr}

The one-loop contributions to $\Delta r$ consist of the $W$~boson
self-energy, vertex and box diagrams, and the corresponding counter terms (CT). The box diagrams are themselves UV-finite in a renormalizable gauge and require no counter terms. Schematically, this can be expressed as
\begin{equation}
\begin{split}
 \Delta r^{(\alpha)}\text{\hspace{0.1cm}}&= 
\text{\hspace{0.1cm}{$W$ Self-energy}\hspace{0.01cm}} 
+\text{\hspace{0.01cm}{$W$ Self-energy CT}\hspace{0.01cm}}
+\text{\hspace{0.01cm}Vertex\hspace{0.01cm}}
+\text{\hspace{0.01cm}{Vertex CT}\hspace{0.01cm}}
+\text{\hspace{0.01cm}Box\hspace{0.1cm}}\\
 &={\frac{\Sigma_T^{WW}\left(0\right)}{\MW^2}}
+{\left(-\delta Z_W-\frac{\delta \MW^2}{\MW^2}\right)}
+\text{\hspace{0.05cm}Vertex\hspace{0.05cm}}\\
&+{\left(2 \delta e-2\frac{\delta \sw}{\sw}
+\delta Z_W+\frac{1}{2}\left(\delta Z^{\mu}+\delta Z^{e}
+\delta Z^{\nu_{\mu}}+\delta Z^{\nu_{e}}\right)\right)}
+\text{\hspace{0.05cm}Box\hspace{0.05cm}}.
\label{herleitungdeltar}
\end{split}
\end{equation}
Here $\Sigma_T^{WW}(0)$ denotes the transverse part the $W$ boson
self-energy (evaluated at vanishing momentum transfer), 
$\delta \MW$ is the renormalization constant for the $W$ boson mass, 
$\delta e$ and $\delta \sw$ are the renormalization constants for the
electric charge and $\sw \equiv \sin \theta_W$, respectively. 
The $\delta Z$ denote different field renormalization constants.
Since the $W$~boson occurs in the muon decay amplitude only as a virtual particle, its field
renormalization constant $\delta Z_W$ cancels in the expression for $\Delta r$. 

We employ the on-shell renormalization scheme. The one-loop renormalization constants of the $W$ and $Z$ boson masses are then given by
\BE
M^2_{W/Z,0}=M^2_{W/Z}+\delta M^2_{W/Z}\;,\;\;\; \delta M^{2}_{W/Z}  = \text{Re} \,\Sigma_{T}^{WW/ZZ}(M^{2}_{W/Z}),
\label{eq:renmw}
\EE
where bare quantities are denoted with a zero subscript.
The renormalization constant of the electric charge is 
\begin{equation}
e_0=(1+\delta e)e\;,\;\;\;\delta e
=\frac{1}{2}\Pi^{AA}(0)+\frac{\sw}{\cw}\frac{\Sigma_{T}^{AZ}(0)}{M^{2}_{Z}}\;,
 \label{eq:ladungsren}
\end{equation}
with
\begin{equation}
\Pi^{AA}(k^2)=\frac{\Sigma_{T}^{AA}(k^{2})}{k^2}\;,\;\;\; \Pi^{AA}(0)=\frac{\partial\Sigma_{T}^{AA}(k^{2})}{\partial k^{2}}\vert_{k^{2}=0}.
 \label{eq:pi2}
\end{equation}
Note that the sign appearing in front of $\sw$ in \refeq{eq:ladungsren} 
depends on convention chosen for the $SU(2)$ covariant
derivative.\footnote{We adopt the sign conventions for the $SU(2)_L$
covariant
derivative used in the code {\tt FeynArts}~\cite{Kublbeck:1990xc,Denner:1992me,Denner:1992vza,Kublbeck:1992mt,Hahn:2000kx,Hahn:2001rv}, where (for historical reasons) the $SU(2)_L$ covariant
derivative is defined by $\partial_{\mu} - \mathrm{i} g_2 I^a W^a_{\mu}$ for the SM and $\partial_{\mu} + \mathrm{i} g_2 I^a W^a_{\mu}$ for the (N)MSSM. The expressions given here correspond to the (N)MSSM convention.}
The sine of the weak mixing angle is not an independent parameter in the on-shell renormalization scheme. 
Its renormalization constant 
\begin{equation}
s_{\text{w},0}=\sw+\delta \sw\;,\;\;\;
\frac{\delta \sw}{\sw}  =  -\frac{1}{2}\frac{\cw^{2}}{\sw^{2}}\text{Re}\,\left(\frac{\Sigma ^{WW}_{T}(M_{W}^{2})}{M_{W}^{2}}-\frac{\Sigma ^{ZZ}_{T}(M_{Z}^{2})}{M_{Z}^{2}}\right)
\label{eq:cwswct}
\end{equation}
is fixed by the renormalization constants of the weak gauge boson masses.

Finally, the renormalization constant of a (left-handed) lepton field $l$ (neglecting the lepton mass) is
\BE
l_{0}^{L}  =  \left(1+\frac{1}{2}\delta Z^{l,L}\right)  l^{L}\;,\;\;\; \delta Z^{l,L}  =  -\Sigma^{l}_{L}(0) \,,
\label{eq:renlep}
\EE
where $\Sigma^{l}_{L}$ denotes the left-handed part of the lepton self energy.

Inserting these expressions for the renormalisation constants into \refeq{herleitungdeltar} yields
\begin{equation}
\begin{split}
\Delta r^{(\alpha)} = &\frac{\Sigma_T^{WW}(0)-\text{Re} \left[\Sigma_T^{WW}(\MW^2)\right]}
                 {\MW^2}+\Pi^{AA}\left(0\right)
-\frac{\cw^2}{\sw^2}\,  \text{Re} 
\left[\frac{\Sigma^{ZZ}_T(\MZ^2)}{\MZ^2}
     -\frac{\Sigma^{WW}_T(\MW^2)}{\MW^2}\right]\\
&+ 2 \;\frac{\sw}{\cw} \frac{\Sigma^{AZ}_T(0)}{\MZ^2} 
+\text{Vertex}+\text{Box}-
\frac{1}{2} 
\left( \Sigma^{e}_L(0) +\Sigma^{\mu}_L(0)
+\Sigma^{\nu_{e}}_L(0) +\Sigma^{\nu_{\mu}}_L(0) \right).
\end{split}    
\label{eq:deltar1LOS}
\end{equation} 
The quantity $\Delta r$ is at one loop level conventionally split into three
parts,
\begin{equation}
\Delta r^{(\al)}=\Delta \alpha-\frac{\cw^2}{\sw^2} \Delta \rho + 
\Delta r_{\text{rem}}.
\label{eq:deltarSM1L2}
\end{equation} 
The shift of the fine structure constant $\Delta \alpha$
arises from the charge renormalization
which contains the contributions from light fermions.
The quantity
$\Delta \rho$ contains loop corrections to the $\rho$
parameter~\cite{Veltman:1977kh}, 
which describes the ratio between neutral and charged weak currents, and can be written as
\begin{equation}
\Delta \rho = \frac{\Sigma^{ZZ}_T (0)}{M^2_Z}- 
\frac{\Sigma^{WW}_T (0)}{M^2_W}\;.
 \label{eq:deltarho}
\end{equation} 
This quantity is sensitive to the mass splitting between the isospin partners in a doublet~\cite{Veltman:1977kh},
which leads to a sizeable effect in the SM
in particular from the heavy quark doublet.
While $\Delta \alpha$ is a pure SM contribution, $\Delta \rho$ can
get large contributions also from SUSY particles, in particular 
the superpartners of the heavy quarks.
All other terms, both of SM and SUSY type, are contained in the remainder term $\Delta r_{\text{rem}}$.

We have performed a diagrammatic one-loop calculation of $\Dr$ in the 
NMSSM according to \refeq{eq:deltar1LOS}, using the {\tt Mathematica}-based programs {\tt FeynArts}~\cite{Kublbeck:1990xc,Denner:1992me,Denner:1992vza,Kublbeck:1992mt,Hahn:2000kx,Hahn:2001rv} and {\tt FormCalc}~\cite{Hahn:1998yk}. The NMSSM model file for  {\tt FeynArts}, first used in~\cite{Benbrik:2012rm}, has been adapted from output from {\tt SARAH}~\cite{Staub:2011dp,Staub:2008uz}.

The calculation of the SM-type diagrams (being part of the NMSSM contributions) are not discussed here.
This calculation has been discussed in the literature already many years ago~\cite{Sirlin:1980nh,Marciano:1980pb}, and 
we refer to~\citeres{Freitas:2002ja,Zeune:2011yla} for details. The one-loop
result for $\Delta r$ is also known for the MSSM (with
complex parameters), see \citeres{Heinemeyer:2006px,Heinemeyer:2013dia}. The calculation in the NMSSM follows along the same lines as for the MSSM. However, the result gets modified from differences in the Higgs and the neutralino sectors. Below we outline the NMSSM one-loop contributions to $\Delta r$, for completeness also including the MSSM-type contributions.

\begin{figure}[h]
\centering
\includegraphics[height=0.17\columnwidth]{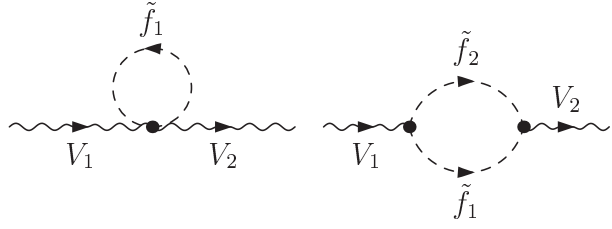}
\caption{Generic (N)MSSM one-loop gauge boson self-energy diagrams with a sfermion loop; 
$V_1\text{, }V_2=\gamma\text{, }Z\text{, }W^{\pm}$ and
$\tilde{f_1}\text{, }\tilde{f_2}=\tilde{\nu}\text{, }\tilde{l}\text{, }\tilde{u}\text{, }\tilde{d}$
.} 
\label{fig:Sfermion_GBSE}
\end{figure}

\begin{figure}[h]
\centering
\includegraphics[height=0.16\columnwidth]{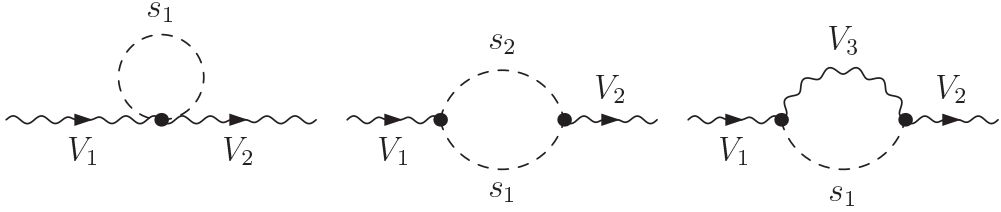}
\caption{Generic NMSSM one-loop gauge boson self-energy diagrams with gauge bosons, Higgs and Goldstone bosons in the loop; $V_1\text{, }V_2\text{, }V_3=\gamma\text{, }Z\text{, }W^{\pm}$ and $s_1\text{, }s_2 = h_1\text{, }h_2\text{, }h_3\text{, }a_1\text{, }a_2\text{, }H^{\pm}\text{, }G\text{, }G^{\pm}$.
 } 
\label{fig:Higgs_GBSE}
\end{figure}

Besides the SM-type contributions with fermions and gauge bosons in the loops (not discussed further here), many additional self-energy, vertex and box diagrams appear in the NMSSM with sfermions, (SUSY) Higgs bosons, charginos and neutralinos in the loop.
Generic examples of gauge-boson self-energy diagrams with sfermions are depicted in~\reffi{fig:Sfermion_GBSE}.
Their contribution to $\Delta r$ is finite by itself.  
The NMSSM Higgs bosons enter only in gauge boson self-energy diagrams, since we have neglected the masses of the external fermions. The contributing diagrams are sketched in~\reffi{fig:Higgs_GBSE}.
These contributions are not finite by themselves. Only if one considers all
(including SM-type) gauge boson and Higgs contributions to the gauge
boson self-energy diagrams, the vertex diagrams and vertex counterterm
diagrams together, the divergences cancel, and one finds a finite result.

\begin{figure}[h]
\centering
\includegraphics[height=0.15\columnwidth]{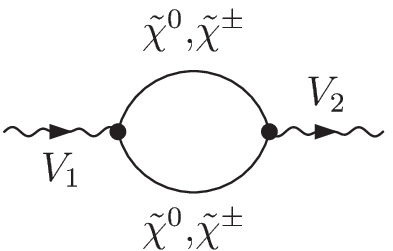}
\caption{Generic NMSSM one-loop gauge boson self-energy diagram with charginos/neutralinos; $V_1\text{, }V_2=\gamma\text{, }Z\text{, }W^{\pm}\text{, }\tilde{\chi}^{\pm}=\tilde{\chi}^{\pm}_{1,2}$ and $\tilde{\chi}^0=\tilde{\chi}^0_{1,2,3,4,5}$.
}
\label{fig:ChaNeu_GBSE}
\end{figure}

\begin{figure}[h]
\centering
\includegraphics[height=0.16\columnwidth]{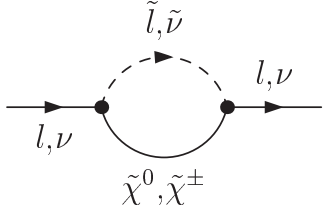}
\caption{Generic NMSSM one-loop fermion self-energy diagram with a chargino/neutralino/sfermion contribution; $\tilde{\chi}^{\pm}=\tilde{\chi}^{\pm}_{1,2}$ and $\tilde{\chi}^0=\tilde{\chi}^0_{1,2,3,4,5}$.
} 
\label{fig:ChaNeu_FSE}
\end{figure}

\begin{figure}[h]
\centering
\includegraphics[height=0.18\columnwidth]{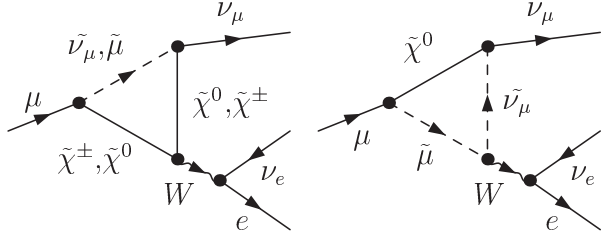}
\caption{Generic one-loop vertex correction diagrams in the NMSSM; $\tilde{\chi}^{\pm}=\tilde{\chi}^{\pm}_{1,2}$ and $\tilde{\chi}^0=\tilde{\chi}^0_{1,2,3,4,5}$. Analoguous diagrams exist for the other $W$ vertex.} 
\label{fig:vertex}
\end{figure}

\begin{figure}[h]
\centering
\includegraphics[height=0.19\columnwidth]{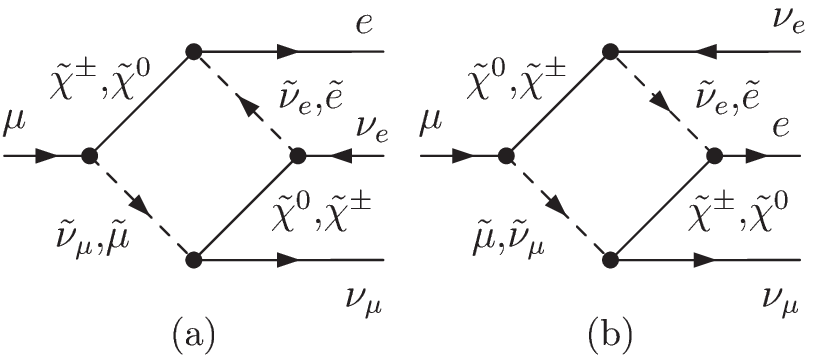}
\caption{Generic one-loop box diagrams contributing to the muon decay amplitude in the NMSSM; $\tilde{\chi}^{\pm}=\tilde{\chi}^{\pm}_{1,2}$ and $\tilde{\chi}^0=\tilde{\chi}^0_{1,2,3,4,5}$.
} 
\label{fig:box}
\end{figure}

Charginos and neutralinos enter in gauge boson self-energy diagrams (depicted in \reffi{fig:ChaNeu_GBSE}), fermion self-energy diagrams (depicted in \reffi{fig:ChaNeu_FSE}), vertex diagrams (depicted in \reffi{fig:vertex}, the analogous vertex corrections exist also for the other vertex) and box diagrams (depicted in \reffi{fig:box}). 
The vertex contributions from the chargino/neutralino sector, together with the chargino/neutralino contributions to the vertex CT
and the gauge boson self-energies are finite. Each box diagram is UV-finite by itself.

In order to determine
the contribution to $\Delta r$ from a particular loop diagram, 
the Born amplitude has to be factored out of the one-loop muon decay amplitude, as shown in~\refeq{eq:drborn}. 
While most loop diagrams directly give a result proportional to the Born
amplitude, more complicated spinor structures that do not occur in
the SM case arise from box diagrams containing neutralinos and
charginos\footnote{The same complication occurs in the MSSM and was discussed in \citere{Heinemeyer:2006px}.}. 
Performing the calculation of the box diagrams in~\reffi{fig:box} in {\tt FormCalc}, the spinor chains are returned in the form
\begin{equation}
\begin{split}
& {\cal M}_{\text{SUSY Box} (a)} = (\bar{u}_e \gamma_{\lambda} \omega_{-} u_{\mu})(\bar{u}_{\nu_{\mu}} \gamma^{\lambda} \omega_{-} v_{\nu_e}) b_{(a)} \\
& {\cal M}_{\text{SUSY Box} (b)} = (\bar{u}_{\nu_e} \omega_{-} u_{\mu})(\bar{u}_{\nu_{\mu}} \omega_{+} v_e) b_{(b)}\;.
\end{split}
 \label{eq:susybox1}
\end{equation}
The expressions for the coefficients $b_{(a)}$ and $b_{(b)}$ are lengthy and not given here explicitly.
In order to factor out the Born amplitude 
\begin{equation}
{\cal M}_{\text{Born}}=\frac{2 \pi \alpha}{\sw^2 M_W^2}\left(\bar{u}_{\nu_{\mu}}\gamma_{\lambda}\omega_{-}u_{\mu}\right)\left(\bar{u}_e\gamma^{\lambda}\omega_{-}v_{\nu_{e}}\right)\,,
\end{equation} 
the spinor chains in \refeq{eq:susybox1} have to be transformed into the same structure.
We modify the spinor chains following the procedure described in \citere{Heinemeyer:2006px} and get
 \begin{equation}
\begin{split}
& {\cal M}_{\text{SUSY Box} (a)} =  -\frac{\sw^2 M_W^2}{2 \pi \alpha}\, b_{(a)}\, {\cal M}_{\text{Born}} \\
& {\cal M}_{\text{SUSY Box} (b)} =  \frac{\sw^2 M_W^2}{4 \pi \alpha}\, b_{(b)}\, {\cal M}_{\text{Born}}\;.
\end{split}
 \label{eq:susybox2}
\end{equation}
The coefficients $b_{(a)}$ and $b_{(b)}$ contain ratios of mass-squared differences of the
involved particles. These coefficients can give rise to numerical
instabilities in cases of mass degeneracies. In the implementation of
our results (which has been carried out in a {\tt Mathematica}) 
special care has been taken of such parameter regions
with mass degeneracies or possible threshold effects. By adding appropriate expansions  a 
numerically stable evaluation is ensured.

\subsection{Higher-order corrections}\label{sec:Fulldeltar}

The on-shell renormalization conditions correspond to the
definition of the $W$ and $Z$ boson masses according to the real part
of the complex pole of the propagator, which from two-loop order on is the
only gauge-invariant way to define the masses of unstable
particles (see \citere{Freitas:2002ja} and references therein). 
The expansion around the complex pole results in a 
Breit-Wigner shape with a fixed width (fw). 
Internally we therefore use this definition (fw) of the gauge boson masses.
The experimentally measured values of the gauge boson masses
are obtained using a mass definition in
terms of a Breit-Wigner shape with a running width (rw).
As the last step of our calculation, we therefore transform the $W$ boson
mass value to the running width definition, $M_{W}^{\text{rw}}$ to
facilitate a direct comparison to the experimental value of $\MW$.
The difference between these two definitions is
\begin{equation}
\begin{split}
& M_{W}^{\text{rw}}=M_{W}^{\text{fw}}+\frac{\Gamma^2_{W}}{2 M_{W}^{\text{rw}}},
 \label{eq:masssdiff2}
\end{split}
\end{equation}
where $M_{W}^{\text{fw}}$ corresponds to the fixed width description, see~\citere{Bardin:1988xt}. For the prediction of the $W$ decay width we use
\begin{equation}
\begin{split}
&\Gamma_{W}=\frac{3 G_{\mu} \left(M_W^{\text{rw}}\right)^3}{2 \sqrt{2}\pi}\left(1+\frac{2 \alpha_s}{3 \pi}\right)\,,
 \label{eq:wbosonwidth}
\end{split}
\end{equation}
parametrized by $G_{\mu}$ and including first order QCD corrections.
The difference between the fixed- and running width definitions amounts
to about $27 \mev$, which is very relevant in view of the current
theoretical and experimental precisions. For the $Z$ boson mass, which is
used as an input parameter in the prediction for $\MW$, 
the conversion from
the running-width to the fixed-width
definition is carried out in the first
step of the calculation. Accordingly, keeping track of the
proper definition of the gauge boson masses is obviously important in the
context of electroweak precision physics.
For the remainder of this paper we will not use the labels ($\text{rw}$,
$\text{fw}$) explicitly; if we do not refer to an internal variable, $\MW$
and $\MZ$
will always refer to the mass definition according to a 
Breit-Wigner shape with a \emph{running} width
(see e.g.~\citere{Freitas:2002ja} for further details;
see also \citeres{Martin:2015lxa,Martin:2015rea}).

We combine the SM one-loop result (which is part of the NMSSM calculation)
with the relevant available higher-order corrections 
of the state-of-the-art prediction
for $M_W^{\text{SM}}$.
As we will describe below in more detail, the higher-order corrections
of SM-type are also incorporated in the NMSSM calculation of $\Delta r$
in order to achieve an accurate prediction for 
$\MW^{\rm NMSSM}$.
For a discussion of the incorporation of higher-order contributions to $\MW$ in the MSSM see~\citeres{Heinemeyer:2013dia,Heinemeyer:2006px}. 

In a first step, we write the NMSSM result for $\Delta r$ 
as the sum of the full
one-loop and the higher-order corrections,
\begin{equation}
\begin{split}
&\Delta r^{\text{NMSSM}}=\Delta r^{\text{NMSSM}(\alpha)}+\Delta r^{\text{NMSSM(h.o.)}},
\end{split}    
\label{eq:deltarsplit1l}
\end{equation}
where $\Delta r^{\text{NMSSM}(\alpha)}$ denotes the NMSSM one-loop contributions 
from the various particle sectors 
\begin{equation}
\begin{split}
\Delta r^{\text{NMSSM}(\alpha)} = \Delta r^{\text{fermion}(\alpha)} + \Delta r^{\text{gauge-boson/Higgs}(\alpha)} +
\Delta r^{\text{sfermion}(\alpha)} + \Delta r^{\text{chargino/neutralino}(\alpha)}
\,,
\end{split}    
\label{eq:deltar1lterms}
\end{equation}
as discussed in the previous subsection.
The term $\Delta r^{\text{NMSSM(h.o.)}}$ denotes the higher-oder
contributions,
which we split into a SM part and a SUSY part,
\begin{equation}
\begin{split}
\Delta r^{\text{NMSSM(h.o.)}} = \Delta r^{\text{SM(h.o.)}}+\Delta
r^{\text{SUSY(h.o.)}}\;.
\end{split}    
\label{eq:deltarsplit}
\end{equation}
The terms $\Delta r^{\text{SM/SUSY(h.o.)}}$ describe the SM/SUSY contributions
beyond one-loop order. For $\Delta r^{\text{SM(h.o.)}}$ we employ the most
up-to-date SM result including all relevant higher-order corrections, while 
$\Delta r^{\text{SUSY(h.o.)}}$ contains the most up-to-date higher-order
contributions of SUSY type (see below). 
The approach followed in \refeq{eq:deltarsplit} has several advantages. It
ensures in particular that the best available SM prediction is recovered in
the decoupling limit, where all
superpartners are heavy, the singlet decouples, and the NMSSM Higgs sector 
becomes SM-like. Furthermore, the approach to combine the most
up-to-date SM prediction with additional ``new physics'' contributions (here from
supersymmetry) allows one to readily compare the MSSM and NMSSM predictions
on an equal footing and it provides an appropriate framework also for an 
extension to other scenarios of physics beyond the SM.

The expression for the higher-order contributions given 
in \refeq{eq:deltarsplit} formally
introduces a dependence of the NMSSM result on the SM Higgs mass, which
enters from the two-loop electroweak corrections onwards.
In those contributions we identify the SM Higgs mass 
with the mass of the NMSSM Higgs boson with the largest coupling to
gauge bosons.

For the higher-order corrections in the SM, $\Delta r^{\text{SM(h.o.)}}$, we
incorporate the following contributions up to the four-loop order
\begin{equation}
\begin{split}
 \Delta r^{\text{SM(h.o.)}}=&\Delta r^{(\alpha \alpha_s)}
+\Delta r^{(\alpha \alpha_s^2)}+\Delta r_{\text{ferm}}^{(\alpha^2)}+\Delta r_{\text{bos}}^{(\alpha^2)}\\
 &+\Delta r^{(G_{\mu}^2 \alpha_s \mt^4)}+\Delta r^{(G_{\mu}^3 \mt^6)}+\Delta r^{(G_{\mu} \mt^2 \alpha_s^3)}\;.
\end{split}
 \label{eq:SMhighercont}
\end{equation} 
The contributions in \refeq{eq:SMhighercont} consist of
the two-loop QCD corrections $\Delta r^{(\alpha \alpha_s)}$ 
\cite{qcd2SMa,qcd2SMb,qcd2SMc,qcd2SMd,qcd2SMe,qcd2SMf},
the three-loop QCD corrections $\Delta r^{(\alpha \alpha_s^2)}$ 
\cite{Avdeev:1994db,qcd3SMa,qcd3SMb,Chetyrkin:1996cf},
the fermionic electroweak two-loop corrections $\Delta r_{\text{ferm}}^{(\alpha^2)}$ \cite{Freitas:2000gg,Freitas:2002ja,2lfermc},
the purely bosonic electroweak two-loop corrections $\Delta r_{\text{bos}}^{(\alpha^2)}$ \cite{2lbos,2Lbosa,2Lbosb},
the mixed QCD and electroweak three-loop contributions $\Delta r^{(G_{\mu}^2 \alpha_s \mt^4)}$ \cite{Faisst:2003px,vanderBij:2000cg}, 
the purely electroweak three-loop contribution $\Delta r^{(G_{\mu}^3 \mt^6)}$ \cite{Faisst:2003px,vanderBij:2000cg}, 
and finally the four-loop QCD correction $\Delta r^{(G_{\mu} \mt^2 \alpha_s^3)}$ \cite{Boughezal:2006xk,Chetyrkin:2006bj,Schroder:2005db}. 

The radiative corrections in the SM beyond one-loop level are numerically
significant and lead to a large downward shift in $M_W$ by more than $100
\mev$.\footnote{The corrections beyond one-loop order are in fact
crucial for the important result that the $M_W$ prediction in the SM
favours a light Higgs boson, whereas the one-loop result alone would favour
a heavy SM Higgs.} 
The largest shift (beyond one-loop) is caused by the two-loop QCD
corrections
\cite{qcd2SMa,qcd2SMb,qcd2SMc,qcd2SMd,qcd2SMe,qcd2SMf,StremplatDiplom} 
followed by the three-loop QCD corrections $\Delta r^{(\alpha \alpha_s^2)}$ 
\cite{Avdeev:1994db,qcd3SMa,qcd3SMb,Chetyrkin:1996cf}. 

Most of the higher-order contributions are known analytically (and we include them in this form), except for
the full electroweak two-loop contributions in the SM which involve numerical
integrations of the two-loop scalar integrals.
These contributions are included in our calculation using a simple parametrization formula given in \cite{Awramik:2006uz}.\footnote{
In \cite{Awramik:2006uz} the electroweak two-loop contributions 
are expressed via
$\Delta r^{(\alpha^2)} \equiv \Delta r_{\text{ferm}}^{(\alpha^2)} + \Delta r_{\text{bos}}^{(\alpha^2)} = (\Delta\alpha)^2 +  2\Delta\alpha \, \Delta  \tilde{r}^{(\alpha)} + \Delta
r^{(\alpha^2)}_{\rm rem}$,
where a simple fit formula for the remainder term  $\Delta
r^{(\alpha^2)}_{\rm rem}$ is given.
The quantity
$\Delta \tilde{r}^{(\alpha)}$  in the second term denotes the full one-loop result without the $\Delta \alpha$ term.
}
This fit formula gives a good approximation to the full result for a light SM Higgs (the agreement is better than $0.4 \mev$ for $\MW$) \cite{Awramik:2006uz}. 
Using this parametrization directly for the SM prediction of 
$\Delta r_{\text{ferm}}^{(\alpha^2)} + \Delta r_{\text{bos}}^{(\alpha^2)}$
(rather than for the full SM prediction of $\MW$ --- an approach followed
for the MSSM case in \citere{Heinemeyer:2006px} and for the NMSSM case
in \citere{Domingo:2011uf}) allows us to evaluate these contributions 
at the particular
NMSSM value for $M_W$ in each iteration step.
The output of this formula approximates the full result of $\Delta r_{\text{ferm}}^{(\alpha^2)} + \Delta
r_{\text{bos}}^{(\alpha^2)}$ using the fixed-width definition, such that it
can directly be combined with other terms of our
calculation.\footnote{It should be noted, however, that the gauge boson masses with running width definition are needed as input for the fit formula given in~\cite{Awramik:2006uz}. This is the only part of our calculation where the running width definition is used internally.}
In our expression for $\Delta r^{\text{SM(h.o.)}}$ we use the
result for
$\Delta r^{(\alpha \alpha_s)}$ given in \citere{qcd2SMc}, which contains
also contributions from quarks of the first two
generations and is numerically very close to the result from 
\citere{StremplatDiplom},\footnote{This is an improvement compared to
\citere{Heinemeyer:2013dia}, where the two-loop QCD
contributions from \citere{qcd3SMb} were employed, which include only the
top und bottom quark contributions.} and the result for 
$\Delta r^{(\alpha \alpha_s^2)}$ from \citere{qcd3SMb}.
Both contributions are parametrized in terms of $G_{\mu}$. A comparison between our evaluation of $M_W^{\rm SM}$ and the fit formula for $M_W^{\rm SM}$ given in \citere{Awramik:2003rn} can be found in \refse{subsec:SMho} below.

For the higher-order corrections of SUSY type, see \refeq{eq:deltarsplit}, 
we take the following contributions into account,
\begin{equation}
\begin{split}
\Delta r^{\text{SUSY(h.o.)}} = 
\Delta r_{\text{red}}^{\rm SUSY (\alpha^2)} 
- \frac{\cw^2}{\sw^2} \Delta \rho^{\text{SUSY},(\alpha \alpha_s)}
- \frac{\cw^2}{\sw^2} \Delta \rho^{\text{SUSY}, (\alpha_t^2,\alpha_t \alpha_b,\alpha_b^2)} 
\;,
\end{split}    
\label{eq:deltarsusyho}
\end{equation}
incorporating all SUSY corrections beyond one-loop
order that are known to date.
The first term in \refeq{eq:deltarsusyho} denotes the 
leading reducible ${\cal O}(\alpha^2)$ two-loop corrections. Those
contributions are obtained by
expanding the resummation formula~\cite{Consoli:1989fg}
\begin{equation}
1+\Delta r =\frac{1}{(1-\Delta \alpha)\left(1+\frac{\cw^2}{\sw^2} \Delta \rho\right)-\Delta r_{\rm{rem}}}\;,
\label{resummation}
\end{equation}
which correctly takes terms of the type $(\Delta \alpha)^2$, $(\Delta \rho)^2$ and $\Delta \alpha \Delta \rho$ into account\footnote{In principle one could also include the term $\Delta \alpha \Delta r_{\text{rem}}$, which however is numerically small.}  
if $\Delta \rho$ is parametrized by $G_{\mu}$. 
The pure SM terms are already included in $\Delta r^{\rm SM(h.o.)}$,
and because of numerical compensations those contributions are small beyond
two-loop order~\cite{StremplatDiplom}.
Thus, we only need to consider the leading two-loop terms with SUSY
contributions,
\begin{equation}
\begin{split}
\Delta r_{\text{red}}^{\rm SUSY (\alpha^2)} = &-\frac{\cw^2}{\sw^2} \Delta \alpha \Delta \rho^{\text{SUSY}}
 + \frac{\cw^4}{\sw^4}( \Delta {\rho^{\text{SUSY}}})^2+ 2 \frac{\cw^4}{\sw^4}\Delta \rho^{\text{SUSY}}\Delta \rho^{\rm SM}.
\end{split}    
\label{deltar1LOS}
\end{equation}

\begin{figure}
\centering
\includegraphics[height=0.13\columnwidth]{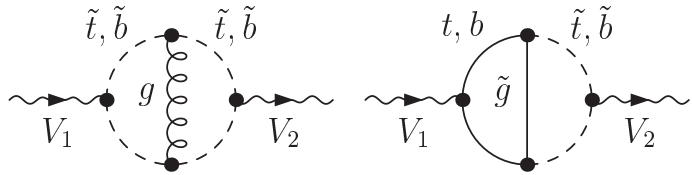}
\caption{Generic ${\cal O}(\alpha \alpha_s)$ two-loop self-energy diagrams in the (N)MSSM. Here
$g$ denotes a gluon and $\tilde{g}$ a gluino; 
$V_1\text{,}V_2=\gamma\text{,}Z\text{,}W^{\pm}$.} 
\label{fig:GG2L}
\end{figure}

The other two terms in \refeq{eq:deltarsusyho} denote irreducible
two-loop SUSY contributions.
The two-loop ${\cal O}(\alpha \alpha_s)$ SUSY
contributions~\cite{Djouadi:1996pa,Djouadi:1998sq}, 
$\Delta \rho^{\text{SUSY},(\alpha \alpha_s)}$,
contain squark
loops with gluon exchange and quark/squark loops with gluino exchange (both
depicted in~\reffi{fig:GG2L}).
While the formula for the gluino contributions is very lengthy, a compact
result exists for the gluon contributions to 
$\Delta \rho$ \cite{Djouadi:1996pa,Djouadi:1998sq}. 
Using these two-loop results for the SUSY contributions to $M_W$ requires the on-shell (physical) values for the squark masses as input.
The $SU(2)$ relation $M_{\tilde t_L}=M_{\tilde b_L}$ implies that one of the stop/sbottom masses is not independent, but can be expressed in terms of the other parameters. 
Therefore, when including higher-order contributions, one cannot choose independent renormalization conditions for all four (stop and sbottom) masses.
Loop corrections to the relation between the squark masses must be taken
into account in order to be able to insert the proper on-shell values
for the squark masses into our calculation.
This one-loop correction to the relation between the squark masses is
relevant when inserted into the one-loop SUSY contributions to $\MW$, while
it is of higher order for the two-loop SUSY contributions.
In our evaluation of $\MW$ this is taken into account by a 
``mass-shift'' correction term. For more details see \citere{Djouadi:1998sq}.
The gluon, gluino and mass-shift corrections, which are identical
in the MSSM and the NMSSM, are included in our NMSSM result for $\MW$.

\begin{figure}
\centering
\includegraphics[height=0.13\columnwidth]{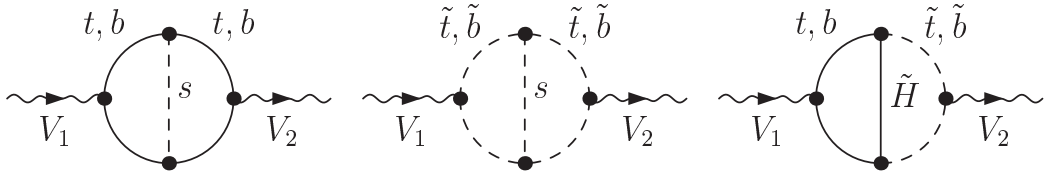}
\caption{Generic ${\cal O}(\alpha_t^2+\alpha_t \alpha_b+\alpha_b^2)$ two-loop diagrams, where
$\tilde{H}$ denotes a Higgsino and is $s$ either an NMSSM Higgs or a Goldstone boson; 
$V_1\text{,}V_2=\gamma\text{,}Z\text{,}W^{\pm}$.} 
\label{fig:Higgsino}
\end{figure}

The third term in \refeq{eq:deltarsusyho} denotes 
the dominant Yukawa-enhanced electroweak two-loop corrections to 
$\Delta \rho$ of 
${\cal O}(\alpha_t^2)$, ${\cal O}(\alpha_t \alpha_b)$ and 
${\cal O}(\alpha_b^2)$~\cite{Heinemeyer:2002jq,Haestier:2005ja}.
These contributions consist of heavy quark ($t/b$) loops with Higgs exchange, 
squark ($\tilde{t}/\tilde{b}$) loops with Higgs 
exchange, and mixed quark-squark loops with Higgsino exchange,
see~\reffi{fig:Higgsino}.
The corrections of this kind, which depend on the specific form of the
Higgs sector, are only known for the MSSM so 
far~\cite{Heinemeyer:2002jq,Haestier:2005ja}. It is nevertheless possible to
take them into account also for the NMSSM in an approximate form. To this
end, the considered NMSSM parameter point needs to be related to appropriate
parameter values of the MSSM.
Besides the values for $\tan \beta$, the sfermion trilinear couplings $A_f$, 
and all the soft mass parameters, which can be directly taken over from the 
parameter point in the  NMSSM, we determine the parameters in the following
way: we set the MSSM $\mu$ parameter equal to $\mu_{\text{eff}}$ and we use
the physical value of the charged Higgs mass as calculated in the NMSSM (see below) 
as input for the calculation of the MSSM Higgs masses. This prescription is
motivated by the fact that in this way the value of the
mass of the charged Higgs boson, which is the only Higgs boson
appearing without mixing in both models, is the same in the NMSSM and the
MSSM. The MSSM Higgs masses
are calculated with {\tt
FeynHiggs}~\cite{Hahn:2009zz,Frank:2006yh,Degrassi:2002fi,Heinemeyer:1998np,Heinemeyer:1998yj},
using the calculated physical mass value of $M_{H^{\pm}}$ as 
on-shell input
parameter. The MSSM Higgs masses and the Higgsino parameter $\mu$ determined
in this way
are then used
as input for the calculation of $\Delta \rho$ to 
${\cal O}(\alpha_t^2)$, ${\cal O}(\alpha_t \alpha_b)$, ${\cal O}(\alpha_b^2)$. 
In order
to avoid double-counting the dominant
Yukawa-enhanced electroweak two-loop corrections in the
SM~\cite{Barbieri:1992dq,Fleischer:1993ub} have been subtracted according to
\refeq{eq:deltarsplit}. 
We find that the impact of the Yukawa-enhanced electroweak corrections
of SUSY type on $\MW$ is relatively small (typically $\lesssim 1
\mev$), and their numerical evaluation is rather time-consuming.
In our numerical code for $\MW$ in the NMSSM, we therefore leave it as an option to choose whether these contributions should be included or not. For the results presented in~\refse{sec:numMWNMSSM} they have not been included, unless stated otherwise.

\section{Numerical results}\label{sec:mwsetup}

\subsection{Framework for the numerical analysis}\label{subsec:numsetup}

For the evaluation of the $W$ boson mass prediction, the masses of the NMSSM particles are needed. 
We use the NMSSM on-shell parameters as input to calculate the 
sfermion, chargino and neutralino masses.
For the calculation of the Higgs boson
masses we use  {\tt NMSSMTools}  (version
4.6.0)~\cite{Ellwanger:2004xm,Ellwanger:2005dv,Belanger:2005kh,Ellwanger:2006rn}.\footnote{
In two plots below the NMSSM Higgs boson masses at tree-level are used.
They are calculated using the tree-level relations given in \refse{sec:NMSSM}.}
For other tools that are available to calculate the NMSSM Higgs 
masses including higher-order radiative corrections see
\citeres{Baglio:2013iia,Goodsell:2015ira,Peternmssm}. The implementation of
the Higgs mass results of \citere{Peternmssm} (using directly the on-shell
parameters as input) is in progress.

In {\tt NMSSMTools} the input parameters are assumed to be $\overline{\rm{DR}}$ parameters at the SUSY breaking scale.
In order to use the {\tt NMSSMTools} Higgs masses in our result, a transformation from the on-shell parameters, needed for our evaluation, to the $\overline{\rm{DR}}$ 
parameters, needed as {\tt NMSSMTools} input, is necessary. 
This effect is approximately taken into account by transforming the on-shell
$X_t$ parameter into its $\overline{\rm{DR}}$ value by the relation given in
\citere{Williams:2011bu} (equations (60) ff in \citere{Williams:2011bu}).
The shift in the other parameters is significantly smaller and therefore neglected here.

We use a setup where the NMSSM parameter space can be tested against a broad set of 
experimental and theoretical constraints. 
Besides the constraints already implemented in {\tt NMSSMTools},\footnote{
{\tt NMSSMTools} contains a number of theoretical and experimental constraints, 
e.g.\ constraints from collider experiments (such as LEP mass limits on SUSY particles), $B$-physics and astrophysics. 
More details on the constraints included in {\tt NMSSMTools} can be found in~\citeres{Ellwanger:2004xm,nmssmtoolswebpage}.} 
further direct constraints on the Higgs sectors are evaluated using the code {\tt HiggsBounds}  (version 4.2.0)~\cite{Bechtle:2013wla,Bechtle:2013gu,Bechtle:2011sb,Bechtle:2008jh}.
All programs used for the numerical evaluation are linked through an interface to the NMSSM {\tt Mathematica} code for the $W$ boson mass prediction.\footnote{The {\tt Mathematica} code is linked to a {\tt Fortran} driver 
program, calling {\tt NMSSMTools} and {\tt HiggsBounds}. 
The calculation of the SUSY particle masses is also included in the {\tt Fortran} driver. 
Similarly to the MSSM case \cite{Heinemeyer:2013dia}, we plan to additionally implement the $M_W$ calculation directly into {\tt Fortran} 
in order to increase the speed of the $M_W$ evaluation. This will be useful in particular for large scans of the parameter space.}

\subsection{Theoretical uncertainties}

Before moving on to our numerical results for the $W$ boson mass prediction 
in the NMSSM, we discuss the remaining
theoretical uncertainties in the $M_W$ calculation.

The dominant theoretical uncertainty of the prediction for $\MW$ 
arises from the parametric uncertainty induced by
the experimental error 
of the top-quark mass. Here one needs to take into account both the
experimental error of the actual measurement and the systematic uncertainty
associated with relating the experimentally determined quantity to a
theoretically well-defined mass parameter, see the discussion above. A total
experimental error of 1 GeV on $\mt$ causes a parametric 
uncertainty on $\MW$ of about $6 \mev$, while the parametric
uncertainties induced by the current
experimental error of the hadronic contribution to the shift in the
fine-structure constant, $\De\al_{\rm had}$, and by the experimental
error of $\MZ$ amount to about $2\mev$ and $2.5\mev$, respectively.
The uncertainty of the SM $\MW$ prediction caused by the experimental 
error of the Higgs boson 
mass, $\de\MH^{\rm exp} = 0.24 \gev$ \cite{Aad:2015zhl}, is
significantly smaller ($\lesssim 0.2 \mev$).
In \citere{Baak:2013fwa} the impact of improved accuracies of 
$\mt$ and $\De\al_{\rm had}$ has been discussed.
With a precise top mass measurement of $\De m_t = 0.1 \gev$ 
(anticipated ILC precision) the associated parametric uncertainty in $M_W$ 
is about $0.6 \mev$.

The uncertainties from unknown higher-order corrections have
been estimated to be around $4$~MeV in the SM for a light Higgs
boson ($\MHSM < 300 \gev$)~\cite{Awramik:2003rn}.  
The prediction for $\MW$ in the NMSSM is affected by additional
theoretical uncertainties from unknown higher-order corrections of SUSY
type. While in the decoupling limit those additional uncertainties
vanish, they can be important if some SUSY particles, in particular in
the scalar top and bottom sectors, are relatively light. The combined
theoretical uncertainty from unknown higher-order corrections of SM- and
SUSY-type has been estimated (for the MSSM with real parameters) in 
\citeres{Haestier:2005ja,Heinemeyer:2006px} as
$\delta \MW \sim (4-9)$~MeV, depending on the SUSY mass scale.\footnote{
The lower limit of 4 MeV corresponds to the SM uncertainty, which
applies to the decoupling limit of the MSSM.
For the upper limit of 9 MeV very light SUSY particles were considered.
In view of the latest experimental bounds from the SUSY searches at the
LHC, the (maximal) uncertainty
from missing higher orders is expected to be somewhat smaller than 9 MeV.}
Since we include the same SUSY higher-order corrections in our NMSSM calculation
as were considered for the uncertainty estimate in the MSSM, the uncertainty from unknown higher-order corrections is estimated to be of similar size.

\subsection{SM higher-order corrections}\label{subsec:SMho}

We compare our evaluation of $M_W^{\rm SM}$ to the result from the fit formula for $M_W^{\rm SM}$ given in \citere{Awramik:2003rn}. 
In the latest version of \citere{Awramik:2003rn} all the corrections of \refeq{eq:SMhighercont} are included. 
The $\MW$ fit formula incorporates the  
${\cal O}({\alpha \alpha_s})$ from \citere{StremplatDiplom}, whereas we use
the ${\cal O}({\alpha \alpha_s})$ from \citere{qcd2SMc}. These results
are in good numerical agreement with each other if in both cases the
electric charge is parametrized in terms of the fine structure constant
$\alpha$. The ${\cal O}({\alpha^2 \alpha_s})$ three-loop corrections
included in \refeq{eq:SMhighercont} are parametrized in terms of $G_{\mu}$.
We therefore choose to parametrize the ${\cal O}({\alpha \alpha_s})$
contributions also in terms of $G_{\mu}$. 
The difference between the $G_{\mu}$ parametrization of the QCD
two-loop corrections that we use here and the $\alpha$ parametrization used in
\citere{Awramik:2003rn} leads 
to a prediction for $\MW^{\rm SM}$ that is $\sim 2 \mev$ lower than the 
result 
given in \citere{Awramik:2003rn}.

The numerical values of the different SM-type
contributions to $\Delta r$ are given in \refta{tab:delrcontribs} for $\MW =80.385$~GeV and $M_H^{\rm SM}= 125.09$~GeV. 
The other relevant input parameters that we
use are
\begin{eqnarray}
&& \mt = 173.34 \gev, \quad \mb = 4.7 \gev, \quad \MZ = 91.1876 \gev, \quad
\Ga_{\rm{Z}} = 2.4952 \gev, \non \\
&& \De\al_{\rm lept} =0.031497686, \quad \De\al^{(5)}_{\rm had} =0.02757,\quad  \al^{-1} = 137.035999074, \non \\
&& \alpha_s(\MZ) = 0.1184, \quad G_{\mu} = 1.1663787 \times 10^{-5} \gev^{-2}.
\label{eq:inputs}
\end{eqnarray}
As explained above, the values for the $W$ and $Z$ boson 
masses given above, which correspond to a Breit-Wigner shape with running
width, have been transformed internally to the definition of a 
Breit-Wigner shape with fixed width associated with
the real part of the complex pole.
\begin{table}[t]
\centering
\ra{1.5}
\small
\begin{tabular}{c|c|c|c|c|c}
$\De r^{(\al)}$ & 
$\De r^{(\al\alpha_s)}$ & 
$\De r^{(\al\alpha_s^2)}$ &
$\De r^{(\al^2)}_{\rm ferm} + \De r^{(\al^2)}_{\rm bos}$  &
$\De r^{(G_{\mu}^2 \alpha_s m_t^4)} + \De r^{(G_{\mu}^3 m_t^6)}$ &
$\De r^{(G_{\mu} \mt^2 \alpha_s^3)}$\\
\hline
297.17 & 36.28 & 7.03 & 29.14  & -1.60 & 1.23 \\
\end{tabular}
\caption{The numerical values ($\times 10^4$) of the different 
contributions to $\De r$ specified in \refeq{eq:SMhighercont} are given 
for $\MW =80.385$~GeV and $M_H^{\rm SM}= 125.09$~GeV. 
\label{tab:delrcontribs}}
\end{table}

\subsection{Results for the $M_W$ prediction in the NMSSM}\label{sec:numMWNMSSM}

We now turn to the discussion of the prediction for $M_W$ in the NMSSM.
Our evaluation has been carried out for the case of real parameters, consequently for all parameters given in this section the phases are set to zero and will not be listed as separate 
input parameters.

An earlier result for $\MW$ in the NMSSM was presented in
\citere{Domingo:2011uf}.
Concerning SUSY two-loop contributions, 
in this result only the part of the contributions to
$\Delta \rho^{\text{SUSY},(\alpha \alpha_s)}$, see \refeq{eq:deltarsusyho},
arising from squark loops with gluon exchange is taken into account. 
As we will show below in the discussion of our improved result for 
$\MW$ in the NMSSM, the two-loop contributions that have been neglected in  
\citere{Domingo:2011uf} can have a sizeable impact. A further improvement of
our results for the MSSM and the NMSSM is that they are based on contributions to 
$\Delta r$ that can all be evaluated at the correct input value for $\MW$
(using an iterative procedure), i.e.\ $\MW^{\rm(N)MSSM}$, while 
the evaluation in \citere{Domingo:2011uf} makes use of the fitting formula for 
$\MW^{\rm SM}$~\cite{Awramik:2003rn}. The corresponding contribution to
$\Delta r$ extracted from the fitting formula for $\MW^{\rm SM}$ is
determined at the input value $\MW^{\rm SM}$ rather than $\MW^{\rm(N)MSSM}$,
while it is the latter that is actually needed for the evaluation in the
(N)MSSM (see \citere{Heinemeyer:2006px} for a discussion how to remedy this
effect). We have compared our result with the one given
in \citere{Domingo:2011uf}%
\footnote{We thank the authors of \citere{Domingo:2011uf} for providing us
with numerical results from their code.}
taking into account only those contributions in our result that are also 
contained in the result of \citere{Domingo:2011uf}.
We found good agreement in this case, at the level of $1$--$2\mev$ on $\MW$.

Throughout this section, we only display parameter points that 
are allowed by the 
LEP limits on SUSY particle masses~\cite{Beringer:1900zz},
by all theoretical constraints in {\tt NMSSMTools} (checking e.g.\ that the Higgs potential has a viable physical minimum and that no Landau pole exists below the GUT scale),
and have the neutralino as LSP.
Unless stated otherwise, we choose the masses of the first and second generation squarks and the gluino to be large enough to not be in conflict with 
the limits from the searches for these particles at the LHC.\footnote{
The most stringent limits from SUSY searches at the LHC are set on the masses of the first and second generation squarks and the gluino, which go beyond $\sim 1 \tev$. However these limits depend on the model assumptions. Relaxing these assumptions, squarks can still be significantly lighter \cite{Mahbubani:2012qq}. 
Substantially weaker limits have been reported for the particles of the other sectors, so that third-generation squarks, stops and sbottoms, as well as the uncoloured SUSY particles, are significantly less constrained by LHC searches.
}
We make use of the code {\tt
HiggsBounds}~\cite{Bechtle:2013gu,Bechtle:2011sb,Bechtle:2008jh} to check
each parameter point against the limits from the 
Higgs searches at LEP, the Tevatron and
the LHC.

\subsubsection{Results in the MSSM limit of the NMSSM}\label{sec:mssmlimsec}
Before turning to the discussion of the genuine NMSSM effects, we show the NMSSM $M_W$ prediction in the MSSM limit
\begin{align}
\la \to 0, \; \ka \to 0, \qquad K\equiv\ka/\la = \mathrm{constant},
\label{eq:mssmlim}
\end{align}
with all other parameters (including $\mueff$) held fixed (such that the MSSM is recovered).
In this limit one $\cp$-even, one $\cp$-odd Higgs boson (not necessarily the heaviest ones) and one neutralino become completely singlet and decouple.
In the discussion of  $M_W^{\rm NMSSM}$ in the MSSM limit, the setup for the numerical evaluation is introduced and 
the comparison to the MSSM $M_W$ prediction serves as validation of our implementation.

\begin{figure}[t!]
\centering
\includegraphics[height=0.45\columnwidth]{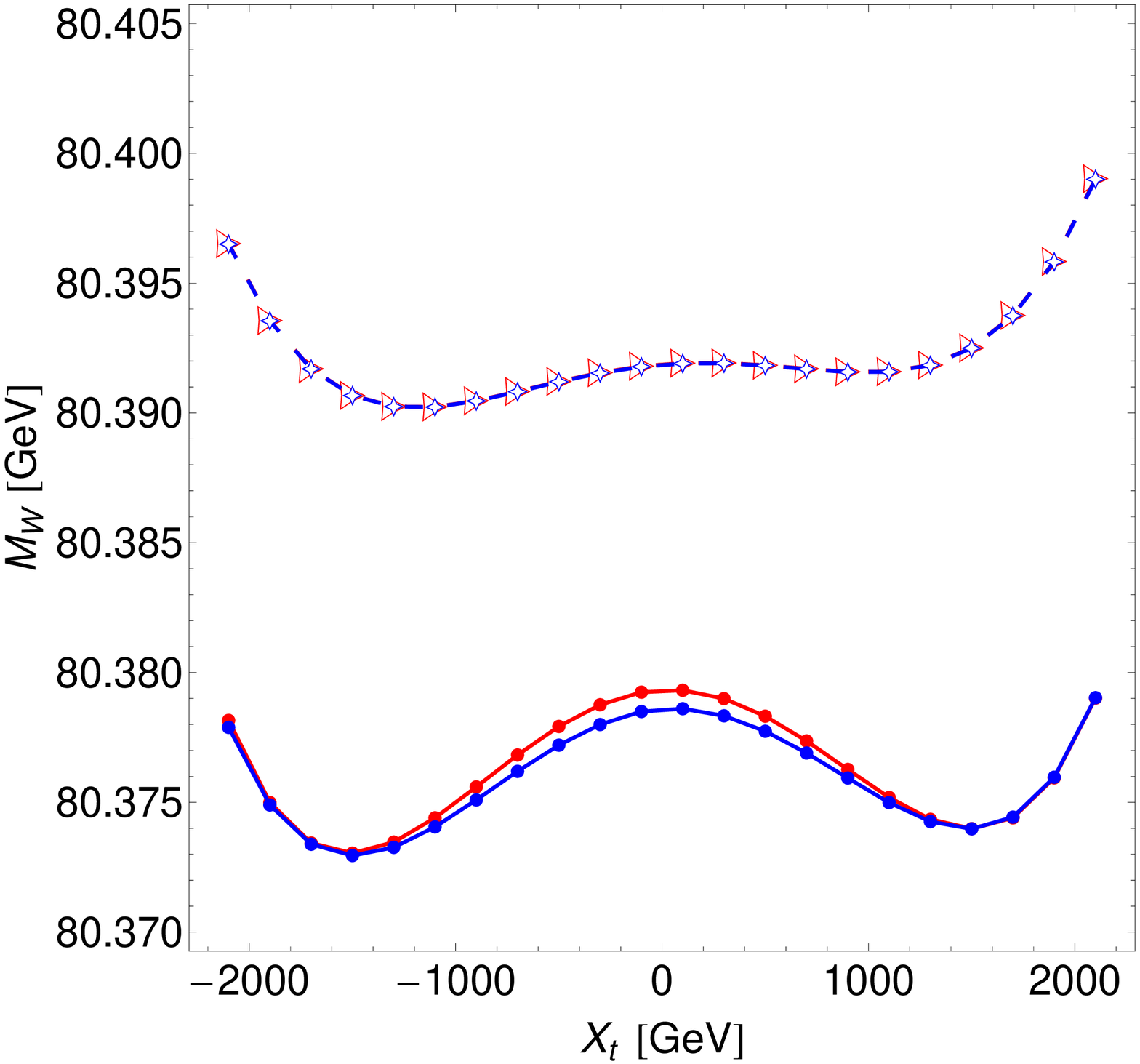}
\includegraphics[height=0.45\columnwidth]{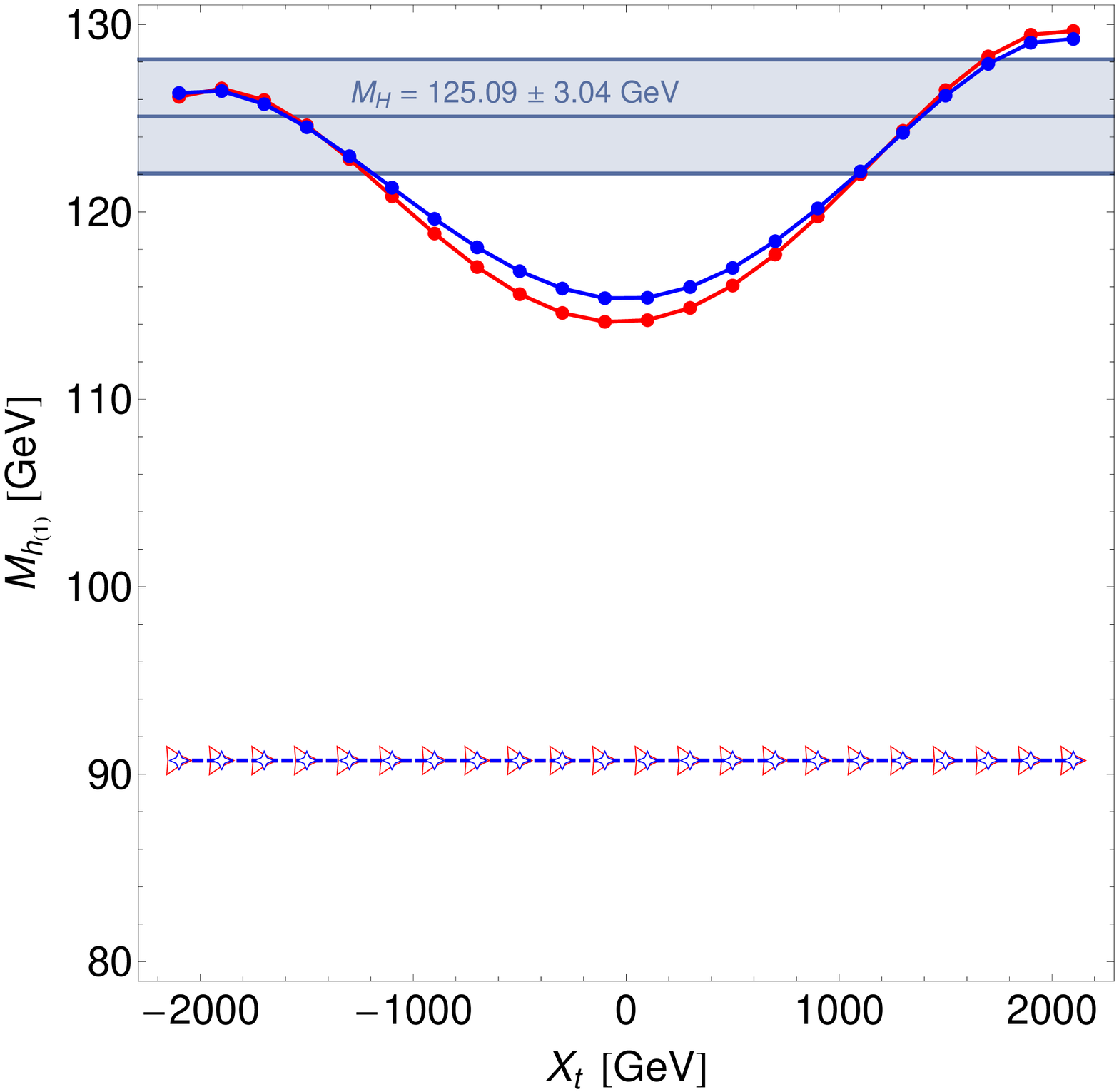}
\caption{
Comparison of the NMSSM predictions in the MSSM limit (blue curves) for the $W$ boson mass (left plot) and the lightest $\cp$-even Higgs mass (right plot) with the MSSM predictions (red curves) plotted against the stop mixing parameter $X_t$. 
The parameters are given in the text.
For the two dashed curves (small blue diamonds for the NMSSM predictions in the MSSM limit, and red triangles for the MSSM predictions) the tree-level Higgs masses are used.
For the solid curves (with filled dots) loop-corrected Higgs masses are used:
the NMSSM Higgs masses are calculated with {\tt NMSSMTools},
and the MSSM Higgs masses calculated with {\tt FeynHiggs}. 
}
\label{fig:nmssm_mssmlim_xt}
\end{figure}

The left plot of \reffi{fig:nmssm_mssmlim_xt} shows the NMSSM predictions in the MSSM limit (blue curves) as well as the MSSM predictions (red curves) for $M_W$ as a function of the stop mixing parameter $X_t$.\footnote{The $X_t$ parameter that we plot here is the on-shell parameter.
As described in \refse{subsec:numsetup} the on-shell value is transformed into a $\overline{\rm{DR}}$ value, 
which is used as input for {\tt NMSSMTools} to calculate the Higgs masses. All numerical values given for $X_t$ in this section refer to the on-shell parameters.}
The parameters in \reffi{fig:nmssm_mssmlim_xt} are 
$m_t=173.34\gev$, 
$\tan \beta=20$, 
$\mu_{(\text{eff})} = 200 \gev$, 
$M_{\tilde{L}/\tilde{E}}=500 \gev$, 
$M_{\tilde{Q}/\tilde{U}/\tilde{D}_{1,2}}=1500 \gev$, 
$M_{\rm SUSY}=M_{\tilde{Q}_{3}}=M_{\tilde{U}_{3}}=M_{\tilde{D}_{3}}=1000 \gev$, 
$A_{\tau}=A_b=A_t$,
$M_2=200 \gev$ and $m_{\tilde{g}}=1500 \gev$.
For the additional NMSSM parameters we choose $\hat{m}_A = 1000 \gev$, $\lambda \rightarrow 0$, $K=\kappa/\lambda=0.5$, $A_{\kappa}=-100 \gev$
(the impact of $A_{\kappa}$ on $M_W$ in the MSSM limit is negligible).
Here, and in the following the prediction for $M_W$ includes all
higher-order corrections described above 
(besides the Higgsino two-loop corrections).

Our approach here 
is the following: 
We start from a NMSSM parameter point. 
We take the effective $\cp$-odd doublet mass $\hat{m}_A$ or the parameter $A_{\lambda}$ (here $\hat{m}_A = 1000 \gev$) as input to calculate the NMSSM Higgs boson spectrum.
The physical value of the charged Higgs mass (calculated in the NMSSM) is used as input for the calculation of the MSSM Higgs masses.
As discussed in \refse{sec:Fulldeltar}, this procedure ensures that the mass of the charged Higgs boson used in our $M_W$ calculation is the same
in the NMSSM and the MSSM, since we calculate the MSSM Higgs masses in {\tt FeynHiggs} (version
2.10.4) where the input parameter $M_{H^{\pm}}$ is interpreted as an on-shell mass parameter.
The other parameters which occur in both models ($\tan \beta$, the sfermion trilinear couplings $A_f$, and the soft mass parameters) are
used with the same values as input for the calculation of the physical masses in the MSSM and the NMSSM. 
For the Higgs mass calculation with {\tt NMSSMTools}
the parameter $X_t$ is transformed into a $\overline{\rm{DR}}$ parameter, while for the $\MW^{\rm (N)MSSM}$ calculations its on-shell value is used.
The MSSM parameter $\mu$ is identified with the NMSSM effective value $\mu_{\text{eff}}$.\footnote{From here on we will leave out the subscript 'eff' for the $\mu$ parameter in the NMSSM}

For the two dashed curves in \reffi{fig:nmssm_mssmlim_xt} (small blue diamonds for the NMSSM predictions in the MSSM limit and red open triangles for the MSSM predictions) the tree-level Higgs masses are used.
For the solid curve (with filled dots) loop-corrected Higgs masses are used:
the NMSSM Higgs masses are calculated with {\tt NMSSMTools} and the MSSM Higgs masses calculated with {\tt FeynHiggs}. 

The corresponding predictions for the lightest $\cp$-even Higgs mass in the (N)MSSM are displayed in the right plot of \reffi{fig:nmssm_mssmlim_xt}.
For illustration, in the plots for the Higgs mass predictions the theoretical uncertainty on the SUSY Higgs mass is combined with the experimental error into an allowed region for the Higgs boson mass, rather than displaying the theoretical uncertainty in the Higgs mass prediction as a band around the theory prediction. Consequently, the
blue band in the right plot shows the region 
$M_H = 125.09 \pm 3.04 \gev$, which was obtained by adding a
theoretical uncertainty of $3 \gev$  quadratically to the experimental
$2\sigma$ error.
Here $M_H$ represents the corresponding mass parameter in the MSSM and
the NMSSM (in the considered case $M_h$ in the MSSM and $M_{h_1}$ in the
NMSSM).
The position of the curves relative to the blue $M_H$ band depends strongly on the other parameters, which are fixed here.
The range in which the NMSSM parameter points (with {\tt NMSSMTools} Higgs masses) are allowed by {\tt HiggsBounds} coincides (approximately) with the region in with the lightest Higgs mass is heavy enough to be interpreted as the signal at $\MHexp \gev$ ($|X_t| \gtrsim 1000 \gev$ and $X_t<1900 \gev$).
While the tree-level Higgs masses agree exactly in the MSSM and the NMSSM in the MSSM limit, we observe a small difference between the masses for the lightest $\cp$-even Higgs calculated with {\tt FeynHiggs} and with {\tt NMSSMTools}.
This discrepancy arises because of differences in the 
higher-order corrections implemented in the two codes%
\footnote{
In {\tt NMSSMTools} the user can set a flag determining the precision for
the Higgs masses.
The result from \citere{Degrassi:2009yq} containing contributions up to
the two-loop level is used if
the flag is set equal to 1 or 2, where the two flags correspond to the
result without (flag 1) and including (flag 2) contributions from non-zero
momenta in the one-loop self-energies. While in {\tt FeynHiggs} this
momentum dependence is taken into account, we nevertheless find better
numerical agreement with flag 1 of the {\tt NMSSMTools} result. For the sake
of comparison between the NMSSM and the MSSM predictions for $\MW$ it is
useful to keep those differences arising from different higher-order
corrections in the MSSM limit of the Higgs sector as small as possible. We
have therefore chosen flag 1 for the Higgs-mass evaluation with {\tt
NMSSMTools} in our numerical analyses presented in this paper. As mentioned
above, an implementation of our predictions using the Higgs-mass evaluation
of \citere{Peternmssm} is in progress.}.
The tree-level Higgs masses are only used in \reffi{fig:nmssm_mssmlim_xt} for illustration. 
In all following plots (if nothing else is specified) the 
full loop-corrected results for the Higgs masses are used.

Going back to the left plot of \reffi{fig:nmssm_mssmlim_xt}, we see that the $M_W^{\text{NMSSM}}$ predictions in the MSSM limit and 
the $M_W^{\text{MSSM}}$ prediction coincide exactly if tree-level Higgs masses are used (which is an important check of our implementation). However, using loop-corrected masses, 
the difference between the {\tt FeynHiggs} and {\tt NMSSMTools} predictions for the lightest $\cp$-even Higgs mass leads to a difference in $M_W$ 
of $\sim 0.8 \mev$ for small $| X_t |$. 
The effect of the difference in the $M_W$ prediction induced by the different Higgs mass predictions is contained in the following plots in this section.
This should be kept in mind when comparing $\MW^{\text{NMSSM}}$ with $\MW^{\text{MSSM}}$.

The dependence of the $\MW$ predictions in
\reffi{fig:nmssm_mssmlim_xt} on $\Xt$ is influenced both by the loop
contributions to $\Delta r$ involving stops and sbottoms, which are
identical at the one-loop level in the MSSM and the NMSSM, and indirectly
via the behaviour of the lightest $\cp$-even Higgs mass. In the chosen
example the impact of the former contributions is relatively small as a
consequence of the relatively high mass scale in the stop and sbottom
sector. The effect of the higher-order corrections in the
Higgs sector is clearly visible in \reffi{fig:nmssm_mssmlim_xt} by comparing
the full predictions with the ones based on the tree-level Higgs masses. As
expected from the behaviour of the $\MW$ prediction in the SM on the
Higgs boson mass, the upward shift in the mass of the lightest $\cp$-even 
Higgs boson caused by the loop corrections 
gives rise to a sizeable downward shift in the predictions for
$\MW$. The local maximum in the $\MW$ predictions at about 
$X_t=0$ is in accordance with the local minimum in the Higgs-mass
predictions. The fact that the local minima in the $\MW$ predictions are
somewhat shifted compared to the local maxima in the Higgs-mass predictions
is caused by the stop-loop contributions to $\Delta r$, whose effect can be
directly seen for the curves based on the tree-level predictions for the 
mass of the lightest $\cp$-even Higgs boson in the left plot of 
\reffi{fig:nmssm_mssmlim_xt}.
The main contribution of the stop/sbottom sector can be associated with $\Delta \rho$ and hence depends strongly on the squark mixing.
$\Delta \rho$ contains terms sensitive to the splitting between the squarks
of one flavour and terms sensitive to the splitting between stops and
sbottoms. These two contributions enter with opposite signs, which tend to
compensate each other for small and moderate values of $\Xt$.

\subsubsection{SUSY higher-oder corrections}\label{sec:susyhosec}
\begin{figure}[t!]
\centering
\includegraphics[height=0.3\columnwidth]{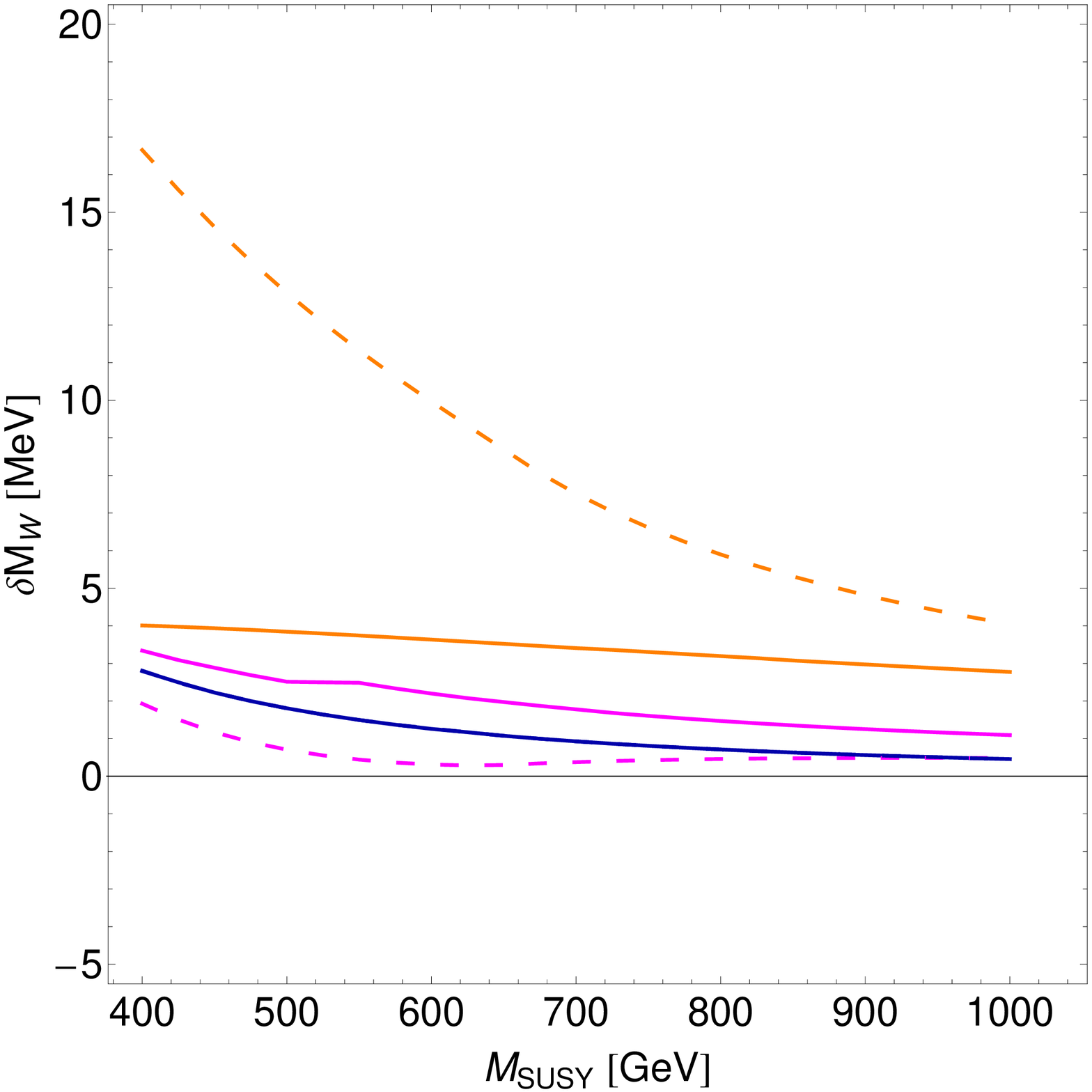}
\includegraphics[height=0.3\columnwidth]{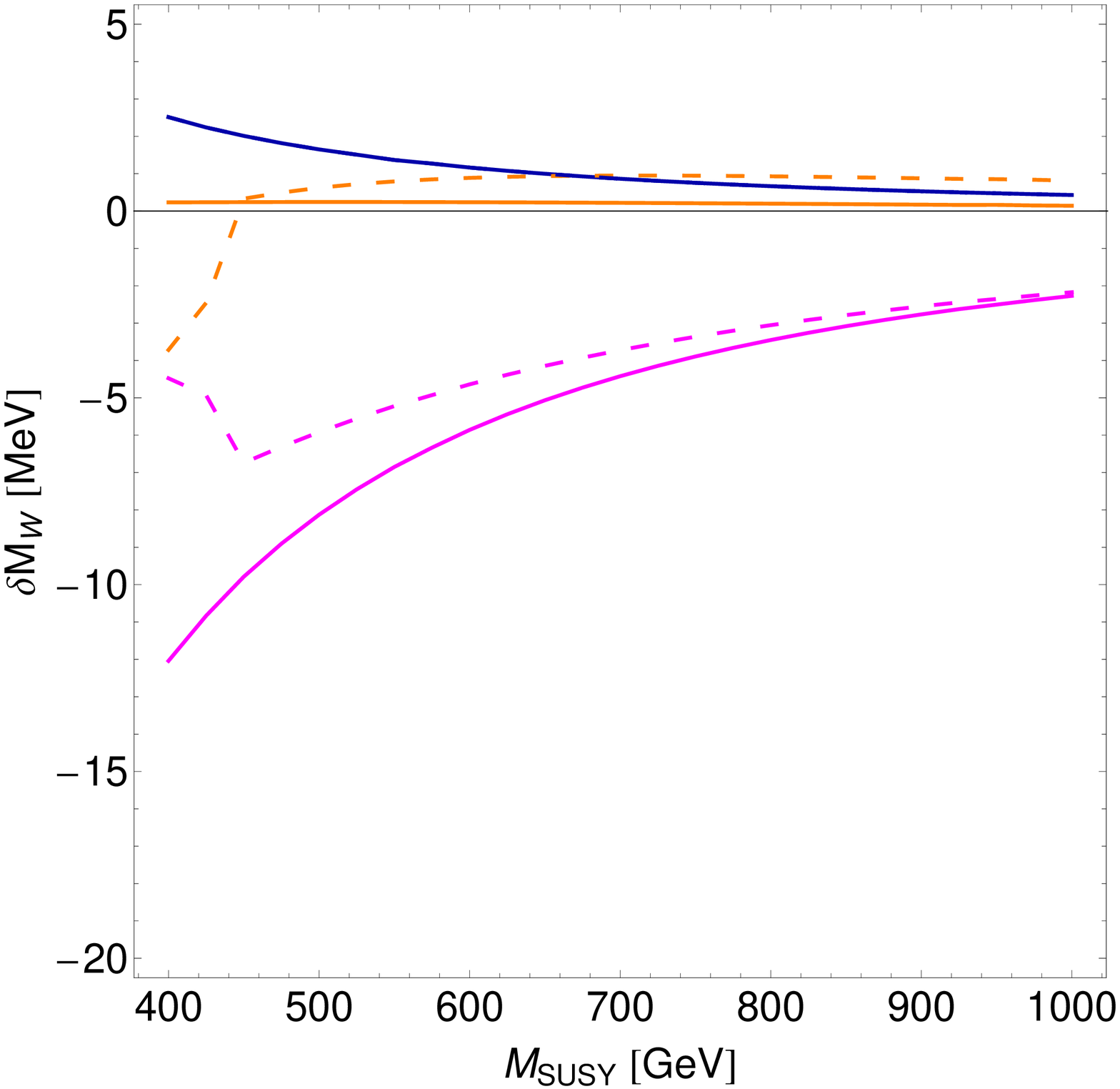}
\includegraphics[height=0.3\columnwidth]{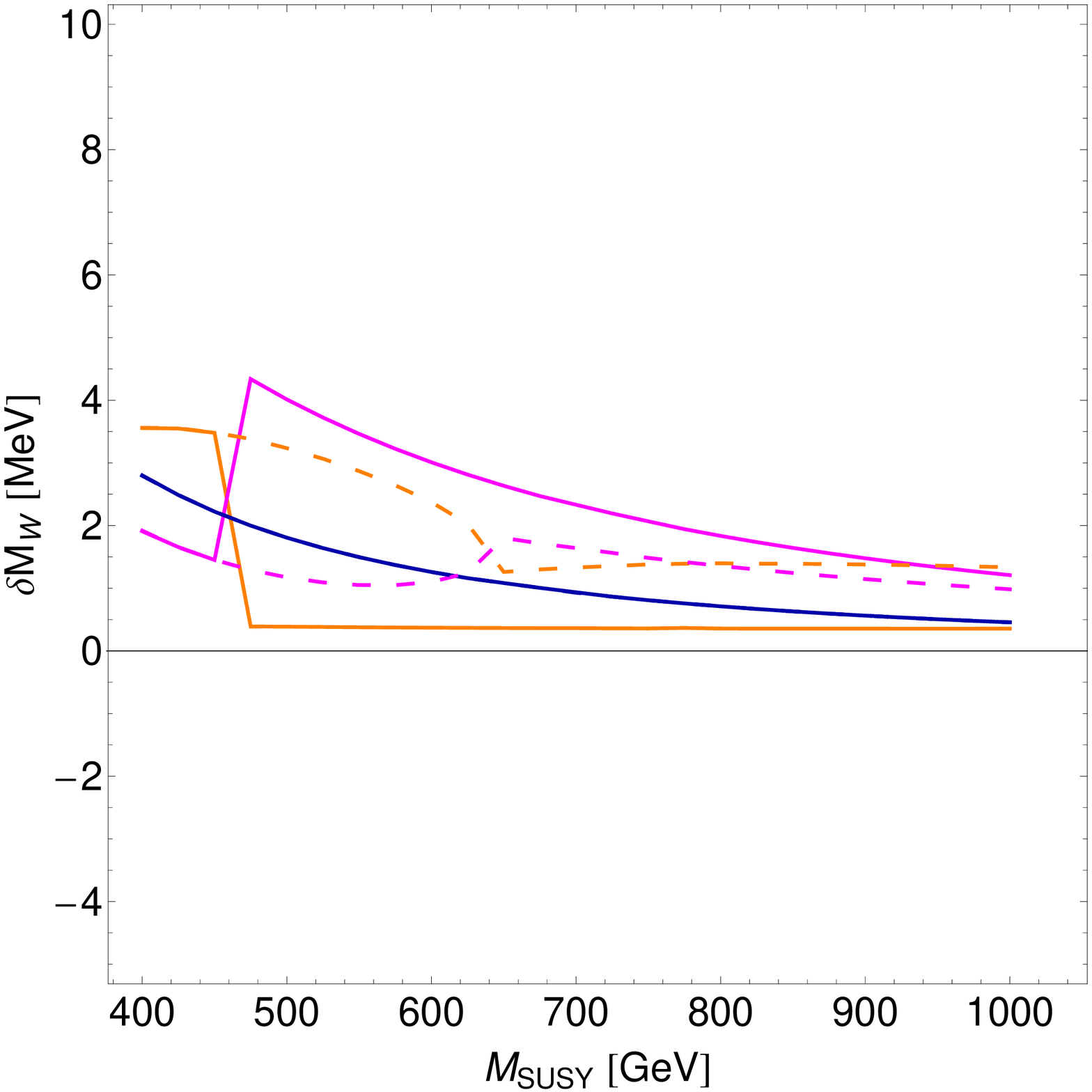}
\caption{
Size of the ${\cal O}(\alpha \alpha_s)$ two-loop corrections with gluon and gluino exchange.
The solid curves correspond to $m_{\tilde{g}}=1500 \gev$ while the dashed curves correspond to $m_{\tilde{g}}=300 \gev$. 
In the left plot we set $X_t=2 \,M_{\text{SUSY}}$, in the middle one $X_t=0$ and in the right one $X_t=-2 \,M_{\text{SUSY}}$.
The plots show the contribution to the $W$ boson mass, $\delta M_W$, from the ${\cal O}(\alpha \alpha_s)$ 
two-loop corrections with gluon exchange (dark blue curves), with gluino exchange (orange curves), and the mass-shift correction (pink curves) as a function of $M_{\text{SUSY}}$. 
The parameter points with $X_t=2 \,M_{\text{SUSY}}$ are {\tt HiggsBounds} allowed for $M_{\text{SUSY}} \lesssim 800 \gev$, whereas
the points with $X_t=0$ and with $X_t=-2\,M_{\text{SUSY}}$ predict too low Higgs masses and are {\tt HiggsBounds} excluded for most $M_{\text{SUSY}}$ values. Note the different scales at the y-axis.
The parameters used 
are given in the text.
}
\label{fig:nmssm4}
\end{figure}

Now we turn to the discussion of the size and parameter dependence of the SUSY two-loop corrections.
\reffi{fig:nmssm4} shows the size of the ${\cal O}(\alpha \alpha_s)$ two-loop corrections.
The parameters used here 
are 
$m_t=173.34 \gev$, 
$\tan \beta=2$, $\mu = 200 \gev$, 
$M_{\tilde{L}/\tilde{E}}=1000 \gev$,
$M_{\tilde{Q}/\tilde{U}/\tilde{D}_{1,2}}=1500 \gev$,
$A_{\tau}=A_b=1000 \gev$,
$M_2=600 \gev$,
$m_{\tilde{g}}=1500 \gev$ (solid curves) and $m_{\tilde{g}}=300 \gev$ (dashed curves),
$A_{\lambda}=395 \gev$, 
$\lambda =0.57$, 
$\kappa=0.2$, 
$A_{\kappa}=-80 \gev$ and we vary $M_{\text{SUSY}}=M_{\tilde{Q}_{3}}=M_{\tilde{U}_{3}}=M_{\tilde{D}_{3}}$.
We show the results for three values of $X_t$:
$X_t=2 \,M_{\text{SUSY}}$ (left), $X_t=0$ (middle) and $X_t=-2 \,M_{\text{SUSY}}$ (right).
It should be stressed here that the parameters for these plots are chosen to demonstrate the possible size and the parameter dependence of the SUSY two-loop corrections,
however they are partially excluded by experimental data:
The parameter points in the left plots with $X_t=2 \,M_{\text{SUSY}}$ are {\tt HiggsBounds} allowed for $M_{\text{SUSY}} \lesssim 800 \gev$ (apart from a small excluded island around $M_{\text{SUSY}} \sim 550 \gev$), whereas
in the middle and the right plots, the chosen parameters are {\tt HiggsBounds} excluded for most $M_{\text{SUSY}}$ values.
A gluino mass value of $m_{\tilde{g}}=300 \gev$ is clearly disfavoured by the negative LHC search results.
\reffi{fig:nmssm4} shows the contribution to the $W$ boson mass, $\delta M_W$, from the ${\cal O}(\alpha \alpha_s)$ 
two-loop corrections with gluon exchange (dark blue curves), with gluino exchange (orange curves) and from the mass-shift correction (pink curves). 
The shift $\delta M_W$ has been obtained by calculating $M_W^{\rm NMSSM}$ twice, once including the corresponding
two-loop corrections, and once without, and the two results have been subtracted from each other.
Starting with the dark blue curves, we find that the gluon contributions lead to a maximal shift of $\sim 3 \mev$ in $M_W$ for all three choices of $X_t$ and that the size 
of the gluon contributions decreases with increasing $M_{\text{SUSY}}$.
Turning to the orange curves, we find that for $m_{\tilde{g}}=1500 \gev$ (solid curves) the $\delta M_W$ shift, induced by the gluino two-loop corrections, is small ($< 1 \mev$) for $X_t=0$, while it is up to $3-4 \mev$ for $X_t=2 \,M_{\text{SUSY}}$ and $X_t=-2 \,M_{\text{SUSY}}$.
Making the gluino light --- choosing $m_{\tilde{g}}=300 \gev$ (dashed curves) --- the gluino corrections can get large. For large positive squark mixing, $X_t=2 \,M_{\text{SUSY}}$, they reach up to 17 MeV for small values of $\msusy$.
The gluino corrections can lead to both a positive and a negative $M_W$ shift, depending on the stop mixing parameter.
Threshold effects occur in the gluino corrections and cause kinks in the orange curves, as can be seen in the middle and the right plots.

\begin{figure}[t!]
\centering
\includegraphics[height=0.3\columnwidth]{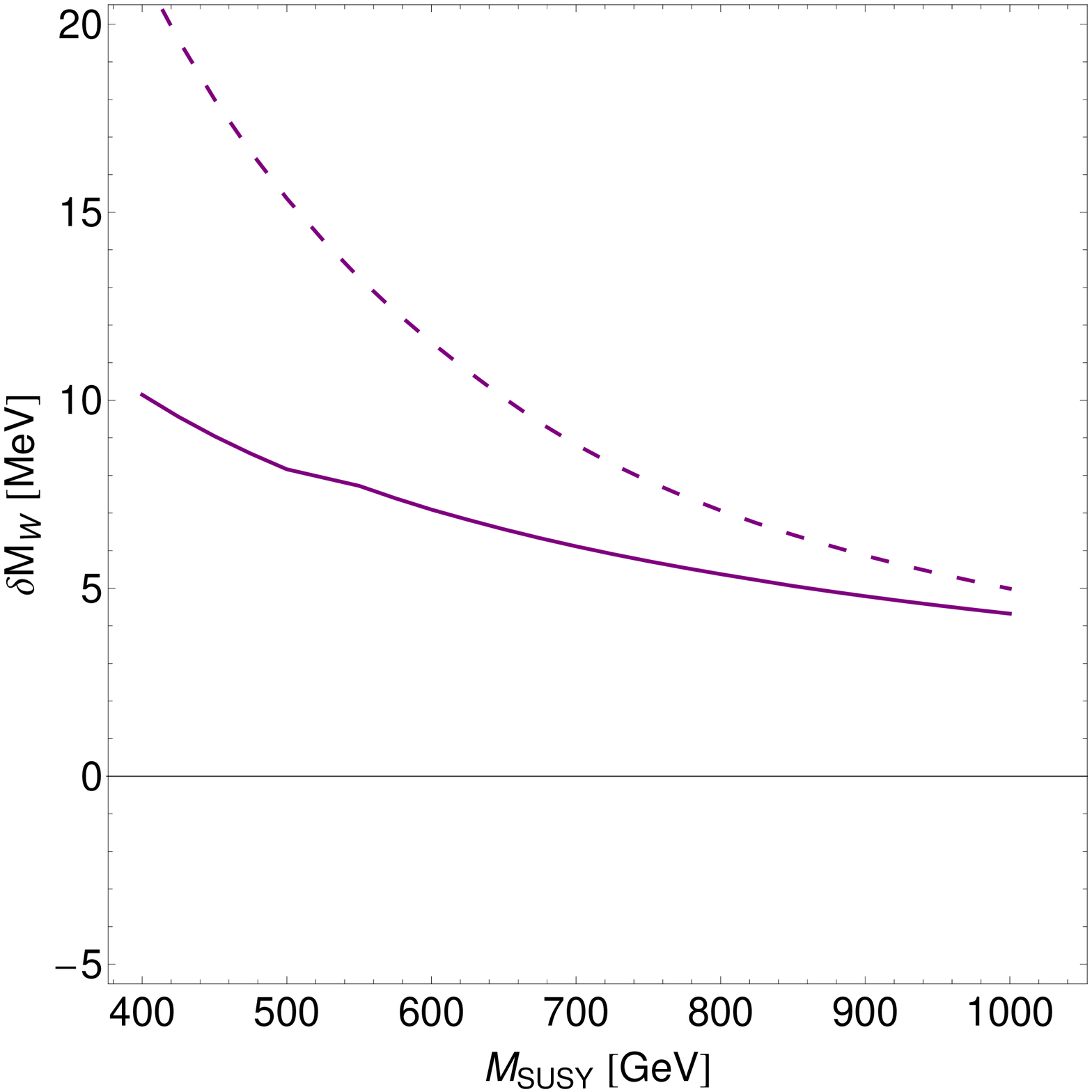}
\includegraphics[height=0.3\columnwidth]{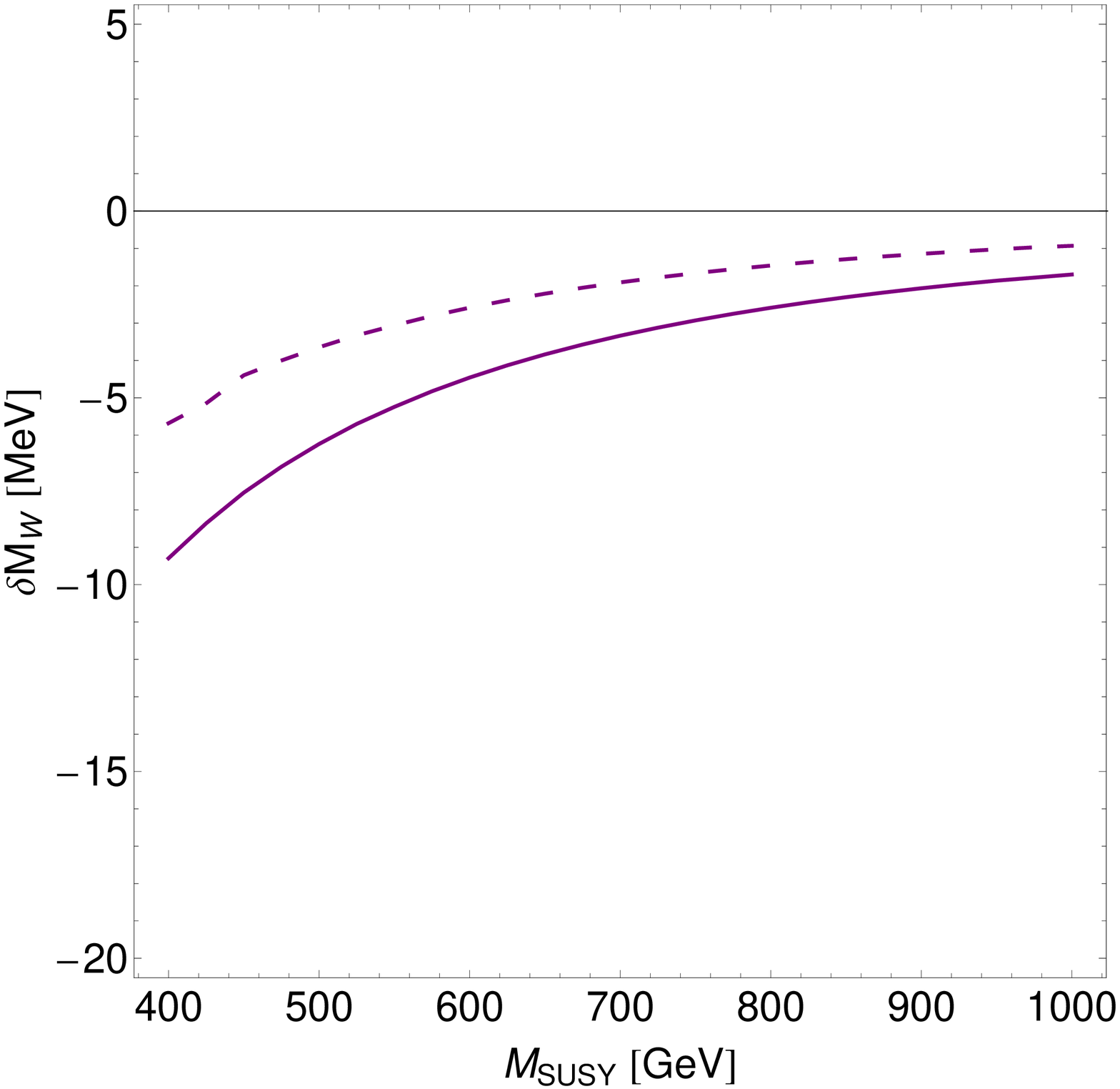}
\includegraphics[height=0.3\columnwidth]{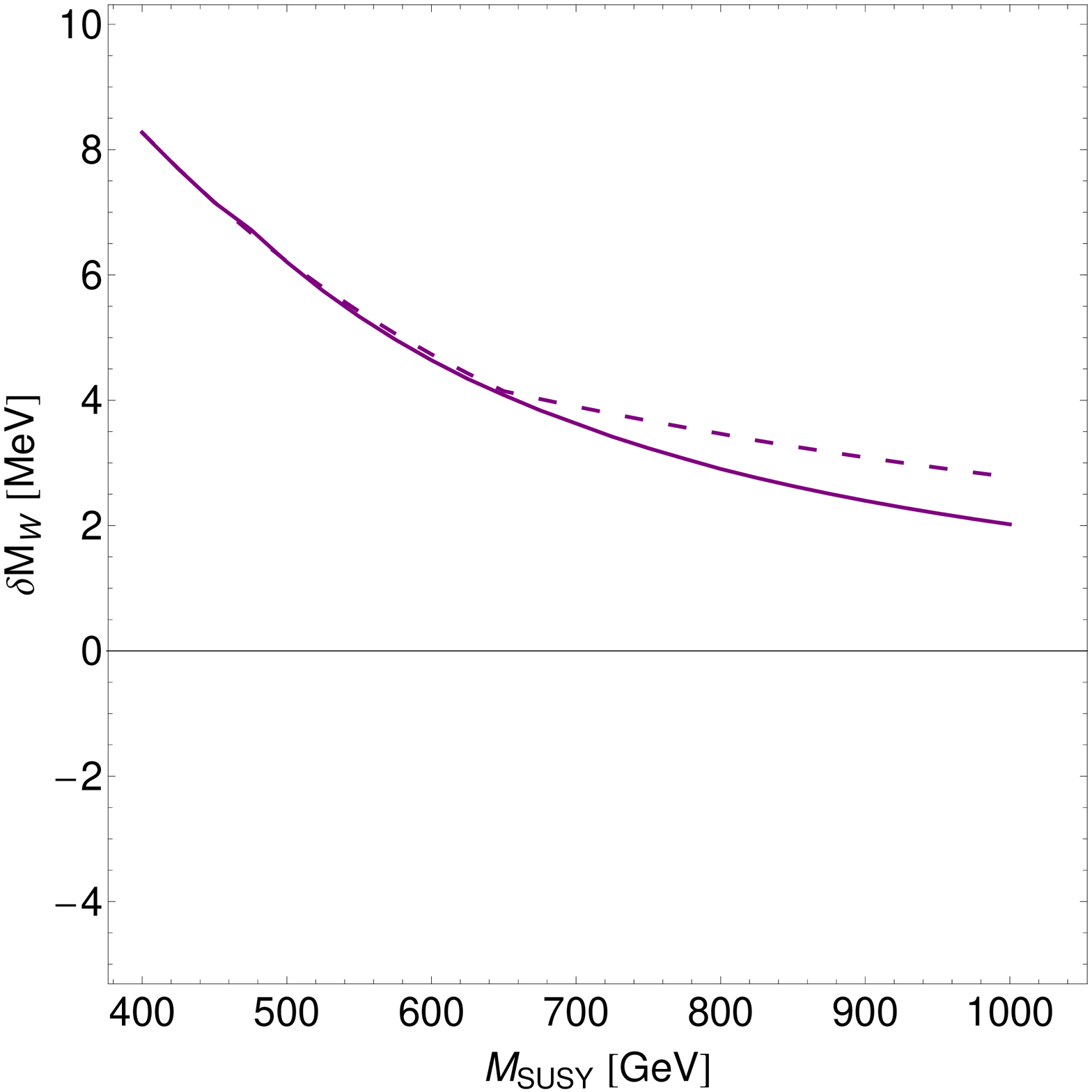}
\caption{
The plots show the full ${\cal O}(\alpha \alpha_s)$ two-loop corrections to $M_W$ (sum of the
corrections shown separately in \reffi{fig:nmssm4}) as a function of $M_{\text{SUSY}}$.
The parameters are the same as in \reffi{fig:nmssm4}.
The solid curves correspond to $m_{\tilde{g}}=1500 \gev$ while the dashed curves correspond to $m_{\tilde{g}}=300 \gev$. 
In the left plot we set $X_t=2 \,M_{\text{SUSY}}$, in the middle one $X_t=0$ and in the right one $X_t=-2 \,M_{\text{SUSY}}$.
}
\label{fig:allnmssm4}
\end{figure}

The gluon and gluino two-loop contributions
are directly related to the mass-shift correction, which has to be incorporated in order to arrive at the complete result for the 
${\cal O}(\alpha \alpha_s)$ contributions to $\Delta \rho^{\text{SUSY}}$.
The pink curves show the impact of this additional correction term.
Starting with the solid curves ($m_{\tilde{g}}=1500 \gev$), we observe that for large stop mixing, $X_t=\pm 2 \,M_{\text{SUSY}}$, the mass-shift corrections are positive and the
maximal shift is $\sim$ 4 MeV.
For zero mixing the mass-shift corrections lead to a large negative shift in $M_W$ (up to $-12$ MeV for small $\msusy$).
For $m_{\tilde{g}}=300 \gev$, the size of the mass-shift correction is smaller. 
The kinks, caused by threshold effects, can be observed (for the same $M_{\text{SUSY}}$ values) also in the mass-shift corrections. 
Adding up the gluino and mass-shift corrections leads to a smooth curve and no kink is found in the full $M_W$ prediction.
This can be seen in \reffi{fig:allnmssm4}, where we plot the sum of the gluon, gluino and mass-shift corrections (all parameters are the same as in \reffi{fig:nmssm4}).
Generally one can see that for large $\msusy$ all contributions decrease, showing the expected decoupling behaviour.
However contributions from the ${\cal O}(\alpha \alpha_s)$ two-loop corrections up to a few MeV are still possible for $\msusy = 1000 \gev$.

\begin{figure}[t!]
\centering
\includegraphics[height=0.3\columnwidth]{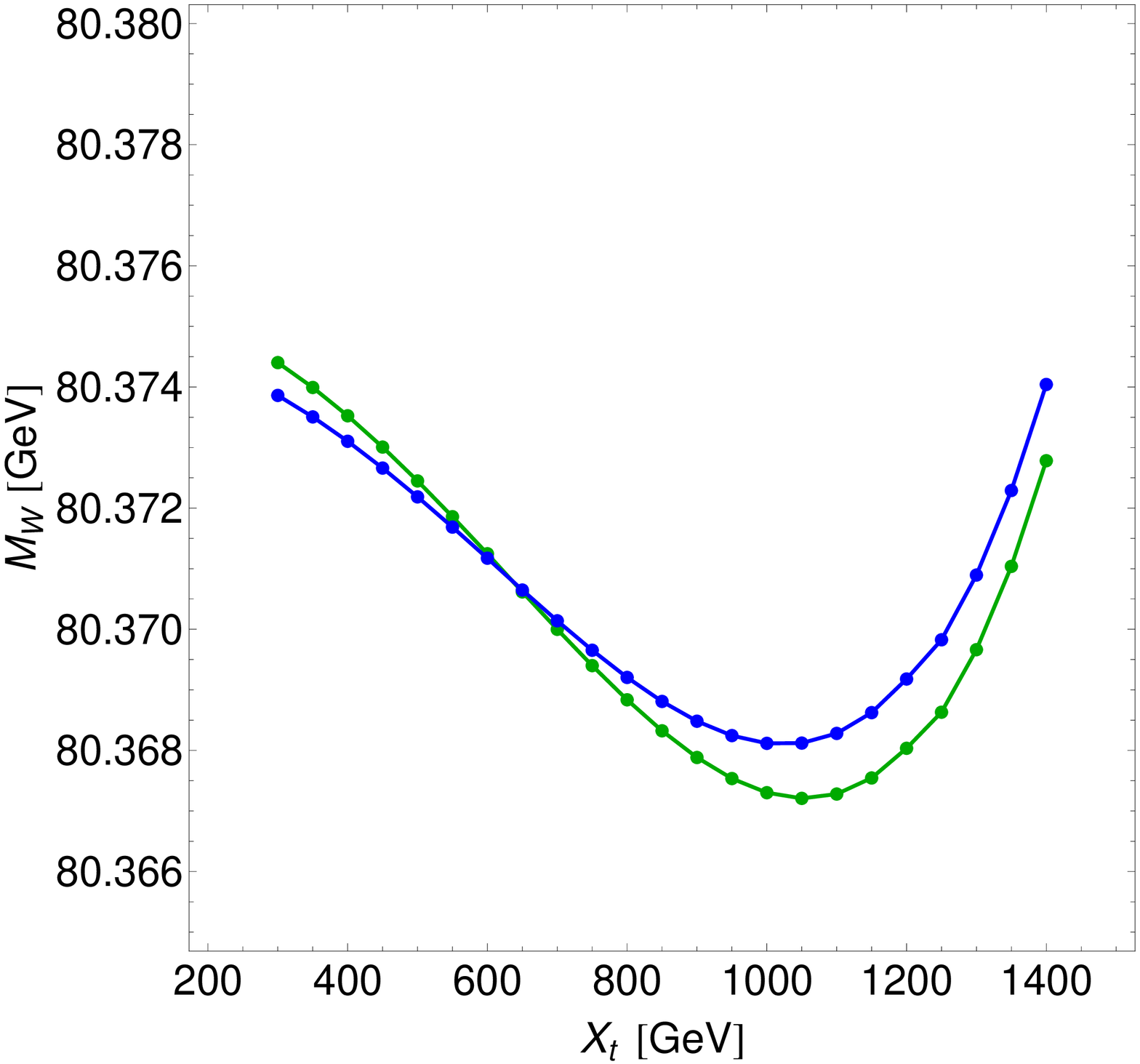}
\includegraphics[height=0.3\columnwidth]{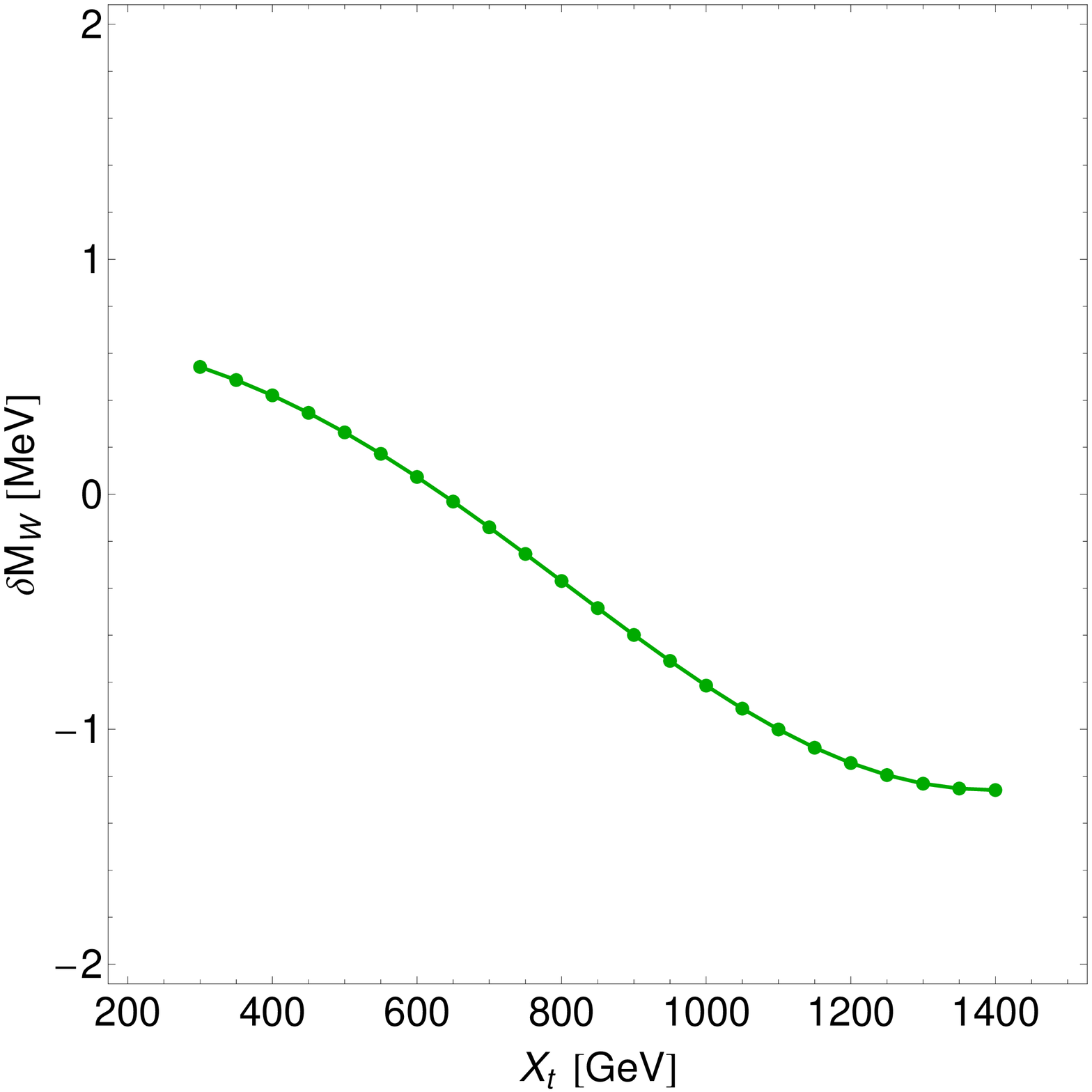}
\includegraphics[height=0.3\columnwidth]{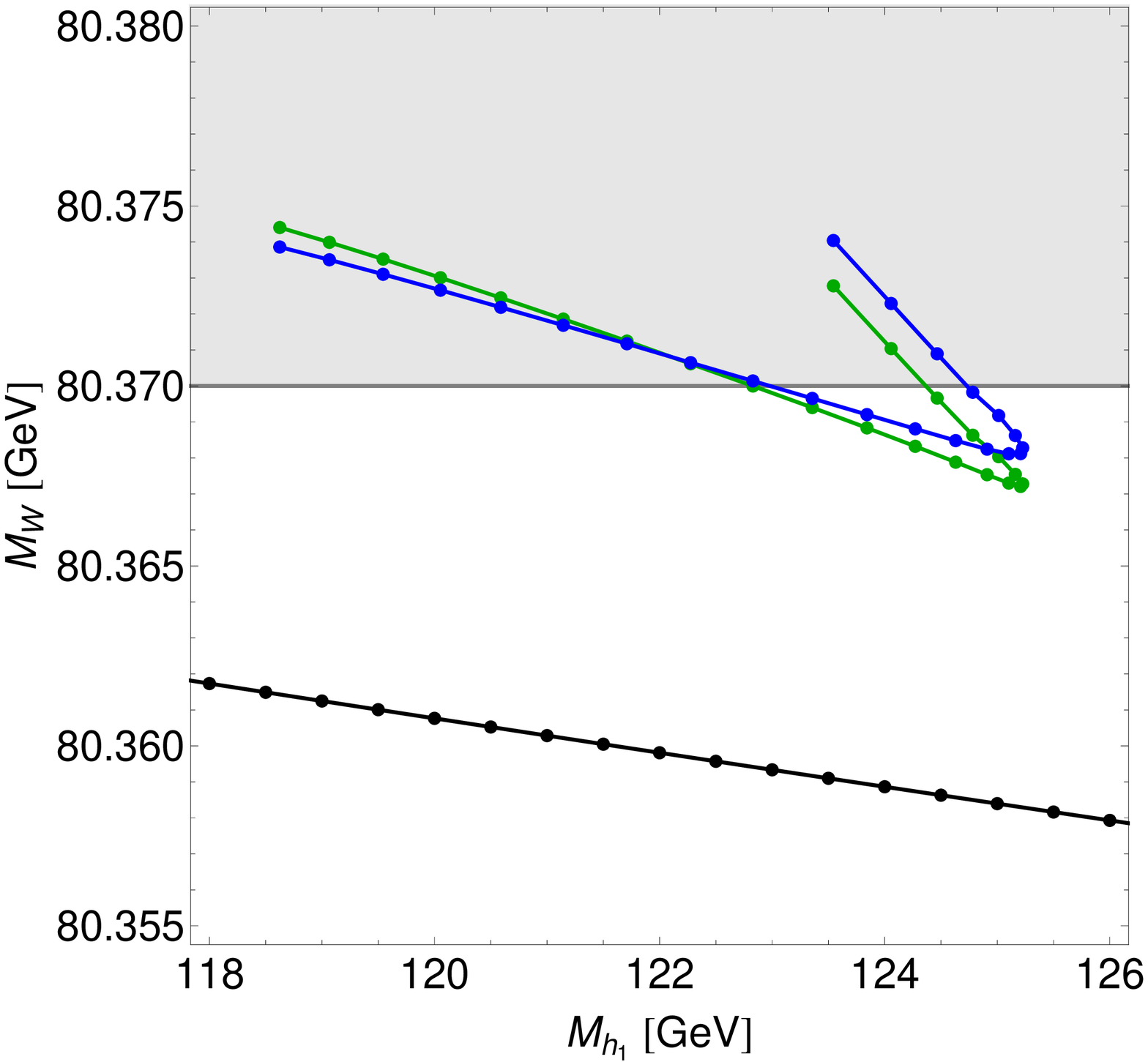}
\caption{
Size of the electroweak ${\cal O}(\alpha_t^2)$, ${\cal O}(\alpha_t \alpha_b)$, ${\cal O}(\alpha_b^2)$ SUSY two-loop corrections.
The left plot shows the NMSSM $M_W$ prediction without Higgsino corrections (blue) and including Higgsino corrections (green).
The middle plot shows the shift $\delta M_W$ induced by the 
Higgsino corrections  (obtained by subtracting the $M_W$ predictions with and without Higgsino corrections as shown in the left plot).
The right plot shows the NMSSM $M_W$ prediction without Higgsino corrections (blue) and including Higgsino corrections (green) 
plotted against the lightest $\cp$-even Higgs mass $M_{h_1}$. The black curve in the right plot indicates the SM $M_W$ prediction with $\MHSM = M_{h_1}$. 
The grey band indicates the $1\,\sigma$ region of the experimental $W$ boson mass measurement.
The parameters used for these plots
are given in the text.
}
\label{fig:nmssm_higgsino}
\end{figure}

The Yukawa-enhanced electroweak two-loop corrections of
${\cal O}(\alpha_t^2)$, ${\cal O}(\alpha_t \alpha_b)$, ${\cal O}(\alpha_b^2)$ 
to $\Delta \rho$ (``Higgsino corrections'') in the MSSM
can be included in our code, as discussed in \refse{sec:Fulldeltar}.
To do so, we calculate the MSSM Higgs masses as described in \refse{sec:Fulldeltar} (taking the NMSSM charged Higgs mass as input for the MSSM Higgs mass calculation) and use them
as input for the $\Delta \rho$ (${\cal O}(\alpha_t^2)$, ${\cal O}(\alpha_t \alpha_b)$, ${\cal O}(\alpha_b^2)$) formula.
The size of these contributions can be seen in \reffi{fig:nmssm_higgsino}.
Here, and in some of the following plots, we choose modified versions of the benchmark points given in~\citere{King:2012is}, which
predict one of the $\cp$-even NMSSM Higgs bosons in the mass range of the observed Higgs signal, as starting point for our study.
Here we take the following parameters: 
$m_t=173.34 \gev$, 
$\tan \beta=2$, $\mu = 200 \gev$, 
$M_{\tilde{L}/\tilde{E}}=1000 \gev$,
$M_{\tilde{Q}/\tilde{U}/\tilde{D}_{1,2}}=1200 \gev$,
$M_{\tilde{Q}_{3}}=M_{\tilde{U}_{3}}=700 \gev$,
$M_{\tilde{D}_{3}}=1000 \gev$,
$A_{\tau}=A_b=1000 \gev$,
$M_2=200 \gev$, $m_{\tilde{g}}=1500 \gev$,  
$A_{\lambda}=405 \gev$,
$\lambda = 0.6$, 
$\kappa=0.18$, $A_{\kappa}=-10 \gev$, and we vary $X_t$.
These parameter points are {\tt HiggsBounds} allowed in the regions $700 \gev<X_t<1000 \gev$ and $1100 \gev<X_t<1400 \gev$.
The left plot shows the NMSSM $M_W$ prediction without Higgsino corrections (blue) and including Higgsino corrections (green) 
plotted against $X_t$.
In the middle plot the shift $\delta M_W$ induced by the 
Higgsino corrections  (obtained by subtracting the $M_W$ predictions with and without Higgsino corrections as shown in the left plot) is plotted against $X_t$.
We see that the Higgsino corrections can enter the $M_W$ prediction with both signs. 
The numerical effect of the $M_W$ shift, induced by the Higgsino corrections, is relatively small ($\sim 1 \mev$).
It was shown in \citere{Haestier:2005ja} that the contributions to $M_W$ from the Higgsino corrections can be slightly larger ($\sim 5 \mev$)
for lighter $\tilde{t}/\tilde{b}$.
The right plot shows the $M_W$ prediction plotted against $M_{h_1}$.
We can clearly see here that this scenario, in which the Higgs signal can be interpreted as the lightest $\cp$-even NMSSM Higgs,
gives a $W$ boson mass prediction in good agreement with the $\MW$ measurement indicated by the grey band.

\subsubsection{NMSSM Higgs sector contributions}
\begin{figure}[t!]
\centering
\includegraphics[height=0.45\columnwidth]{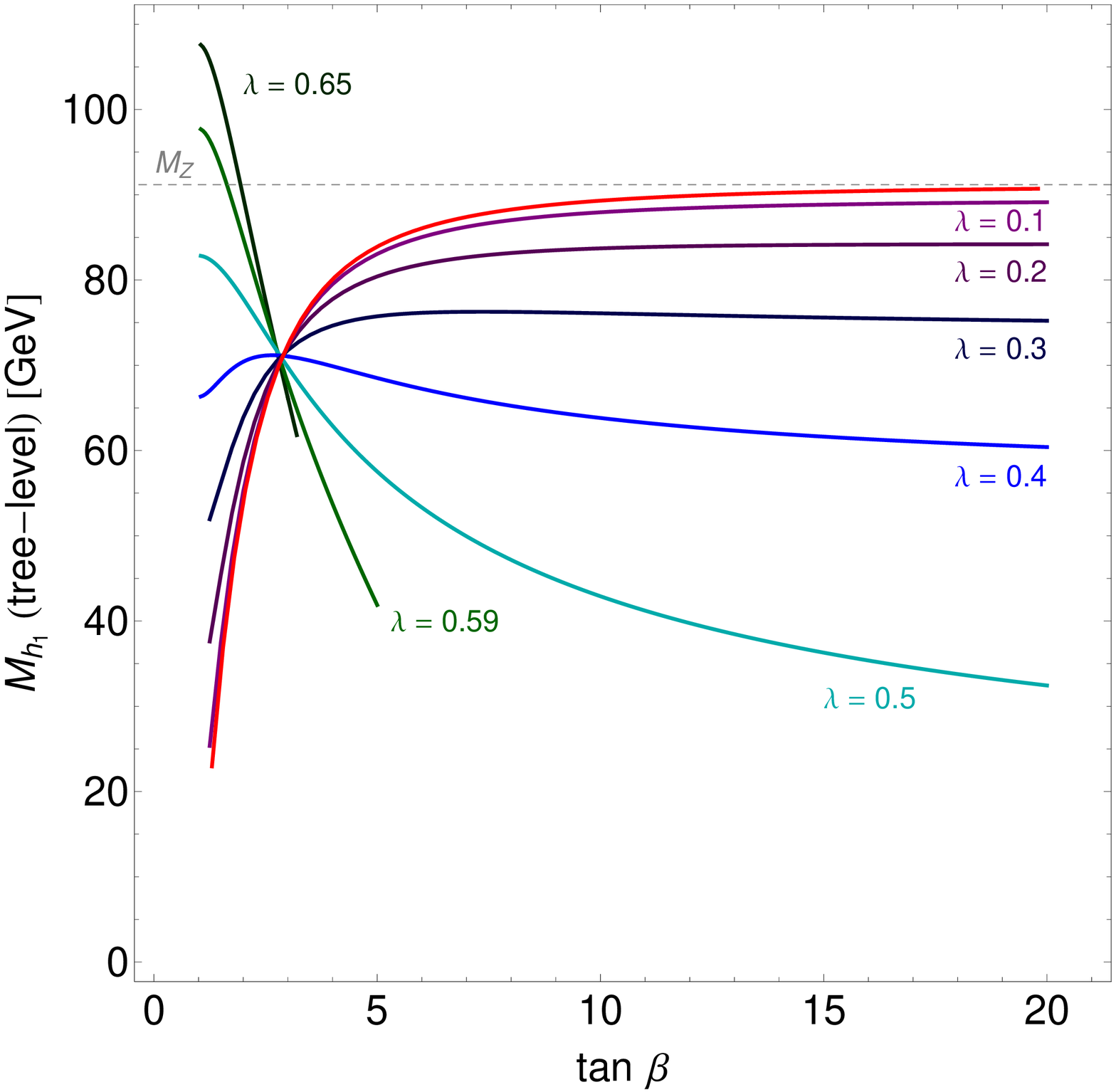}
\includegraphics[height=0.45\columnwidth]{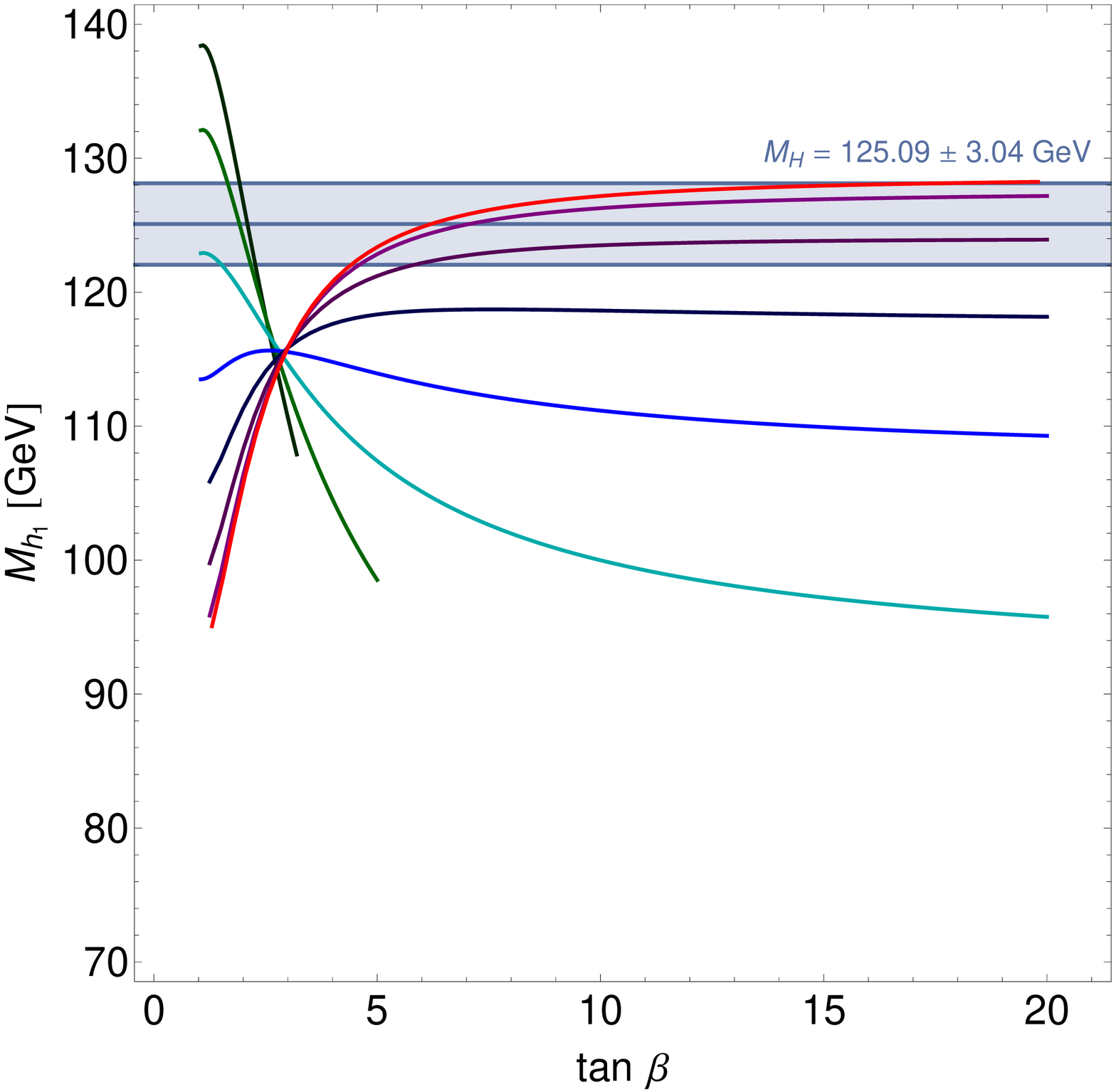}
\includegraphics[height=0.45\columnwidth]{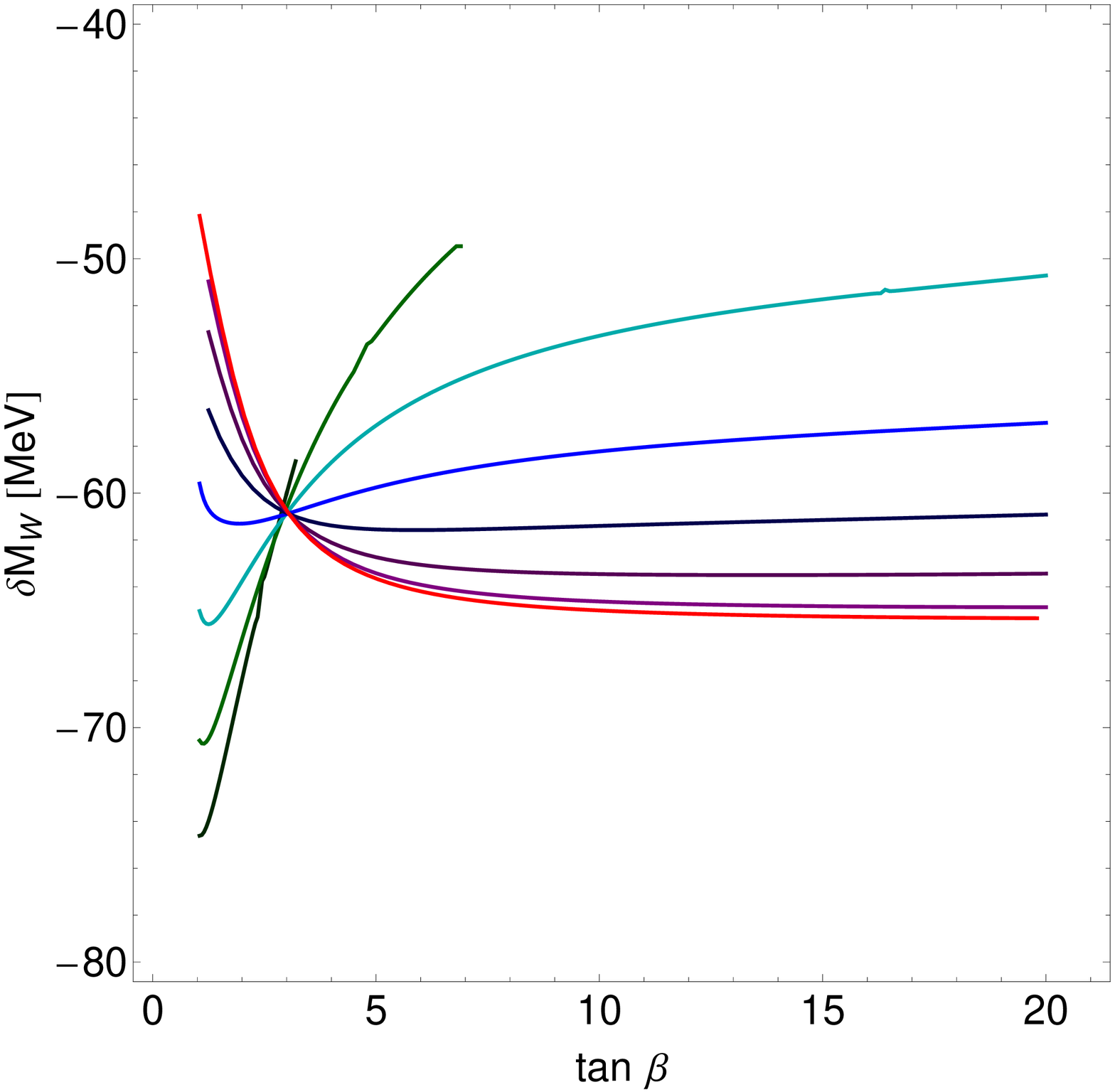}
\includegraphics[height=0.45\columnwidth]{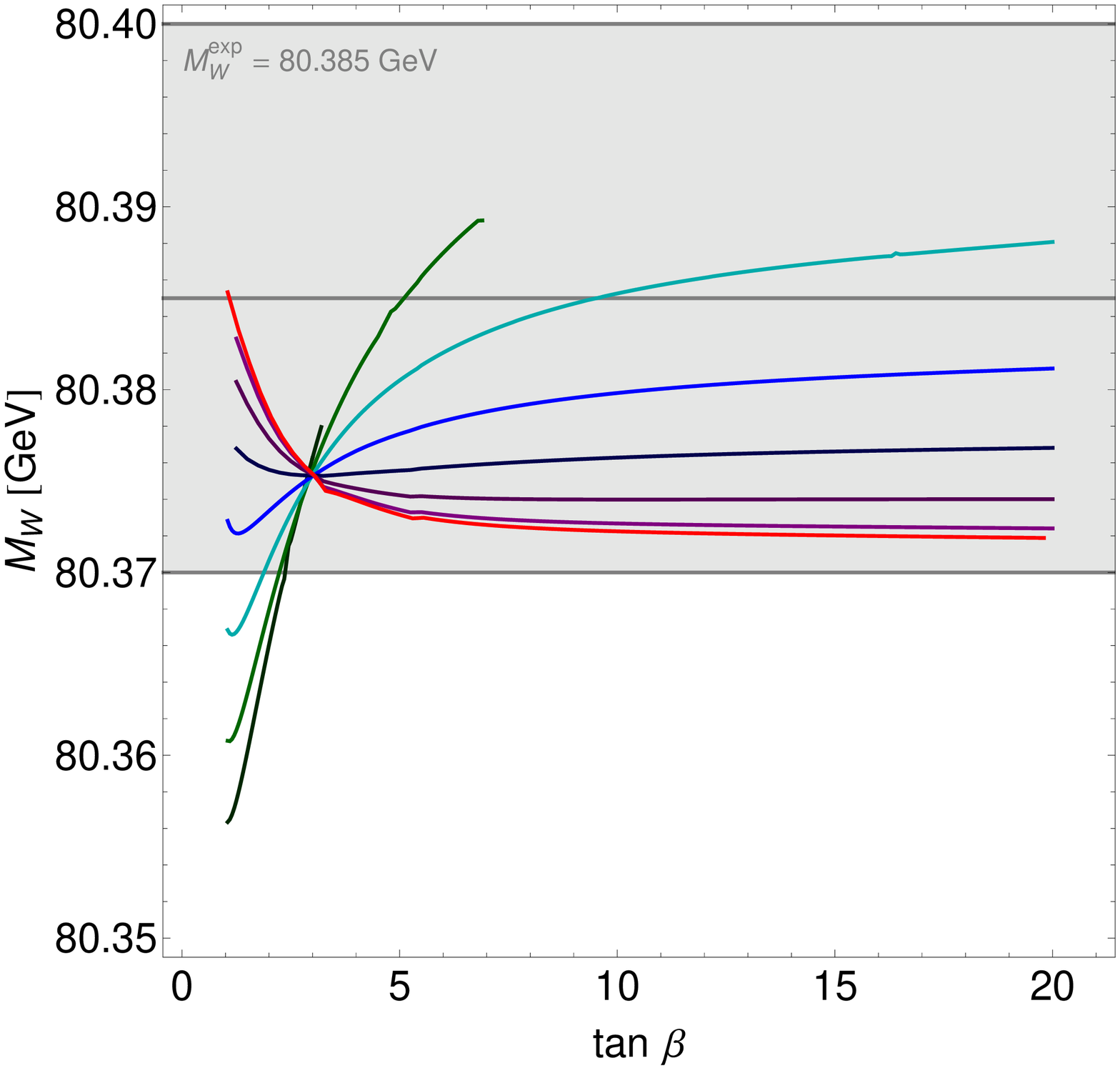}
\caption{
Predictions for $M_{h_1}$ and $\MW$ as a function of $\tan \beta$.
The red curves correspond to the MSSM limit ($\lambda \rightarrow 0$) while for the other curves the $\lambda$ values are given in the figure.
The upper left plot shows the tree-level prediction for the lightest
$\cp$-even Higgs mass $M_{h_1}$, the upper right plot shows $M_{h_1}$
including radiative corrections (calculated with {\tt NMSSMTools} as
described in the text), 
the lower left plot shows the shift $\delta M_W$ (calculated as in \refeq{eq:mwshift}) from diagrams involving Higgs and gauge bosons,
and the lower right plot shows the full $M_W$ prediction.
The parameters used for these plots
are given in the text.
}
\label{fig:nmssm_treelevelhiggs}
\end{figure}
Now we turn  to the discussion of 
effects from the NMSSM Higgs sector.
In the MSSM the maximal value for the tree-level Higgs mass $M_h$ is $M_Z$.
One of the features of the NMSSM Higgs sector is that
the tree-level Higgs mass $M_{h_1}$ gets an additional contribution $\lambda^2 v^2 \sin^2 2 \beta$, which can
shift the tree-level Higgs mass upwards compared to its MSSM value 
(an upward shift can also be caused by singlet--doublet mixing, if the
singlet state is lighter than the doublet state), and thus reduce the size
of the radiative corrections needed to 'push' the lightest Higgs mass up to
the experimental value.
For $\lambda=0.7$ and $\tan \beta = 2$ a tree-level value for $M_{h_1}$ of 112 GeV is possible~\cite{King:2012is}.
This additional tree-level contribution to the Higgs mass, as well as its impact on $M_W$ are shown in \reffi{fig:nmssm_treelevelhiggs}.
The parameters chosen here are
$m_t=173.34 \gev$, $\mu = 500 \gev$, 
$M_{\tilde{L}/\tilde{E}}=500 \gev$, 
$M_{\tilde{Q}/\tilde{U}/\tilde{D}_{1,2}}=1500 \gev$,
$M_{\tilde{Q}_{3}}=M_{\tilde{U}_{3}}=M_{\tilde{D}_{3}}=1000 \gev$, 
$X_t=2000 \gev$, $A_{\tau}=A_b=A_t$, 
$M_2=200 \gev$, $m_{\tilde{g}}=1500 \gev$, 
$\hat{m}_A=450 \gev$,
$\kappa=\lambda$ and $A_{\kappa}=-100 \gev$.
We vary $\tan \beta$ and show the results for different values of $\lambda$.
The red curves correspond to the MSSM limit ($\lambda \rightarrow 0$) while for the other curves the $\lambda$ value is given in the corresponding colour.
The upper left plot shows the tree-level prediction for the lightest $\cp$-even Higgs mass $M_{h_1}$.
As expected, the $M_{h_1}$ prediction in the MSSM limit approaches its maximal value $M_Z$ for large $\tan \beta$.
Increasing $\lambda$, the $M_{h_1}$ prediction decreases for large $\tan
\beta$, caused by doublet--singlet mixing terms.
For small $\tan \beta$ one clearly sees the positive contribution from the term $\lambda^2 v^2 \sin^2 2 \beta$ pushing the tree-level Higgs mass 
beyond $M_Z$ for large $\lambda$.\footnote{ 
The mixing of the $h_1$ state with the heavier singlet leads to a negative
contribution to the tree-level Higgs mass, which pulls the NMSSM Higgs mass
value down (compared to the MSSM case) 
for intermediate and large $\tan \beta$ values (for details see
\citere{Ellwanger:2009dp}). At a specific $\tan \beta$ value this
contributions exactly cancels the positive $\lambda^2 v^2 \sin^2 2 \beta$
shift at the tree level,
and the NMSSM Higgs mass value coincides with the MSSM value. 
In the scenario considered here, this happens for all $\lambda$ at the same
$\tan \beta$ value, since we chose $\kappa=\lambda$. As can be seen in
the upper right plot of \reffi{fig:nmssm_treelevelhiggs}, this behaviour is
approximately retained also in the presence of higher-order corrections in
the Higgs sector.
}
The full $M_{h_1}$ prediction (calculated with {\tt NMSSMTools} as described above) can be seen in the upper right plot.
Now we turn to the $M_W$ contributions from the NMSSM Higgs and gauge boson sector, shown in the lower left plot.
The shift $\delta M_W$ displayed here is based on the approximate
relation~\cite{Heinemeyer:2006px}
\BE
\delta M_W = - \frac{M_W^{\rm ref}}{2}\,\frac{s_W^2}{c_W^2-s_W^2}\, \Delta
r^{x(\alpha)} ,
\label{eq:mwshift}
\EE
where $\Delta r^{x(\alpha)}$ denotes the one-loop contribution from particle
sector $x$ (here $x$=gauge-boson/Higgs), as defined for the NMSSM in \refeq{eq:deltar1lterms}.
The reference $M_W$ value is set here to $M_W^{\rm ref} = M_W^{\rm exp}$.
The overall contribution from the Higgs sector is rather large and negative. 
As we will discuss in more detail below, the Higgs sector contributions here are predominantly SM-type contributions
(with $\MHSM$ set to the corresponding Higgs mass value).
The prediction for $M_W$ in the NMSSM is shown in the lower right plot. 
Larger values for $M_{h_1}$ correspond to a lower predicted value for $M_W$.
Thus, for small $\tan \beta$, where we find a significantly higher predicted value for $M_{h_1}$ for large $\lambda$ than in the MSSM limit (arising from the additional tree-level term),
we get a lower predicted value for $M_W$, which is however still compatible with the experimental $\MW$ measurement at the $2\sigma$ level for the scenario chosen here. For $\tan \beta \sim 1$ the difference between the $W$ boson mass prediction for $\lambda=0.65$ and $\lambda \rightarrow 0$ is 
$\sim$ 25 MeV.
The parameter $\tan \beta$ enters also in the sfermion and in the chargino/neutralino sector. We checked that for the parameters used here, the $\tan \beta$ dependence 
of the contributions from these two sectors is small compared to the Higgs sector contributions, less than $\sim 3 \mev$. 

\begin{figure}[t!]
\centering
\includegraphics[height=0.3\columnwidth]{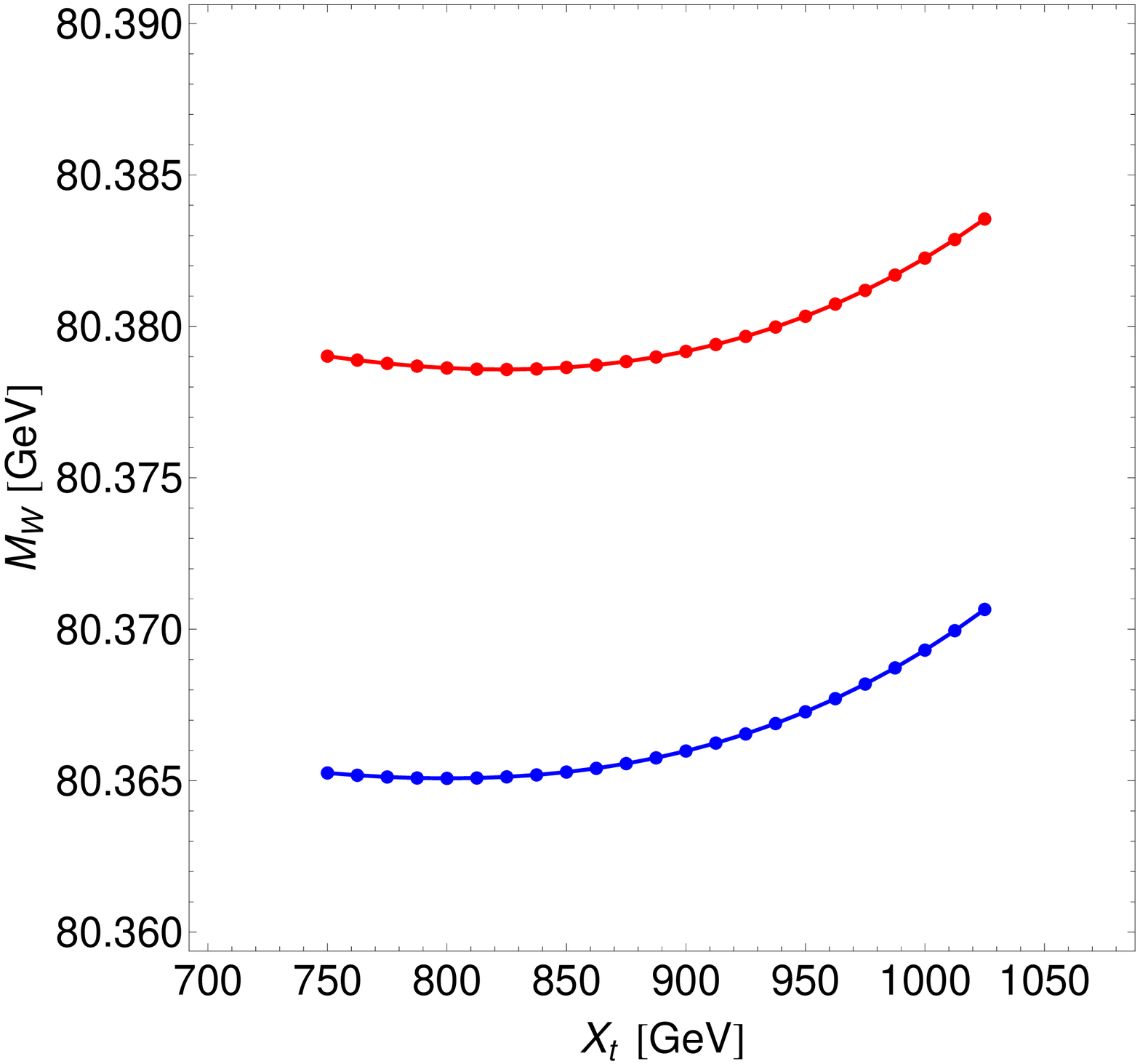}
\includegraphics[height=0.3\columnwidth]{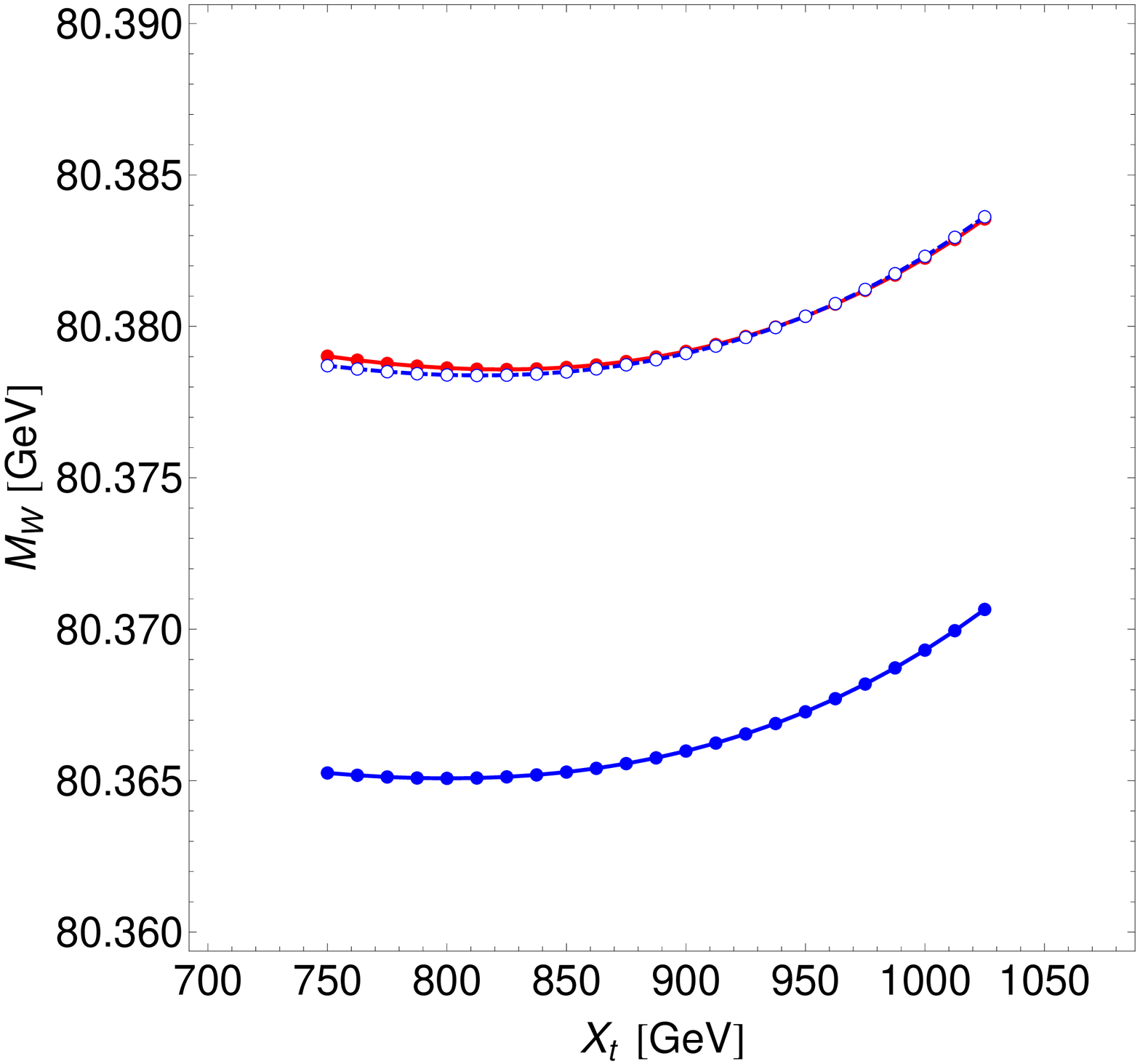}
\includegraphics[height=0.3\columnwidth]{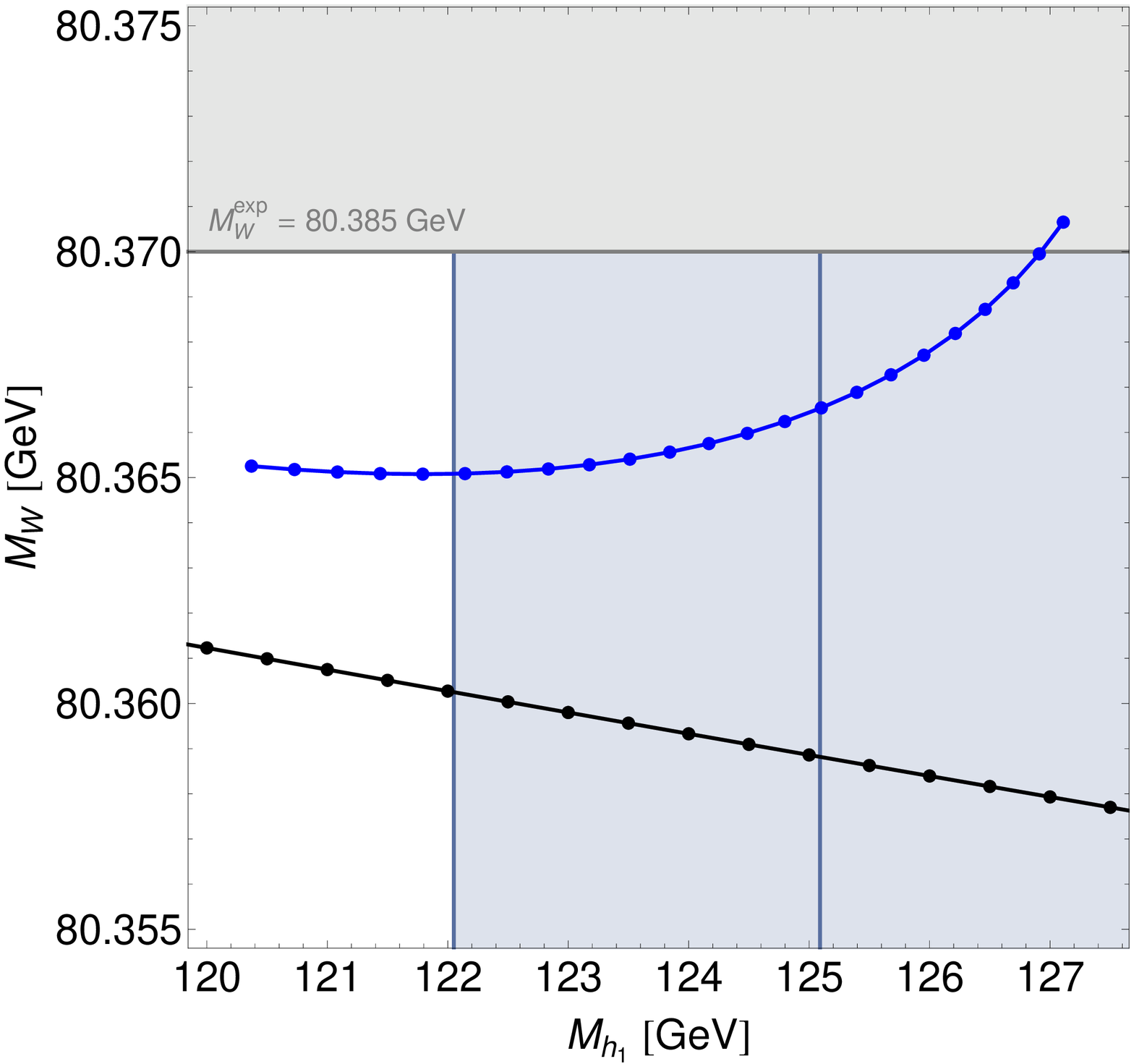}
\caption{
The left plot shows the $M_W^{\text{NMSSM}}$ prediction (blue, solid curve) and the $M_W^{\text{MSSM}}$ prediction (red) plotted against $X_t$. 
In the middle plot, the additional dashed blue curve corresponds to $M_W^{\text{NMSSM}} - M_W^{\text{SM}}(M_{h_1}) +  M_W^{\text{SM}}(M_{h})$ ($M_{h_1}$ is the mass of the lightest $\cp$-even Higgs of the NMSSM, and $M_{h}$ is the mass of the light $\cp$-even Higgs of the MSSM).
The right plots shows the $M_W^{\text{NMSSM}}$ prediction plotted against the lightest $\cp$-even Higgs mass $M_{h_1}$. The black curve in the right plot indicates the SM $M_W$ prediction with $\MHSM = M_{h_1}$. The experimental $M_W$ measurement is indicated by the grey band; the region $M_H=\MHexp \pm 3.04 \gev$ is indicated by the blue band.
The parameters are given in the text.
}
\label{fig:NMP3}
\end{figure}
We continue the study of the NMSSM Higgs sector contributions in \reffi{fig:NMP3}.
In the left plot we compare the NMSSM prediction for $M_W$ (blue curve) with the MSSM prediction (red curve).
The parameters we use here are
$m_t=173.34 \gev$, 
$\tan \beta=2$, 
$\mu = 200 \gev$, 
$M_{\tilde{L}/\tilde{E}}= 1500 \gev$,
$M_{\tilde{Q}/\tilde{U}/\tilde{D}_{1,2}}=1200 \gev$, 
$M_{\tilde{U}_{3}}=M_{\tilde{Q}_{3}}=540 \gev$, 
$M_{\tilde{D}_{3}}=1000 \gev$, 
$A_{\tau}=A_b=1000 \gev$,
$M_2=370 \gev$, 
$m_{\tilde{g}}=1500 \gev$, 
$A_{\lambda}=420 \gev$,
$\lambda = 0.57$, 
$\kappa=0.2$, 
$A_{\kappa}=-10 \gev$, and we vary $X_t$.
The NMSSM parameters are allowed by {\tt HiggsBounds} for $X_t \gtrsim 780 \gev$.
For $X_t \gtrsim 810 \gev$ the mass of the lightest $\cp$-even Higgs 
falls in the range of the observed Higgs signal.
The MSSM prediction is plotted as a comparison to illustrate and discuss the NMSSM effects on $M_W$. 
Here (and in the following) we do not check any phenomenological constraints for the MSSM parameter point (but only for the considered NMSSM scenario). 

The NMSSM prediction for $M_W$ differs from the MSSM prediction by $\sim$ 12 MeV.
The chargino/neu\-tra\-lino contributions can enter with both signs, and we find that
in this scenario the relatively small $\mu$ value causes negative corrections to $\Delta r$.
On the other hand, small $M_2$ values tend to give positive contributions to $\Delta r$. For the chosen parameters, these two effects cancel and 
contributions from the
chargino/neutralino sector are very small, ${\cal O}(0.1 \mev)$.
Consequently, different Higgs sector contributions give rise to the difference between the MSSM and the NMSSM curves.
Any differences in the $\cp$-odd Higgs sector have a negligible impact on the $M_W$ prediction (see also~\citere{Domingo:2011uf}).
Since we set the charged Higgs masses equal to each other in the two models, differences can only come from the $\cp$-even Higgs sector.
For this parameter point the second lightest Higgs ($M_{h_2}=15
0 \gev$) has a large singlet component ($|U^H_{23}|^2 \simeq 95 \%$), 
consequently the singlet components of $h_1$ and $h_3$ are small. $h_3$ is heavy and has no impact on the $M_W$ prediction.
Our procedure to calculate the Higgs masses in the MSSM and the NMSSM leads to the same charged Higgs masses, 
but to different predictions for the lightest $\cp$-even Higgs masses $M_{h_1}$ and $M_{h}$.
This difference arises from the different relations between the charged Higgs mass and the lightest $\cp$-even Higgs mass in the MSSM and the NMSSM.
Further it also
incorporates the (``technical'') difference due to the different radiative corrections included in {\tt FeynHiggs} and {\tt NMSSMTools} (as analysed above in the MSSM limit).
The middle plot of \reffi{fig:NMP3} shows in addition to the NMSSM prediction for $\MW$ (blue) and the MSSM prediction (red), a blue dashed curve (with open dots).
The dashed blue curve corresponds to $M_W^{\text{NMSSM}} - M_W^{\text{SM}}(M_{h_1}) +  M_W^{\text{SM}}(M_{h})$\footnote{The difference in the predictions for the lightest $\cp$-even Higgs masses in the MSSM and the NMSSM, which we subtract this way, includes both the difference between the different mass relations in the MSSM and the NMSSM, as well as the ``technical'' difference between the {\tt FeynHiggs} and the {\tt NMSSMTools} evaluation.}.
As one can see the dashed blue curve is very close to the red MSSM curve, thus here the difference between the 
MSSM and the NMSSM Higgs sector contributions to $M_W$ essentially
arises from the SM-type Higgs sector contributions, in which different Higgs
mass values are inserted.
It should be noted in this context that we have made a choice here by
comparing the
predictions for a particular NMSSM parameter point with an associated MSSM
parameter point having the same value of the 
mass of the charged Higgs boson. Accordingly, the predictions for the other
Higgs boson masses in the two models in general differ from each other, see
above, leading to the effect displayed in the left plot of \reffi{fig:NMP3}.
Instead, one could have chosen, at least in principle, the associated MSSM 
parameter point such that the masses of the lightest 
$\cp$-even Higgs masses, $M_{h_1}$ and $M_{h}$, are equal to each other. 
Also in that case differences in the other parameters in the Higgs sector,
including the mass of the charged Higgs boson, would induce a shift in the 
predictions for $\MW$.

The right plot of \reffi{fig:NMP3}
shows the $M_W^{\text{NMSSM}}$ prediction plotted against the lightest
$\cp$-even Higgs mass $M_{h_1}$. 
In this plot we display both the blue band indicating the region 
$M_{h_1} = \MHexp \pm 3.04 \gev$
as well as the grey band showing the experimental $1\,\sigma$ band from the $W$ boson mass measurement.
The black curve in the right plot indicates the SM $M_W$ prediction for $\MHSM = M_{h_1}$.
It is interesting to note that in the NMSSM it is possible to find both
the predictions for $\MW$ and for the lightest $\cp$-even Higgs mass in the
preferred regions indicated by the blue and grey bands in \reffi{fig:NMP3}.
For the SM, on the other hand, \reffi{fig:NMP3} shows the well-known result
that setting the SM Higgs boson mass to the measured experimental value one
finds a predicted value for $\MW$ which is somewhat low compared to the
experimental value.

\begin{figure}[t!]
\centering
\includegraphics[height=0.3\columnwidth]{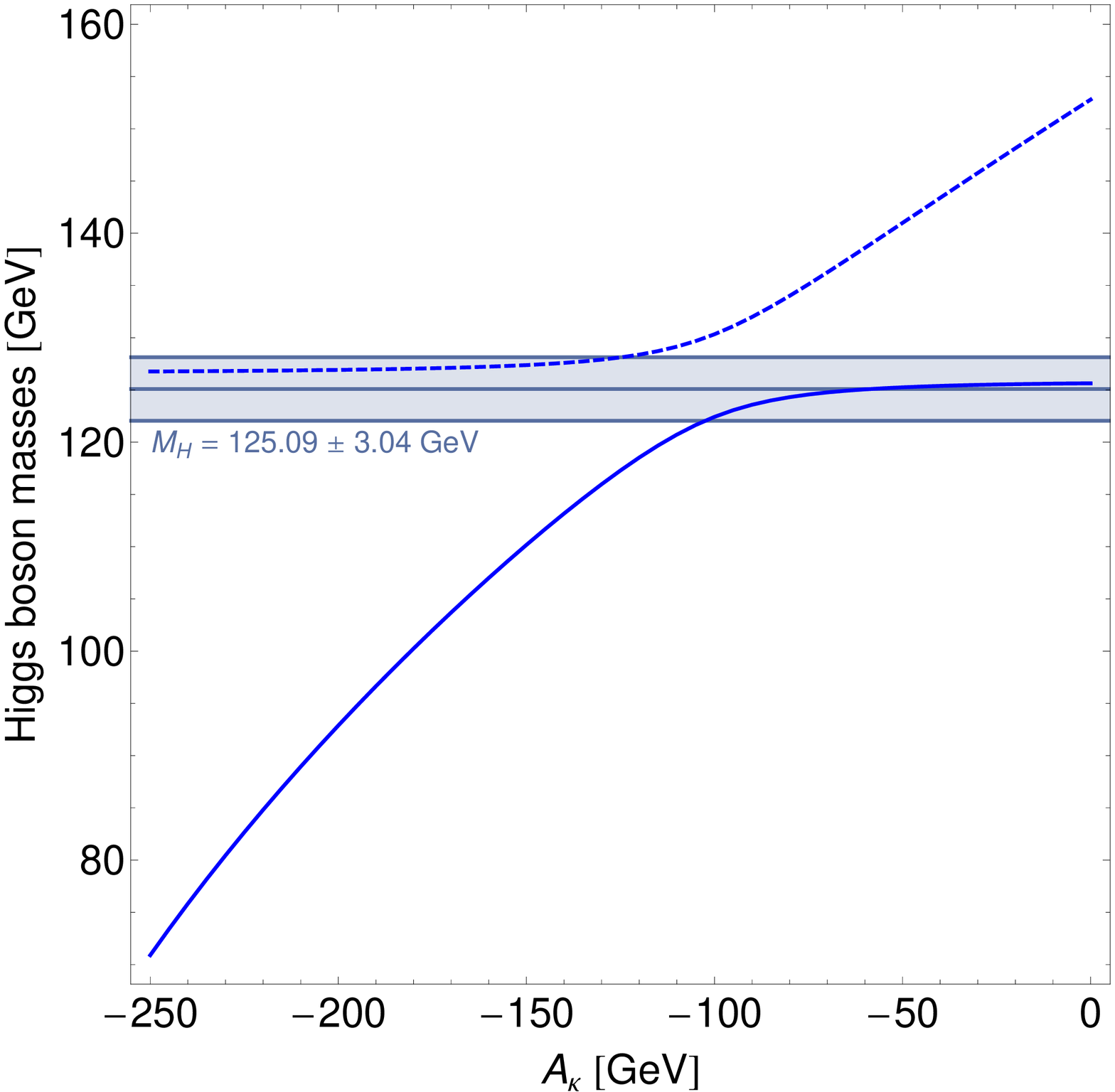}
\includegraphics[height=0.3\columnwidth]{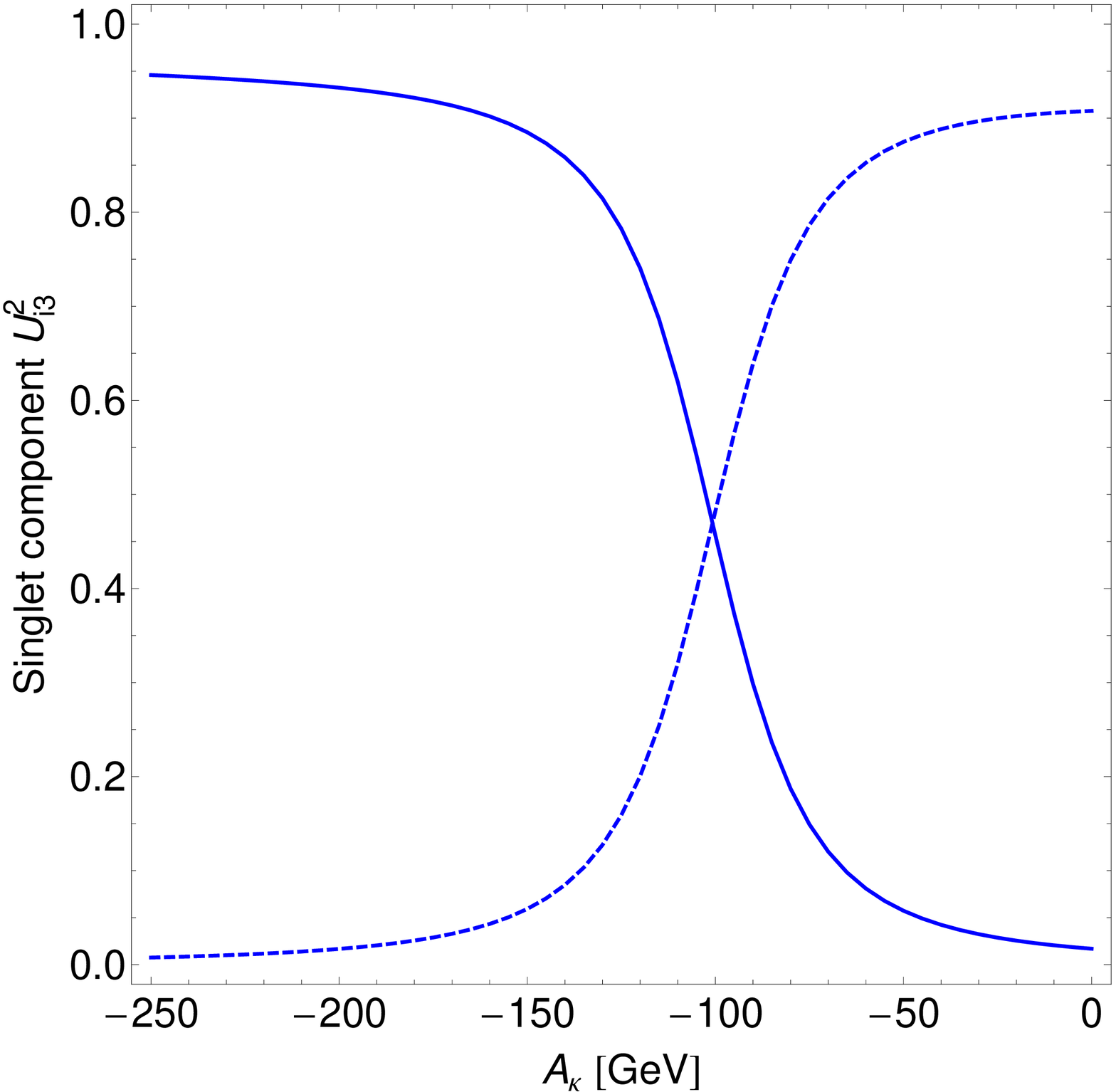}
\includegraphics[height=0.3\columnwidth]{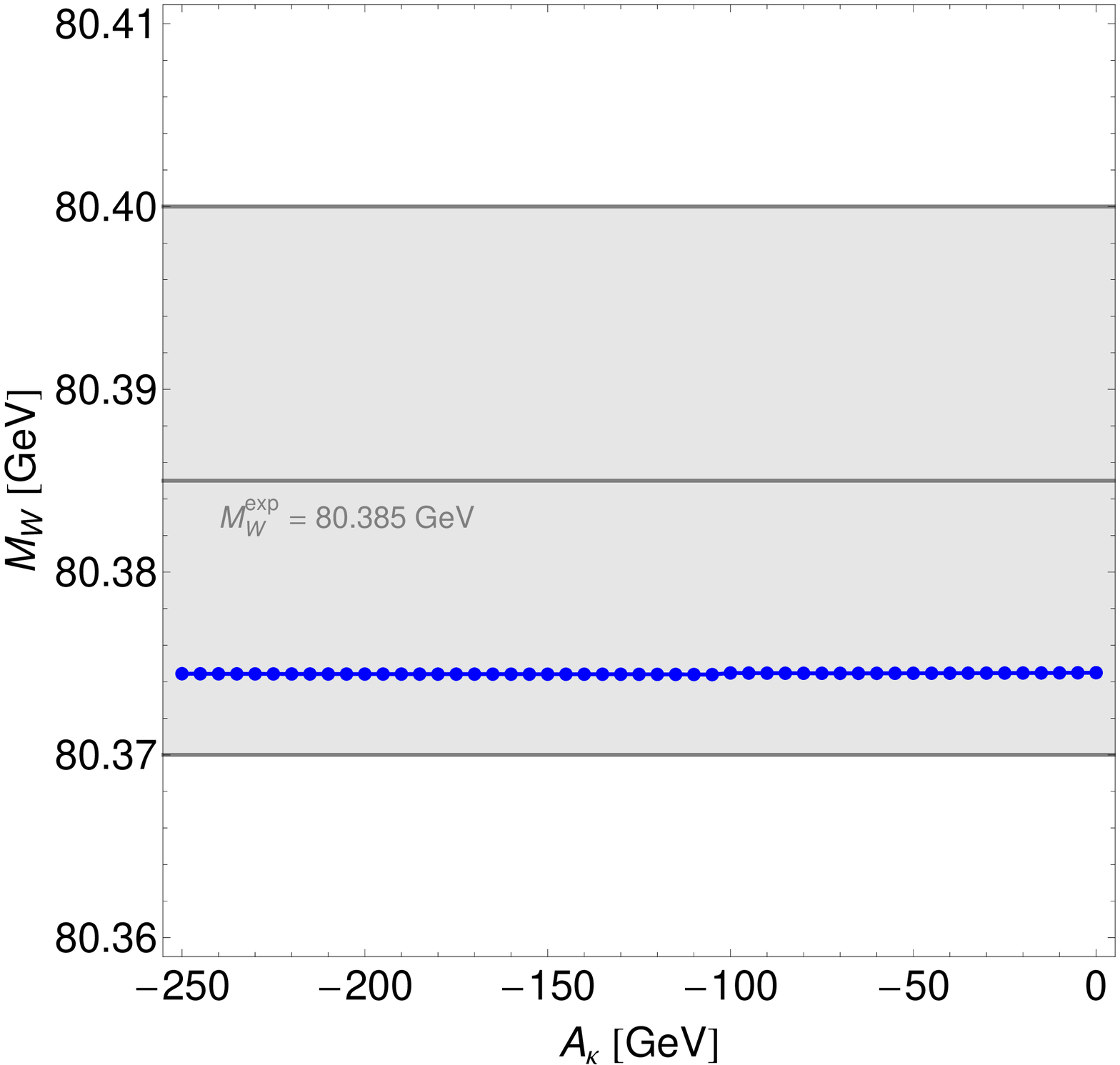}
\caption{
The left plot shows the prediction for $M_{h_1}$ (solid curve) and $M_{h_2}$ (dashed curve) as a function of $A_{\kappa}$. 
The region $\MHexp \pm 3.04 \gev$ is indicated as a blue band.
The middle plot shows the singlet components of $h_1$ and $h_2$,
$U_{13}^2$ (solid) and $U_{23}^2$ (dashed), respectively. 
The right plot shows the $M_W^{\text{NMSSM}}$ prediction. Here the grey band shows the experimental $1\,\sigma$ band from the $W$ boson mass measurement.
The parameters used for these plots
are given in the text.
}
\label{fig:singletdoubletmix}
\end{figure}

Now we want to investigate whether singlet--doublet mixing (a genuine NMSSM
feature) has a significant impact on the $M_W$ prediction.
Such a scenario is analysed in \reffi{fig:singletdoubletmix}.
Our parameters are
$m_t=173.34 \gev$, 
$\tan \beta=2$, 
$\mu = 140 \gev$, 
$M_{\tilde{L}/\tilde{E}} = 130 \gev$
$M_{\tilde{Q}/\tilde{U}/\tilde{D}_{1,2}}=1200 \gev$, 
$M_{\tilde{Q}_{3}}=800 \gev$, 
$M_{\tilde{U}_{3}}=600 \gev$, 
$M_{\tilde{D}_{3}}=1000 \gev$, 
$A_t=1300\gev$
$A_{\tau}=A_b=1000 \gev$,
$M_2=230 \gev$, 
$m_{\tilde{g}}=1500 \gev$, 
$A_{\lambda}=210 \gev$,
$\lambda = 0.55$, 
$\kappa=0.31$, and we vary $A_{\kappa}$.
These parameters are allowed by {\tt HiggsBounds} everywhere apart from $-145 \gev \lesssim A_{\kappa} \lesssim -105 \gev$, and the Higgs signal can be interpreted as either $h_1$ or $h_2$.
The left plot shows the prediction for $M_{h_1}$ (solid curve) and $M_{h_2}$ (dashed). The corresponding singlet components $U_{13}^2$ (solid) and $U_{23}^2$ (dashed) are shown in the middle plot.
The third $\cp$-even Higgs is heavy and has a negligible singlet component.
For $A_{\kappa} \lesssim -120 \gev$, $h_2$ is doublet-like and has a mass in the region of the observed Higgs signal (indicated by the blue band).
In the MSSM, scenarios which allow the interpretation of the Higgs signal as the heavy $\cp$-even Higgs
involve always a (relatively) light charged Higgs (see e.g.\ \citere{Carena:2013qia}).
Due to changed mass relations between the Higgs bosons, it is possible in the NMSSM to have the second lightest $\cp$-even Higgs at $\MHexp \gev$ together with a heavy charged Higgs. 
Therefore in the NMSSM the interpretation of the Higgs signal as the second lightest $\cp$-even Higgs is much less constrained by the LHC results from charged Higgs searches \cite{Aad:2014kga,CMSchargedhiggs}.
The interpretation of the Higgs signal as $h_2$ in this model
is always accompanied by a lighter state with reduced couplings to vector bosons.
In this figure the charged Higgs mass is $\sim 280$ GeV.
For  $A_{\kappa} \gtrsim -100 \gev$, $h_1$ is doublet-like and has a mass in the region of the observed Higgs signal.
In the ``transition" region ($-150 \gev \lesssim A_{\kappa} \lesssim -50 \gev$) the two light $\cp$-even Higgs bosons are close to each other in mass and ``share'' the singlet component.
The right plot shows the NMSSM prediction for $M_W$, which is approximately flat.
Accordingly, the parameter regions of $A_{\kappa}$ corresponding to two
different interpretations of the Higgs signal within the NMSSM lead to 
very similar predictions for the $W$ boson mass, which are in both
cases compatible with the experimental result.
Even a sizeable doublet--singlet mixing has only a minor
effect on the $\MW$ prediction in this case.

\begin{figure}[t!]
\centering
\includegraphics[height=0.3\columnwidth]{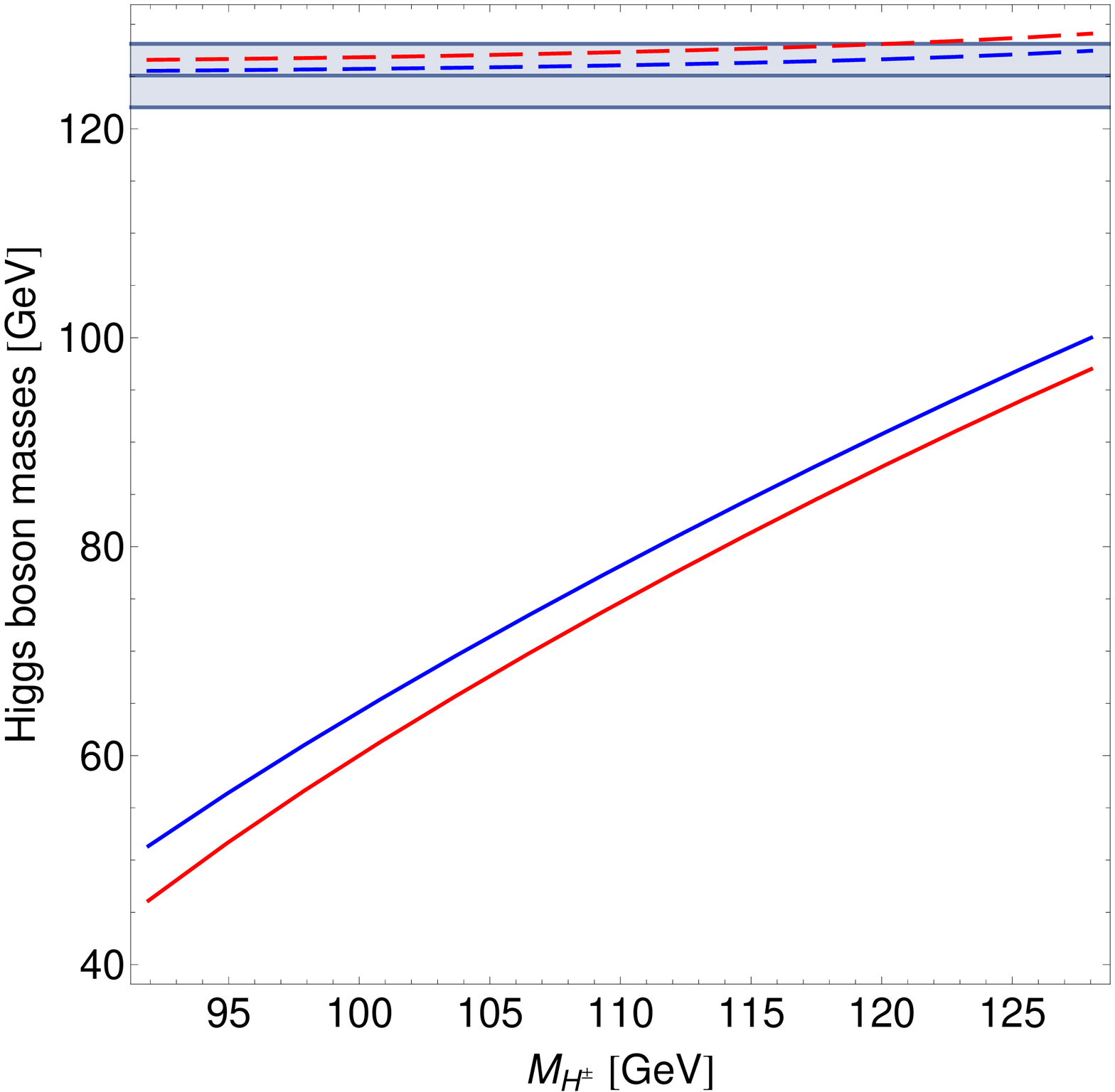}
\includegraphics[height=0.3\columnwidth]{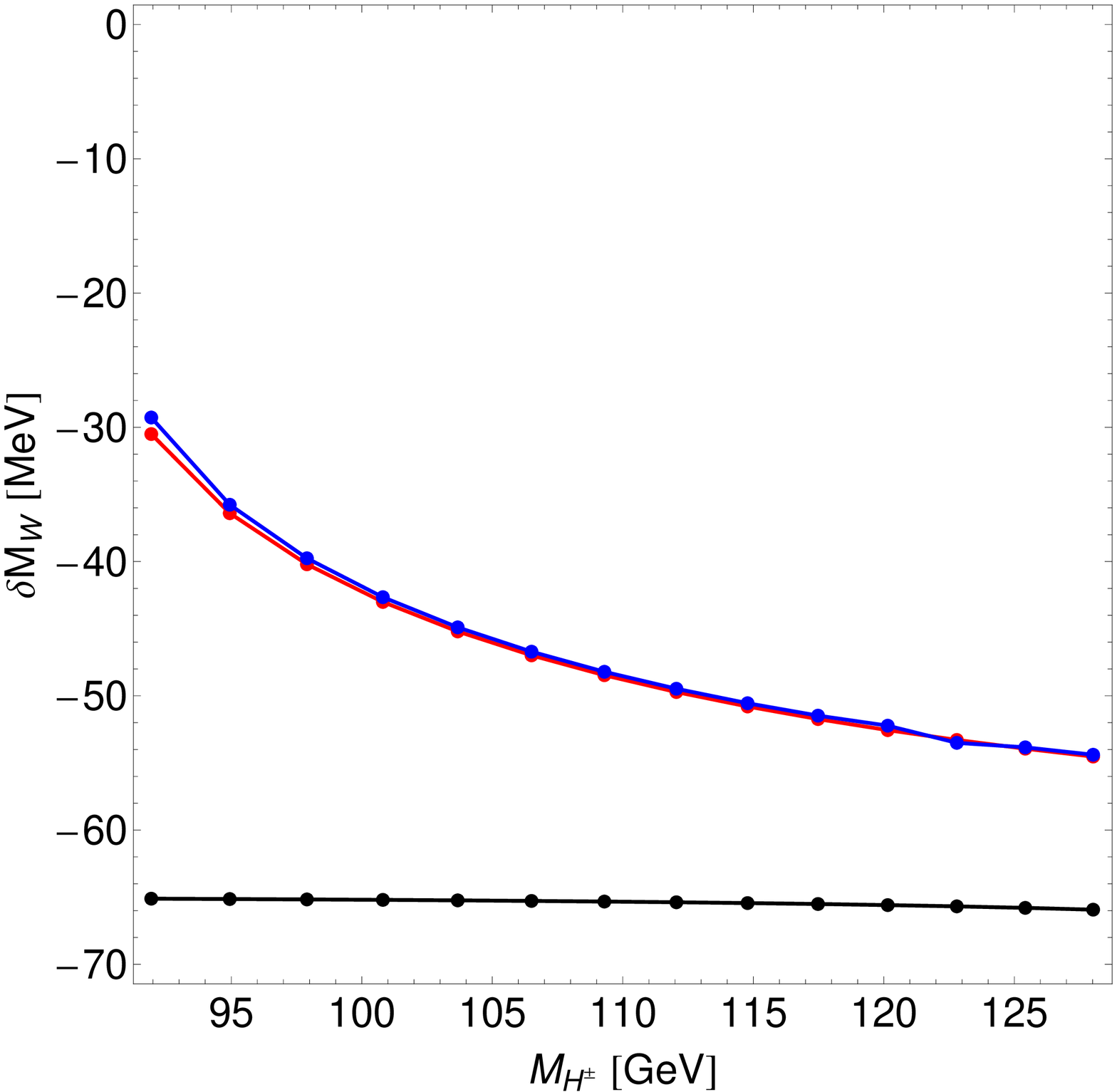}
\includegraphics[height=0.3\columnwidth]{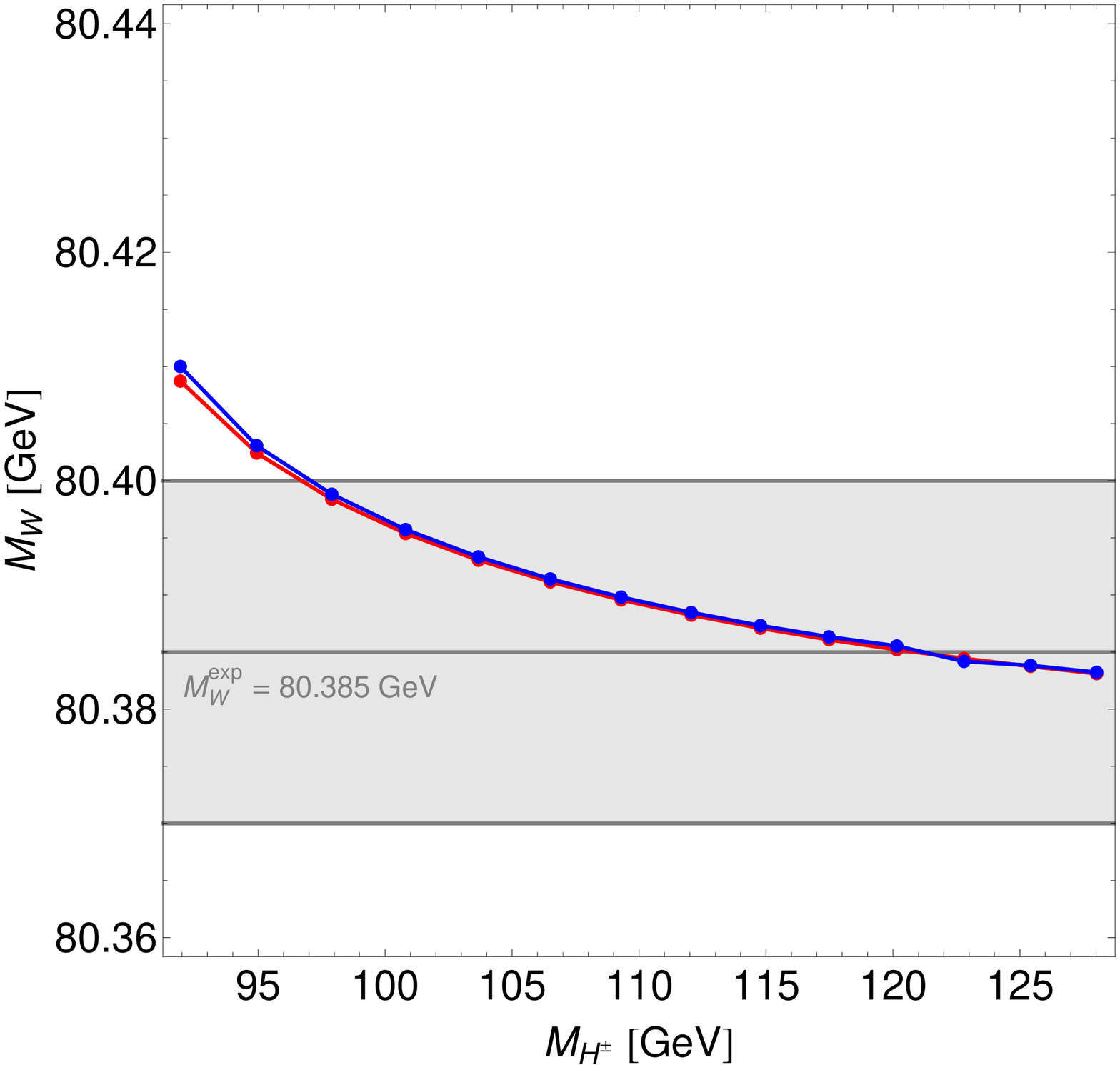}
\caption{
$\MW$ contribution from a light charged Higgs boson.
The left plot shows the prediction for the $\cp$-even Higgs boson masses in the NMSSM and in the MSSM as a function of the charged Higgs mass.
The solid curves correspond to the mass of the lightest $\cp$-even Higgs in the NMSSM (blue) and the MSSM (red).
The dashed curves correspond to the mass of the second lightest $\cp$-even Higgs in the NMSSM (blue) and the MSSM (red).
The middle plot shows the shift $\delta M_W$ (calculated as in \refeq{eq:mwshift}) induced by the Higgs and gauge boson sector in the NMSSM (blue), in the MSSM (red) and in the SM (black) with $\MHSM=M_{h_2}$.
The right plot shows the $W$ boson mass prediction in the NMSSM (blue) and the MSSM (red).
The parameters used for these plots
are given in the text.
}
\label{fig:lightmhp}
\end{figure}

We have demonstrated so far that, taking Higgs search constraints and
the information on the discovered Higgs signal into
account,\footnote{Neglecting those experimental bounds one could
have very light $\cp$-Higgs bosons with only a small singlet component, which would give large contributions to $\MW$.
However this possibility will not be discussed here.
} the genuine NMSSM effects from the extended Higgs sector are quite small,
and the Higgs sector contributions that we analysed so far were dominated 
by SM-type contributions. This is true in the absence of a light charged Higgs boson, as we will discuss now.
Light charged Higgs bosons (together with a light $\cp$-even Higgs with small but non-zero couplings to vector bosons) can lead to sizeable (non SM-like) Higgs contributions to $\MW$.
This effect can also be observed in the MSSM.
Although it is not a genuine NMSSM effect, we want to demonstrate the
impact of such a contribution here. 
For \reffi{fig:lightmhp} we choose the following parameters 
$m_t=173.34 \gev$, 
$\tan \beta=9.25$, 
$\mu = 200 \gev$, 
$M_{\tilde{L}/\tilde{E}} = 300 \gev$
$M_{\tilde{Q}/\tilde{U}/\tilde{D}_{1,2}}=1500 \gev$, 
$M_{\tilde{Q}_{3}}=M_{\tilde{U}_{3}}=M_{\tilde{D}_{3}}=1100 \gev$, 
$A_t=-2300 \gev$
$A_{\tau}=A_b=-1500 \gev$,
$M_2=500 \gev$, 
$m_{\tilde{g}}=1500 \gev$, 
$\lambda = 0.2$, 
$\kappa=0.6$, 
$A_{\kappa}=-1370 \gev$,
and we vary $\hat{m}_A$.
The left plot in \reffi{fig:lightmhp} shows the predictions for the masses of the lightest two $\cp$-even Higgs bosons in the NMSSM (blue) and in the MSSM (red)
as a function of the charged Higgs mass.
In both models the second lightest Higgs falls in the mass range $\MHexp \pm 3.04 \gev$ for the chosen parameters.
This scenario is essentially 
excluded by the latest charged Higgs searches~\cite{Aad:2014kga,CMSchargedhiggs}.
Nevertheless, we include these plots to illustrate the possible size of the contributions from a light charged Higgs.

The middle plot shows the shift $\delta M_W$ calculated as in \refeq{eq:mwshift} with $x$=gauge-boson/Higgs in the NMSSM (blue) and in the MSSM (red) while the right plot shows the full $M_W$ prediction in the NMSSM (blue) and in the MSSM (red). 
As one can see the MSSM and NMSSM contributions to $\MW$ are very similar.
Since the masses of charginos, neutralinos and sfermions stay constant when varying $\hat{m}_A$ (or $\MHp$), the change in $\MW$ with $\MHp$ stems purely from the Higgs sector.
The Higgs sector contribution to $\MW$ comes dominantly from the light charged Higgs, while 
the lightest $\cp$-even Higgs gives only a rather small contribution to $\MW$ due to its
reduced vector boson couplings. 
In the middle plot the SM result for $\delta M_W$ with $\MHSM=M_{h_2}$ is shown in black.
A significant difference between the SM Higgs contribution and the MSSM/NMSSM Higgs contributions can be observed. 
As one can see in the right plot, the displayed variation with the
charged Higgs boson mass corresponds to about a $1 \sigma$ shift in
$\MW$.

\subsubsection{Neutralino sector contributions}
\begin{figure}[t!]
\centering
\includegraphics[height=0.45\columnwidth]{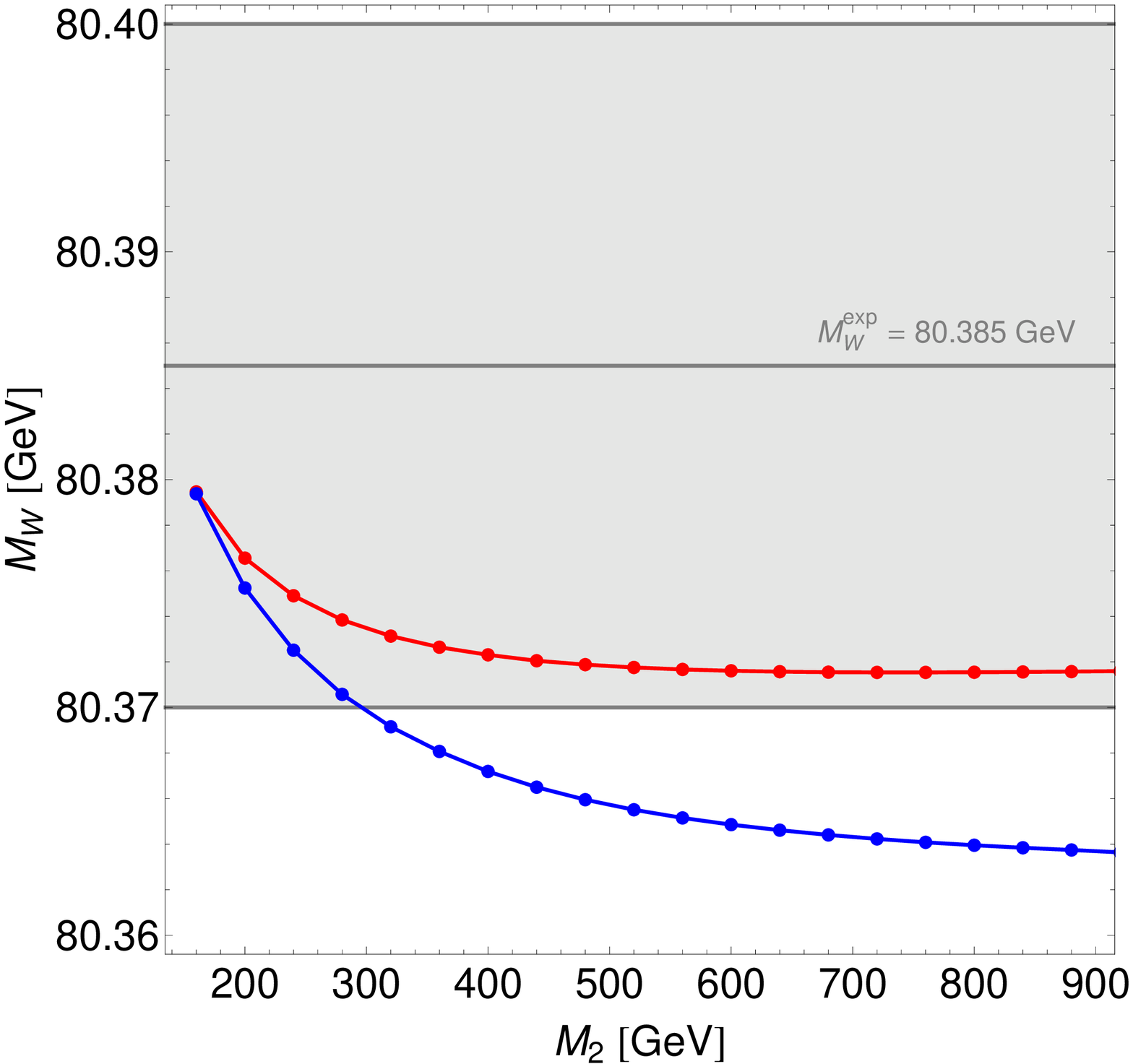}
\includegraphics[height=0.45\columnwidth]{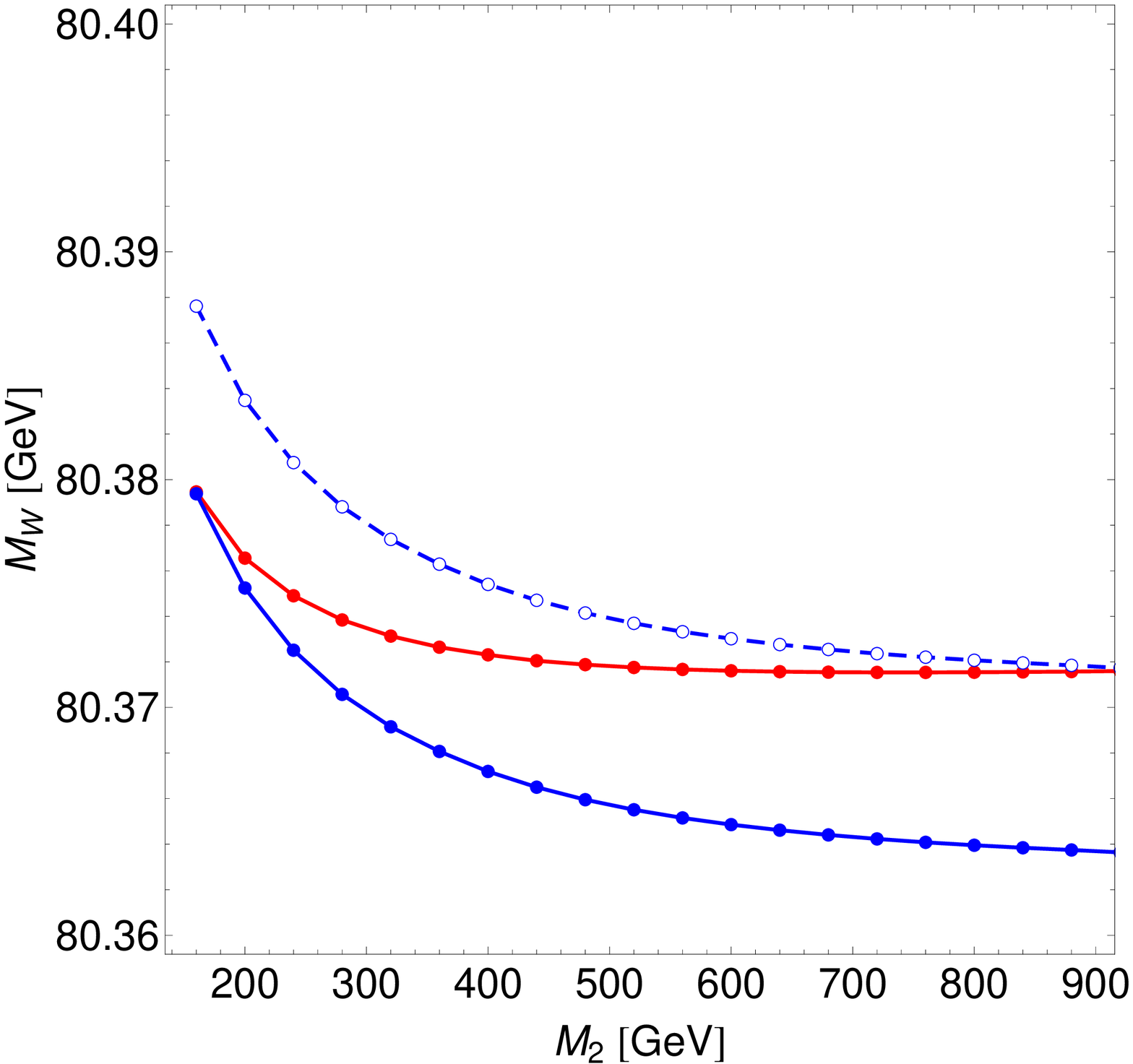}
\includegraphics[height=0.45\columnwidth]{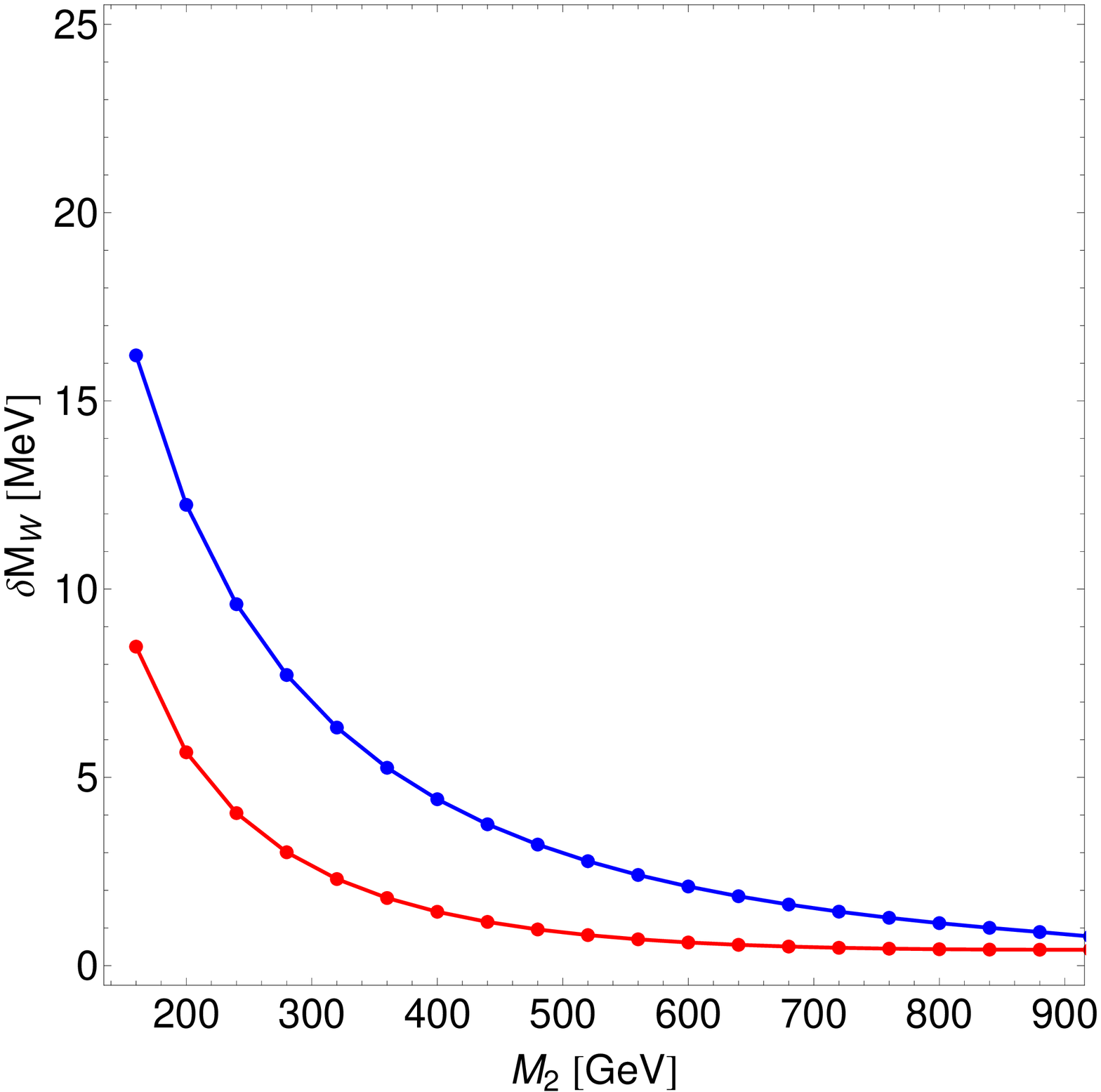}
\includegraphics[height=0.45\columnwidth]{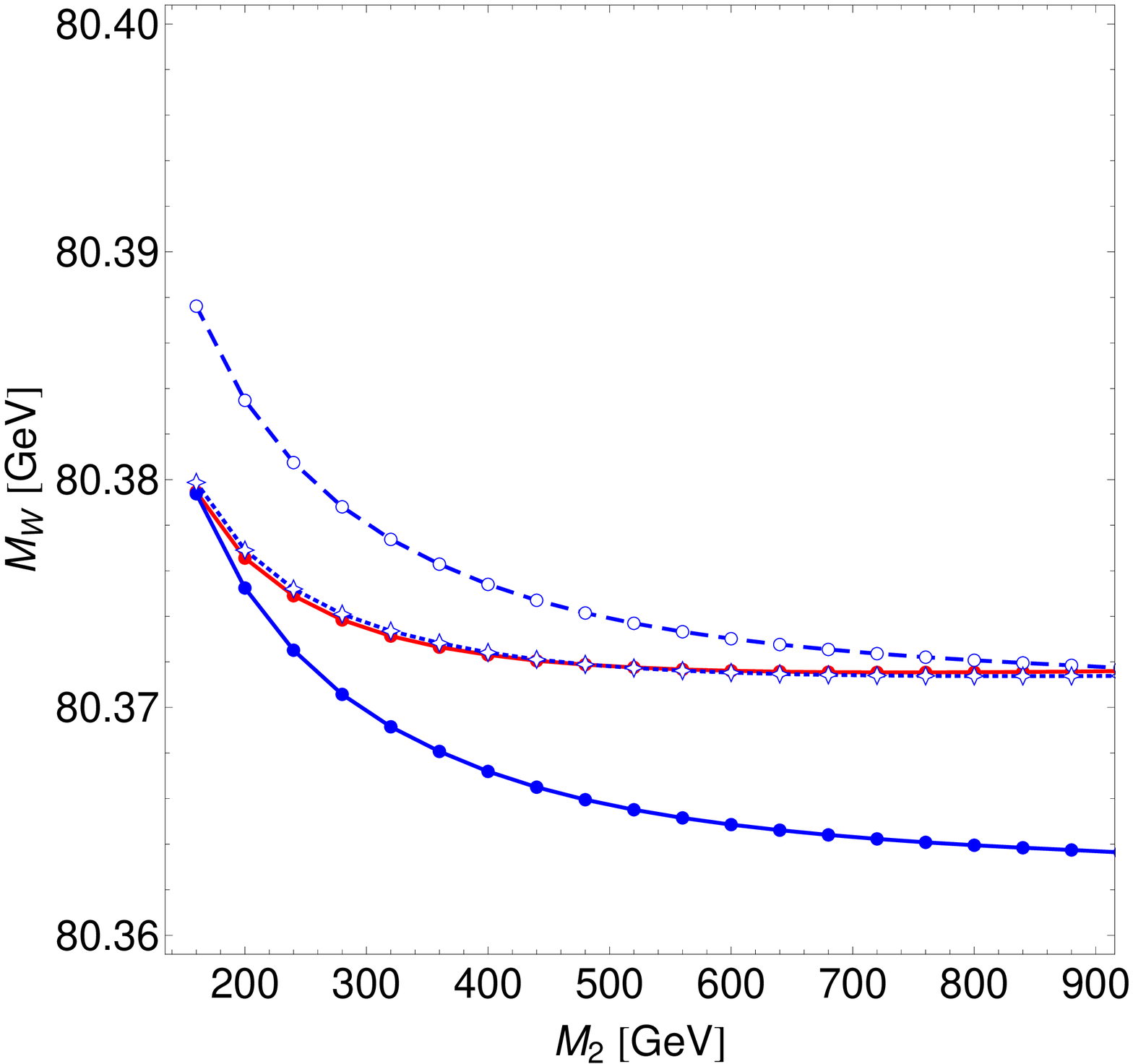}
\caption{
The upper left plot shows the $M_W^{\text{NMSSM}}$ prediction (blue) and the $M_W^{\text{MSSM}}$ prediction (red) as a function of $M_2$.
The experimental $M_W$ measurement is indicated as a grey band. 
The upper right plot shows additionally a dashed blue curve (open dots) corresponding to $M_W^{\text{NMSSM, sub}}=M_W^{\text{NMSSM}} - M_W^{\text{SM}}(M_{h_1}) +  M_W^{\text{SM}}(M_{h})$. 
The lower left plot shows the shift in the $W$ boson mass $\delta M_W$ (calculated as in \refeq{eq:mwshift} with $x$=chargino/neutralino) in the MSSM (red) and in the NMSSM (blue).
The lower right plot is similar to the upper right plot but it additionally contains the dotted blue curve (open diamonds) which corresponds to $M_W^{\text{NMSSM, sub}} - \delta M_W^{\text{NMSSM}}+\delta M_W^{\text{MSSM}}$ where $\delta M_W$ is the shift in $M_W$ induced by the chargino/neutralino contributions.
The NMSSM parameter points are allowed by {\tt HiggsBounds}, and $M_{h_1}$ falls in the range $\MHexp \pm 3.04 \gev$ for $M_2\lesssim725 \gev$.
The parameters used for these plots
are given in the text.
}
\label{fig:nmssm6}
\end{figure}
We start the discussion of the contributions from the NMSSM neutralino sector, which differs from the respective MSSM sector, with \reffi{fig:nmssm6}.
We choose 
the parameters
$m_t=173.34 \gev$, 
$\tan \beta=3$, 
$\mu = 200 \gev$, 
$M_{\tilde{L}/\tilde{E}}= 1000 \gev$,
$M_{\tilde{Q}/\tilde{U}/\tilde{D}_{1,2}}=1500 \gev$, 
$M_{\tilde{Q}_{3}}=M_{\tilde{U}_{3}}=650 \gev$, 
$M_{\tilde{D}_{3}}=1000 \gev$, 
$A_t=A_{\tau}=A_b=1000 \gev$,
$m_{\tilde{g}}=1500 \gev$, 
$A_{\lambda}=580 \gev$,
$\lambda = 0.64$, 
$\kappa=0.25$, 
$A_{\kappa}=-10 \gev$, and we vary $M_2$.
In the upper left plot, the blue curve shows the $M_W^{\text{NMSSM}}$ prediction and the red curve the $M_W^{\text{MSSM}}$ prediction.
The difference between the NMSSM prediction and the MSSM prediction is small for $M_2 \lesssim 200 \gev$ and increases for larger $M_2$ values.
The origin of this difference is investigated in the other three plots of \reffi{fig:nmssm6}.
As before our procedure to identify an MSSM point which can be compared to the NMSSM point implies different predictions for the lightest $\cp$-even Higgs mass.
Here we subtract again the difference in the SM contributions, arising from the different Higgs mass predictions.
The additional blue dashed curve (with open dots) in the upper right plot of \reffi{fig:nmssm6} corresponds to $M_W^{\text{NMSSM, sub}}=M_W^{\text{NMSSM}} - M_W^{\text{SM}}(M_{h_1}) +  M_W^{\text{SM}}(M_{h})$.
For large $M_2$ the difference between the NMSSM and the MSSM prediction for $\MW$ can be fully explained by the difference in the (SM-type) Higgs mass contributions, which arise from inserting different predictions for $M_{h_1}$ and $M_h$.
However after subtracting the difference from the Higgs mass contributions we observe a sizeable difference between $M_W^{\text{NMSSM, sub}}$ and $M_W^{\rm MSSM}$ for small $M_2$.
This difference stems from different sizes of the chargino/neutralino sector contributions between the two SUSY models, which 
tend to compensate the difference between $M_W^{\text{NMSSM}}$ and $M_W^{\text{MSSM}}$ arising from the Higgs sector. 
This can be seen in the lower left plot, where we display the shift
$\delta M_W$ (calculated as in \refeq{eq:mwshift}) induced by the chargino/neutralino contributions in the MSSM (red) and in the NMSSM (blue).
At $M_2 = 160 \gev$ the chargino mass is $108 \gev$ and thus just above the LEP limit.
The $\delta M_W$ contribution from the chargino/neutralino sector in the
MSSM reaches 8.5 MeV in this case.\footnote{This is not the maximal
effect possible for the chargino/neutralino contributions in the MSSM.
The chargino/neutralino contributions depend on the slepton masses (see
diagrams in \reffis{fig:ChaNeu_FSE}-\ref{fig:box}). For lighter slepton
masses the chargino/neutralino contributions in the MSSM can reach up to 20
MeV, as analysed in \citere{Heinemeyer:2013dia}.}
In the NMSSM the maximal $\delta M_W$ contribution from the chargino/neutralino sector is 16.5 MeV --- significantly larger than in the MSSM. 
Both in the MSSM and the NMSSM, the chargino/neutralino contributions decrease when increasing $M_2$ and therewith the chargino and neutralino masses, showing the expected behaviour when decoupling the gaugino sector.
The largest difference between the NMSSM and the MSSM chargino/neutralino contributions is $\sim 8$ MeV (at $M_2 = 160 \gev$).
The difference arises from the neutralino sector, since the chargino sector is unchanged in the NMSSM with respect to the MSSM.
We will discuss in more detail below why the contributions from the neutralino sector are larger in the NMSSM than in the MSSM.
The lower right plot of \reffi{fig:nmssm6} is similar to the upper right plot, but it contains a fourth curve (blue dotted with open diamonds) which was obtained by subtracting the different chargino/neutralino contributions, thus it corresponds to $M_W^{\text{NMSSM,sub}} - \delta M_W^{\text{NMSSM}}+\delta M_W^{\text{MSSM}}$.
This curve lies very close to the MSSM prediction. We have therefore identified the contributions causing the 
difference between the $M_W^{\text{NMSSM}}$ and the $M_W^{\text{MSSM}}$ predictions.

\begin{figure}[t!]
\centering
\includegraphics[height=0.45\columnwidth]{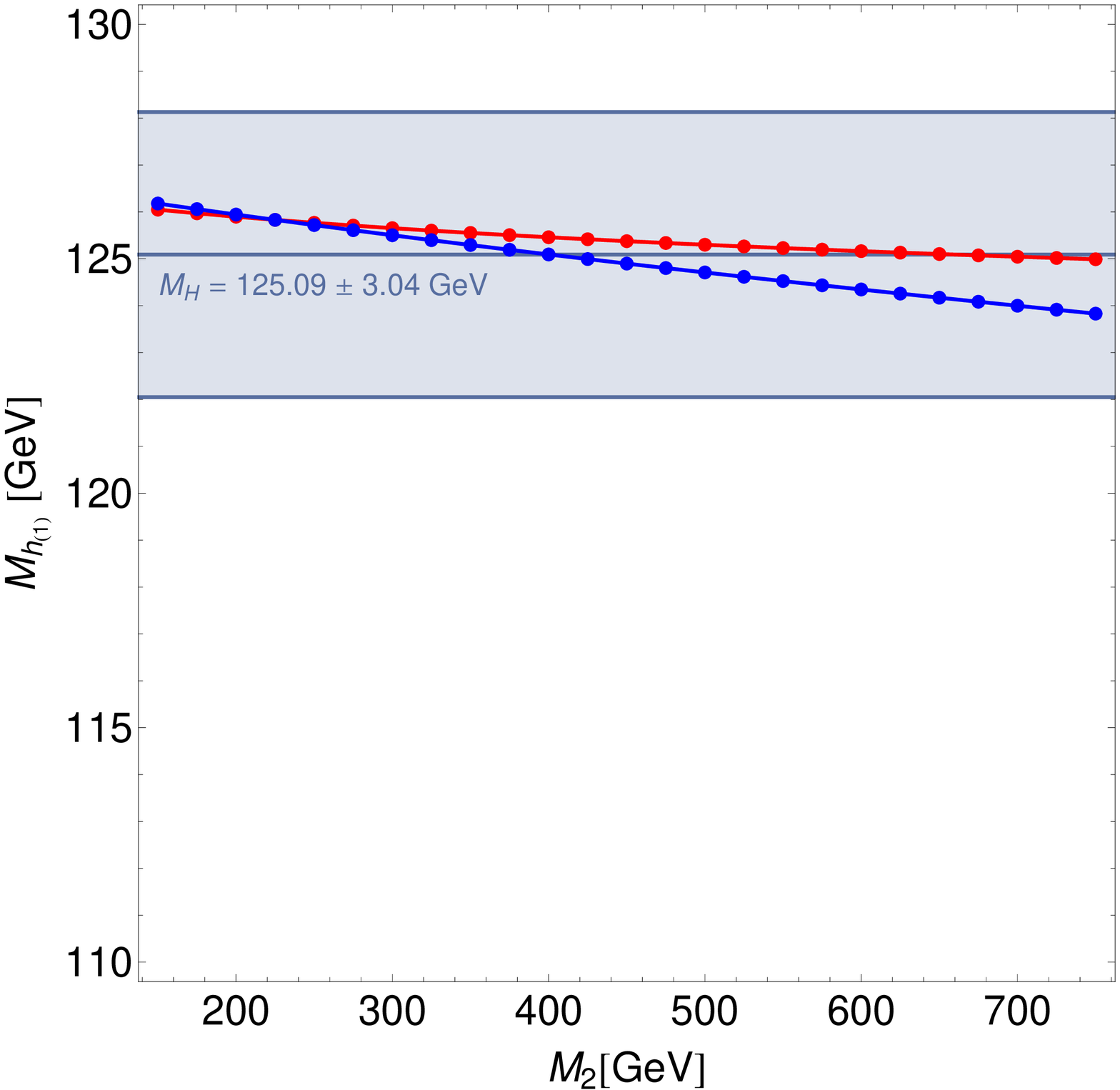}
\includegraphics[height=0.45\columnwidth]{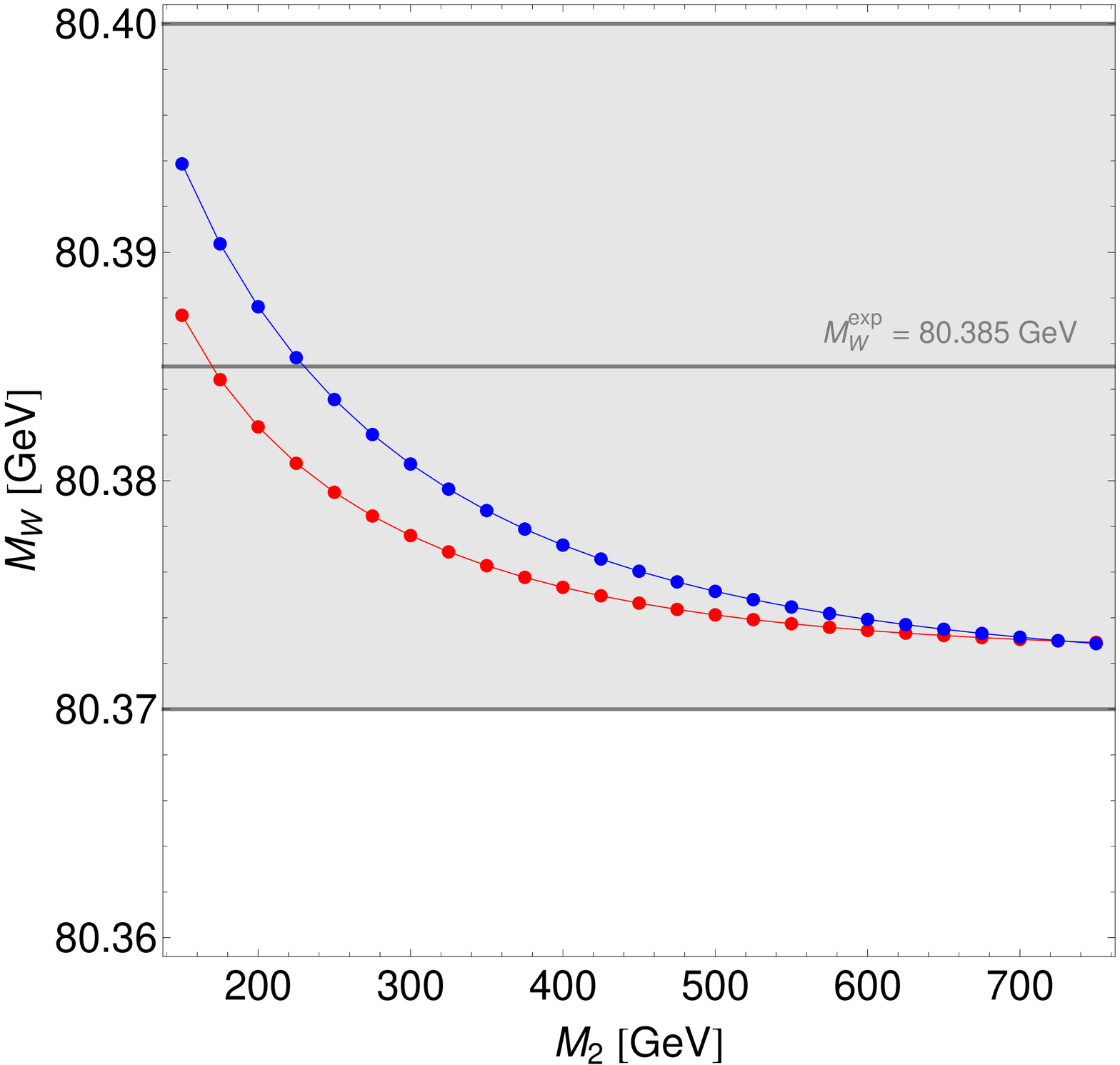}
\includegraphics[height=0.45\columnwidth]{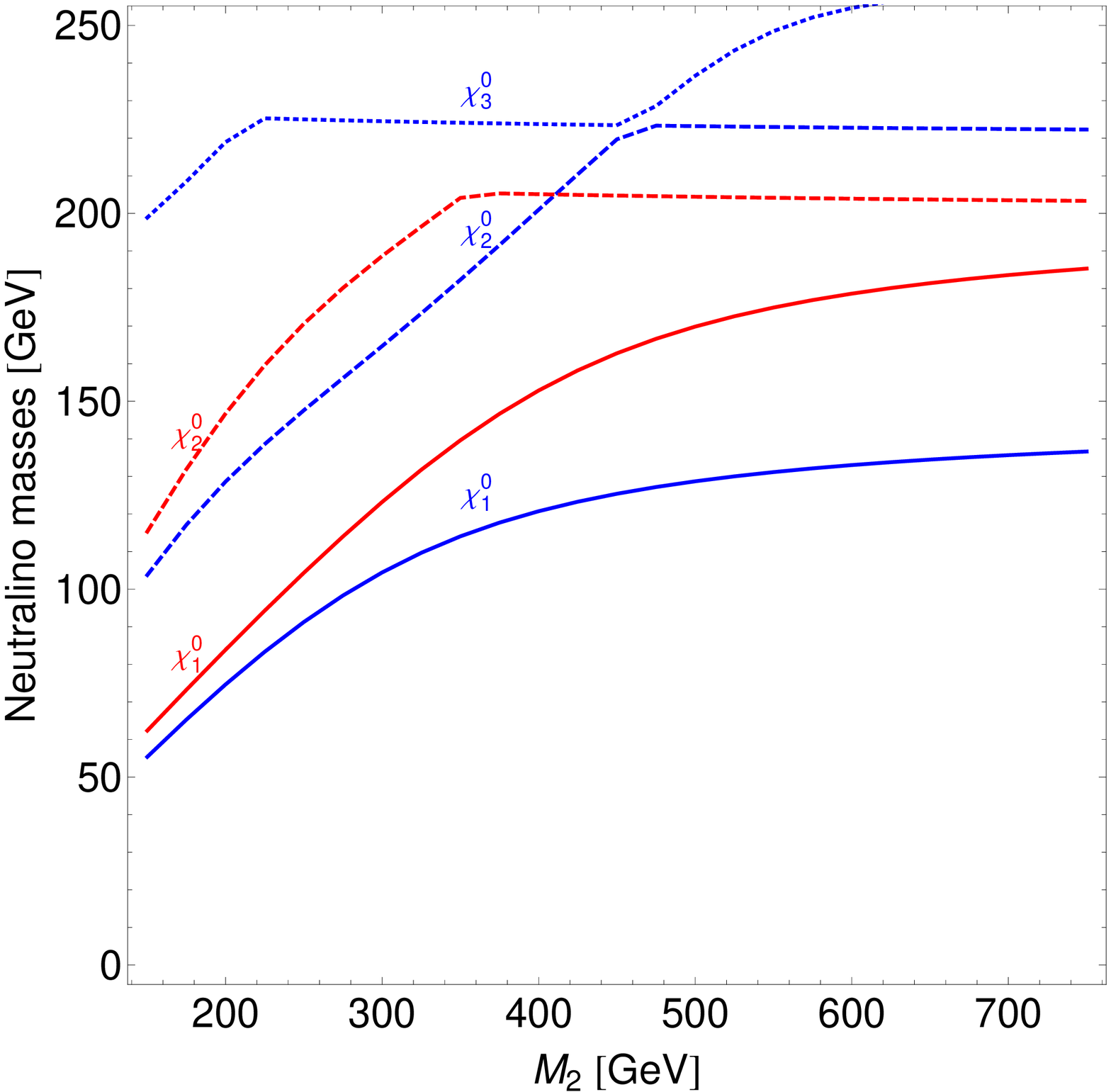}
\includegraphics[height=0.45\columnwidth]{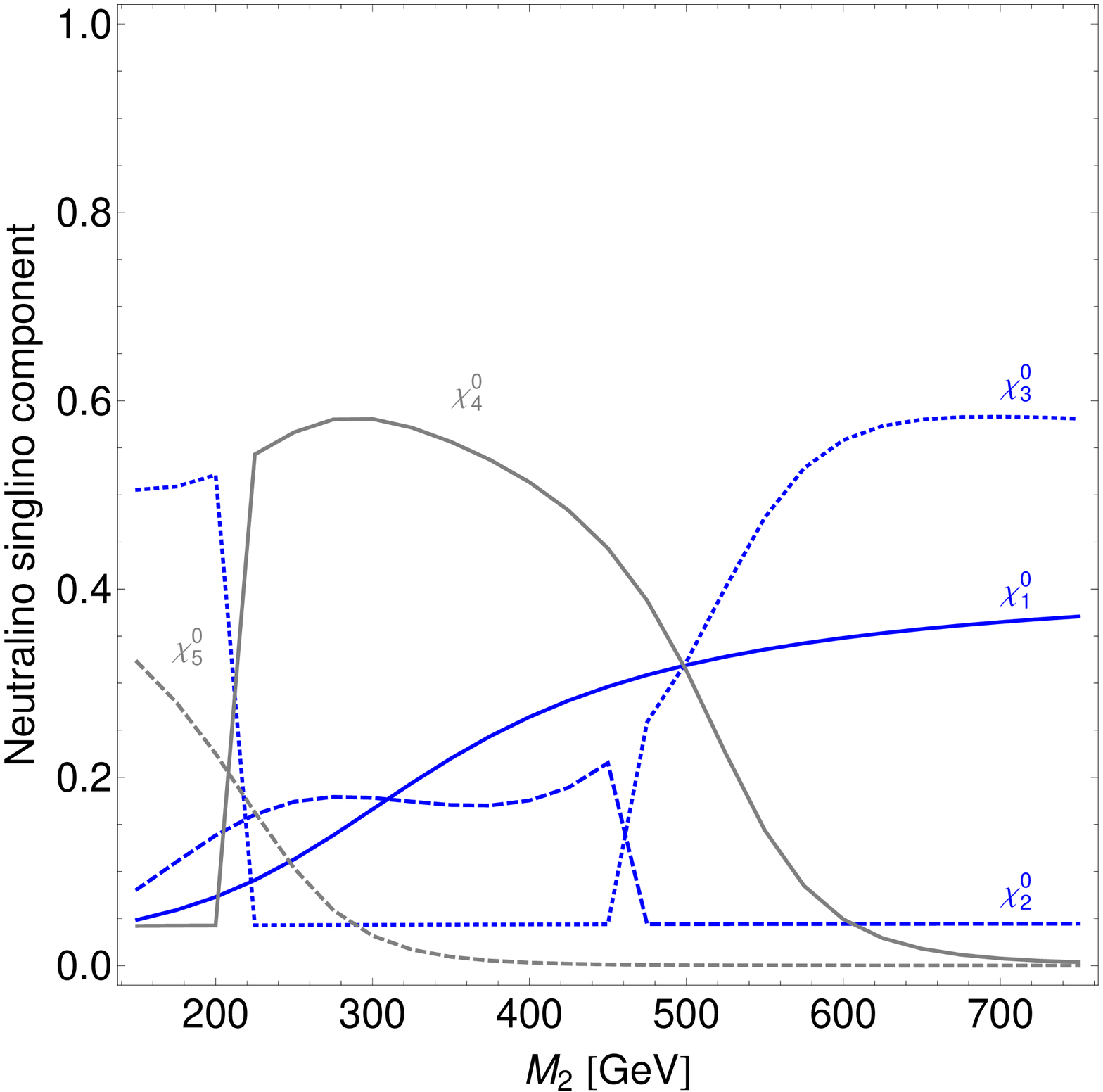}
\caption{
The upper left plot shows the masses of the lightest $\cp$-even Higgs boson in the NMSSM (blue) and the MSSM (red) as a function of $M_2$.
The upper right plot shows the prediction for $M_W^{\text{NMSSM}}$ (blue) and for $M_W^{\text{MSSM}}$  (red). 
The lightest three neutralino masses and the neutralino singlet components are displayed in the lower row.
The parameters (given in the text) are chosen such that the Higgs sectors of
the MSSM and the NMSSM are very similar to each other.
The parameter region in
both models is allowed by {\tt HiggsBounds} and predicts the lightest $\cp$-even Higgs (which is SM-like) close to $\MHexp \gev$.
}
\label{fig:neuttralinopheno}
\end{figure}
We continue with the discussion of the neutralino contributions to $\MW$ in the NMSSM in \reffi{fig:neuttralinopheno}.
The chosen parameters are 
$m_t=173.34 \gev$, 
$\tan \beta=5.5$, $\mu = 200 \gev$, 
$M_{\tilde{L}/\tilde{E}}=245 \gev$, 
$M_{\tilde{Q}/\tilde{U}/\tilde{D}_{1,2}}=1500 \gev$, 
$M_{\tilde{Q}_{3}}=M_{\tilde{U}_{3}}=M_{\tilde{D}_{3}}=1000 \gev$, 
$A_t=A_{\tau}=A_b\simeq1964 \gev$, 
$m_{\tilde{g}}=1500 \gev$, 
$\hat{m}_A=1200 \gev$,
$\lambda =0.62$, 
$\kappa=0.3$, 
$A_{\kappa}=-10 \gev$, and $M_2$ is varied.
All parameter points are {\tt HiggsBounds} allowed.
Again we get the
MSSM prediction by setting the {\tt FeynHiggs} 
$M_{H^{\pm}}$ input to the value of the charged Higgs mass calculated by {\tt NMSSMTools}. 
For this set of parameters this procedure leads to a scenario where the MSSM
and the NMSSM Higgs boson sectors are very similar to each other. Both
models predict the lightest $\cp$-even Higgs close to the experimental value
$\MHexp \gev$, as one can see in the upper left plot of
\reffi{fig:neuttralinopheno} showing the masses of the two states $M_h$ (MSSM, red) and $M_{h_1}$ (NMSSM, blue).
The difference between $M_h$ and $M_{h_1}$ is $\lesssim 1 \gev$, resulting in a small (${\mathcal O} (1 \mev)$) difference in $M_W$ from the Higgs sector contributions.
The upper right plot of \reffi{fig:neuttralinopheno} displays the $W$ boson mass prediction in the NMSSM (blue) and in the MSSM (red).
The difference between these two predictions is largest (7 MeV) for $M_2=150 \gev$ and (almost) vanishes for large $M_2$.
Since differences in the Higgs sector contributions are quite small, the difference between $M_W^{\text{NMSSM}}$ and $M_W^{\text{MSSM}}$ arises predominately from the differences in the neutralino sector.
We note that in this scenario both $M_{h_1}$ and $\MW$ lie within the
preferred regions indicated by the blue and grey bands for the whole
parameter range displayed in the figure.

In order to investigate the reasons for the different predictions for the chargino/neutra\-lino contributions we plot the masses of the three lightest neutralino states in the NMSSM (blue) 
and the MSSM (red) in the lower left plot. The other MSSM/NMSSM neutralinos are heavier than 250 GeV and hardly affect the $M_W$ prediction.
We set here the (unphysical) soft masses $M_1$ and $M_2$ equal in the MSSM and the NMSSM and identify the MSSM $\mu$ parameter with the effective $\mu$ of the NMSSM.
The resulting predictions for the masses of $\tilde{\chi}^0_1$ and $\tilde{\chi}^0_2$ are a few GeV lower in the NMSSM than in the MSSM. 
The singlino components of the NMSSM neutralinos, $|N_{i5}|^2$, where $N$ was defined in \refeq{eq:neuNNMSSM}, are shown in the lower right plot,
and we can observe a strong mixing between the five states.
The singlino components of $\tilde{\chi}^0_1$ and $\tilde{\chi}^0_2$ are below $10\%$ for $M_2 = 150 \gev$ and increase 
up to $40\%$($20\%$) for $\tilde{\chi}^0_1$($\tilde{\chi}^0_2$) for higher $M_2$ values.
The lighter neutralino states (with relatively small singlino component) lead to larger contributions from the neutralino sector to $\MW$ in the NMSSM compared to the MSSM.

\begin{figure}[t!]
\centering
\includegraphics[height=0.45\columnwidth]{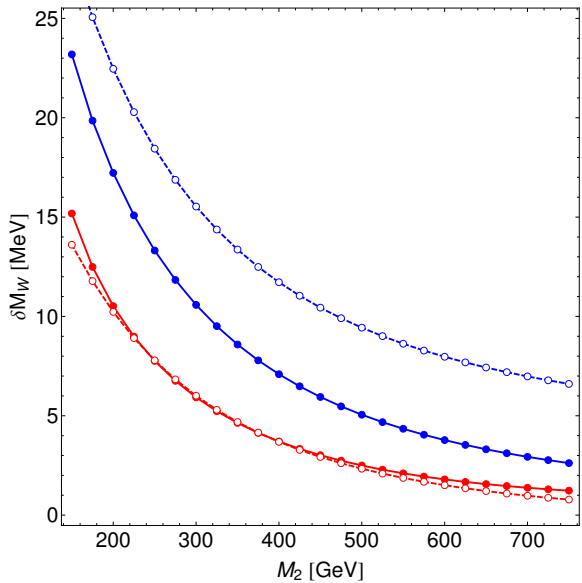}
\caption{
The shifts $\delta M_W$ in the NMSSM (blue) and in the MSSM (red),
calculated taking the full chargino/neutralino contribution to $\Delta r$ into account (solid)
and using only the $\Delta \rho$ approximation (dashed).
The parameters are chosen as in \reffi{fig:neuttralinopheno}.
}
\label{fig:neuttralinodeltarho}
\end{figure}
In the next step we analyse how well the full $\Delta r$ contribution of the chargino/neutra\-lino sector
can be approximated by taking into account only the leading term $-c_W^2/s_W^2\,\Delta \rho$ (defined in \refeq{eq:deltarho}).
The $\Delta \rho$ term contains only the $W$ and $Z$ boson self-energies at zero momentum transfer, thus this approximation neglects in particular 
the contributions from box, vertex and fermion self-energy diagrams containing charginos and neutralinos.
The $\Delta \rho$ term corresponds to the $T$ parameter of the $S, T, U$
parameters \cite{Peskin:1990zt,Marciano:1990dp}, often used to parametrize new physics contribution to electroweak precision observables.
For the plot in \reffi{fig:neuttralinodeltarho} we use the same parameters as in \reffi{fig:neuttralinopheno}. 
Again the blue(red) solid curve shows the 
$\delta M_W$ shift as a function of $M_2$, calculated as in \refeq{eq:mwshift} with $x$=chargino/neutralino in the NMSSM(MSSM)
(the two solid curves are identical to the ones in the upper right plot of \reffi{fig:neuttralinopheno}).
The two dashed curves show the $M_W$ contributions in the NMSSM (blue) and in the MSSM (red) obtained when the full 
$\Delta r^{\text{chargino/neutralino}(\alpha)}$ is approximated by 
the chargino and neutralino contributions to the 
$\Delta \rho$ parameter:
\BE
\delta M_W = - \frac{M_W^{\rm ref}}{2}\,\frac{s_W^2}{c_W^2-s_W^2}\,
\left(-\frac{c_W^2}{s_W^2}\right)\, \Delta \rho^{\text{chargino/neutralino}} .
\label{eq:mwshiftrho}
\EE
In the MSSM the $\Delta \rho$ term containing charginos and neutralinos provides a very good approximation of the full $\Delta r$ term in the intermediate range $200 \gev \lesssim M_2 \lesssim 500\gev$.
In the range of small and large $M_2$ values, $\Delta \rho$ slightly underestimates the full $\Delta r$ contribution, the difference here is 
$\sim 1.5 \mev$ for $M_2=150 \gev$ and $\sim 0.5 \mev$ for $M_2=750 \gev$.
In the NMSSM the $\Delta \rho$ term gives a $\delta M_W$ contribution which is larger ($\gtrsim 4 \mev$) than the full $\Delta r$ result for the full $M_2$ range plotted here.
It should be noted that the chargino/neutralino sector does not completely
decouple for large $M_2$ in this case, which is a consequence of the presence of a 
light Higgsino, $\mu=200 \gev$.
For $M_2=750\gev$ the lightest neutralino has a mass of $M_2=140 \gev$, with a singlino component of $\sim 40\%$ and a Higgsino component of $\sim 60\%$.
In this scenario the singlino-higgsino mixing leads to a positive contribution to $\Delta \rho$, but to a negative contribution to the $\Delta r$ terms beyond $\Delta \rho$ (we checked that the contribution from the box diagrams is negligible for large $M_2$ values).
We also checked that going to large $\mu$ values, the chargino/neutralino sector decouples and all terms vanish.
In this scenario the two effects largely cancel each other and for large $M_2$ one finds a small positive value for the full $\Delta r$ result.
This however depends on the chosen parameters and the admixture of the light
neutralino, e.g.\ in the scenario discussed in \reffi{fig:NMP3} the negative contributions 
exceed the positive ones so that the full $\Delta r$ result 
is negative for large $M_2$.
Thus, we have shown that the $\Delta \rho$ approximation for the
chargino and neutralino contributions works quite well in the MSSM, whereas sizeable corrections to $\MW$ beyond the $\Delta \rho$ approximation can occur in the NMSSM.

\begin{figure}[t!]
\centering
\includegraphics[height=0.3\columnwidth]{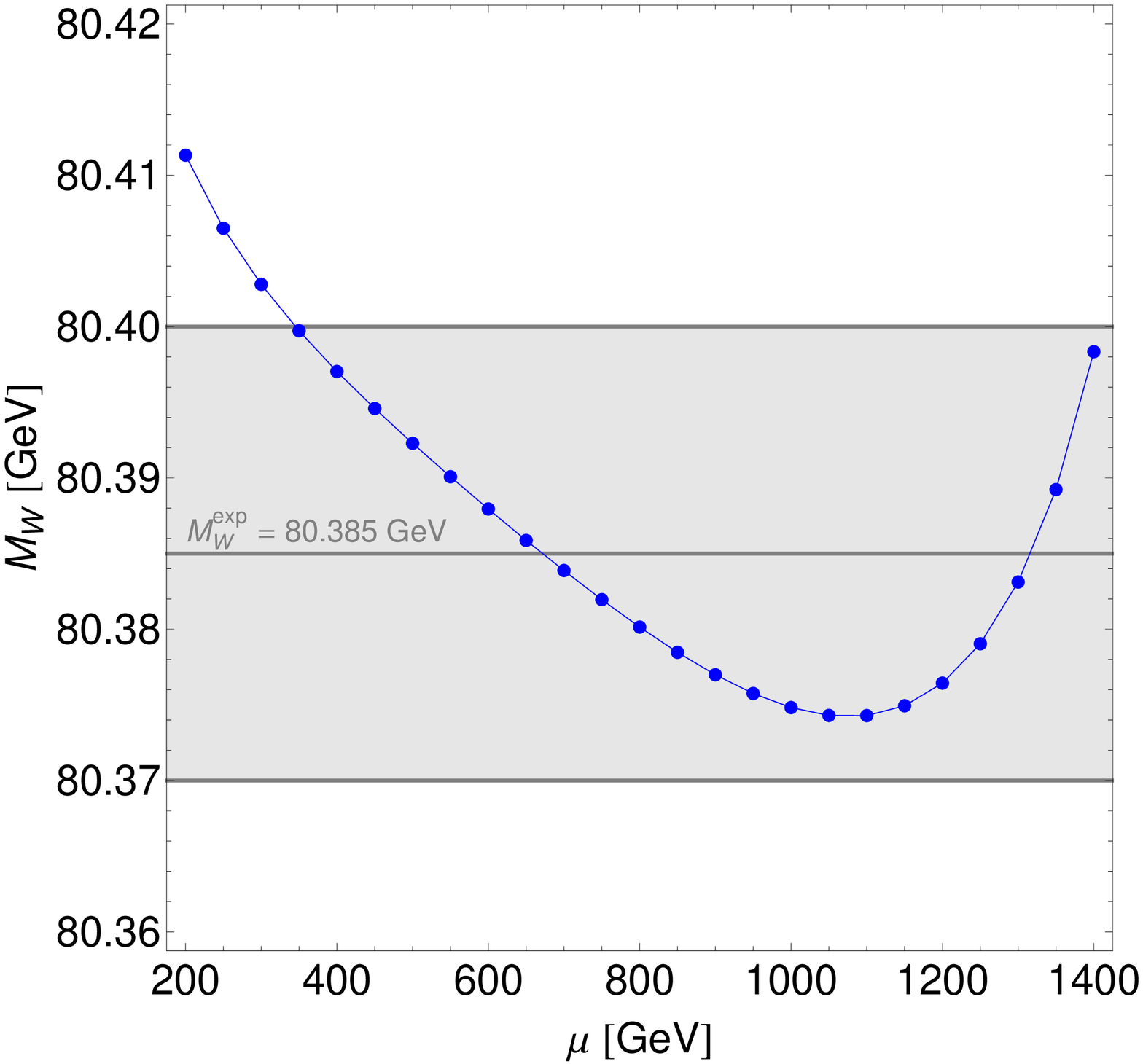}
\includegraphics[height=0.3\columnwidth]{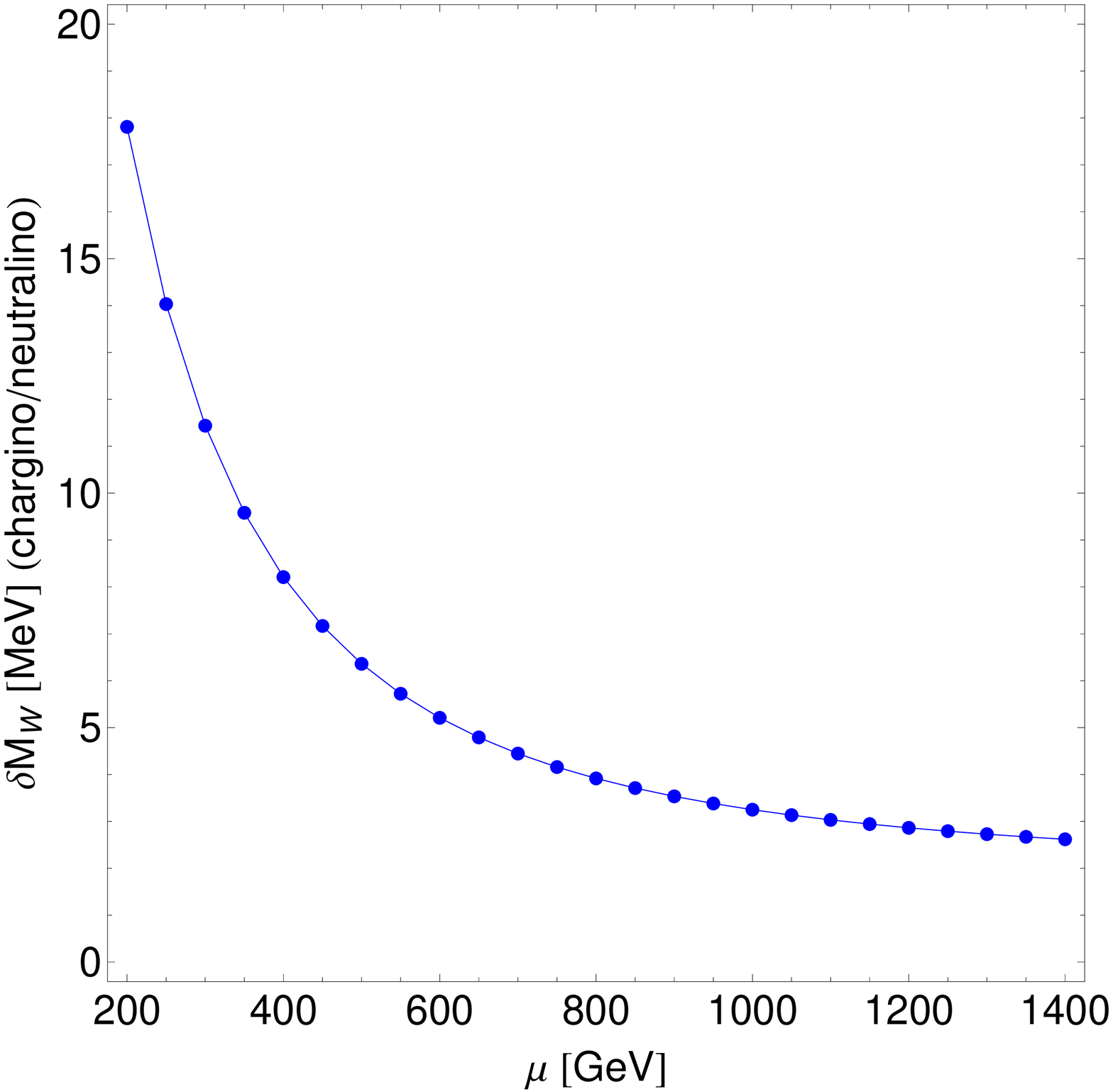}
\includegraphics[height=0.3\columnwidth]{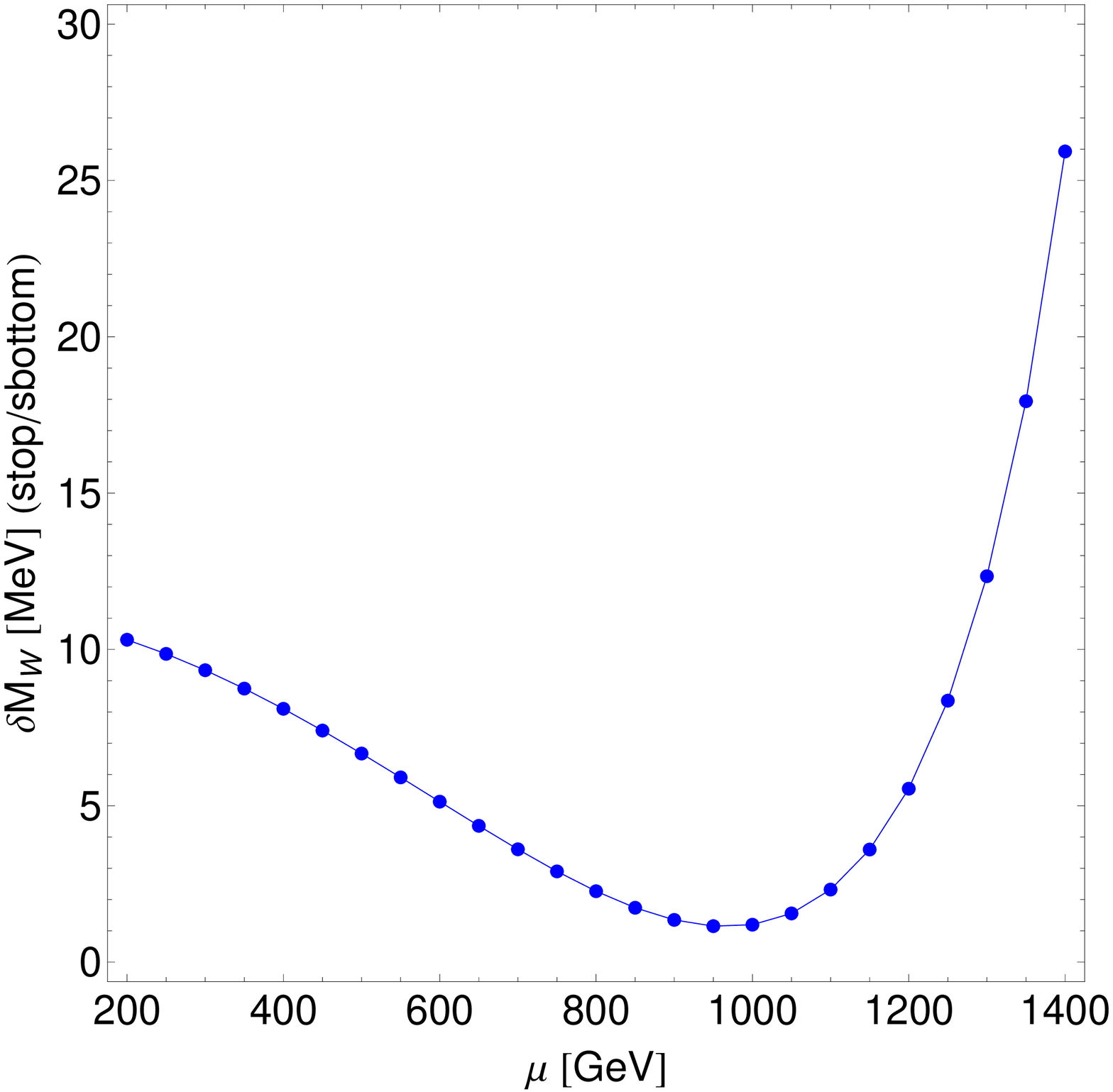}
\caption{
Dependence of the $W$ boson mass prediction in the NMSSM on the $\mu$ parameter.
The left plot shows the $M_W^{\text{NMSSM}}$ prediction, the middle one the $\delta M_W$ contribution from the chargino/neutralino sector and the right one 
shows the $\delta M_W$ contribution from the stop/sbottom sector.
The parameters are given in the text. 
}
\label{fig:mwmudependence}
\end{figure}
As a final step we want to discuss the dependence of the $\MW$ prediction in the NMSSM on the $\mu$ parameter, which enters both in the sfermion and in the chargino/neu\-tralino sectors.
The left plot of \reffi{fig:mwmudependence} shows the $W$ boson mass prediction in the NMSSM as a function of $\mu$, with 
the parameters chosen as
$m_t=173.34 \gev$, $\tan \beta=20$, 
$M_{\tilde{L}/\tilde{E}}=250\gev$, 
$M_{\tilde{Q}/\tilde{U}/\tilde{D}_{1,2}} = 1500\gev$, 
$M_{\tilde{Q}_{3}}=500 \gev$
$M_{\tilde{U}_{3}}=1500 \gev$, $M_{\tilde{D}_{3}}= 300\gev$, 
$A_{\tau}=0 \gev$, $A_t=A_b=-2185 \gev$, 
$M_2 =150 \gev $, $m_{\tilde{g}}= 1500\gev$, 
$\hat{m}_A= 1500\gev$,
$\lambda =0.2$, $\kappa=0.6$, $A_{\kappa}= -1370\gev$.
The parameter points are {\tt HiggsBounds} allowed, and $h_1$ falls in the mass range $\MHexp \pm 3.04 \gev$.
When increasing $\mu$, the $M_W^{\text{NMSSM}}$ prediction decreases first, reaches its minimum for $\mu \sim 1100 \gev$ and then rapidly increases.
This behaviour can be explained by looking at the contributions to $\MW$ from the chargino/neutralino sector (here we take again the full $\Delta r$ contributions into account) and from the stop/sbottom sector.
The shift $\delta M_W$ arising from charginos and neutralinos is shown in the middle plot of \reffi{fig:mwmudependence}.
The chargino/neutralino contribution is largest for small $\mu$ and decreases with increasing $\mu$.
Going to larger $\mu$ the masses of the (higgsino-like) chargino and neutralino states increase and the $\MW$ contribution decreases.
The shift $\delta M_W$ arising from the stop/sbottom sector is shown in the right plot of \reffi{fig:mwmudependence}.
The contributions from the stop/sbottom sector (dominated by the $\Delta \rho$ contributions) get smaller when $\mu$ is increased 
up to $\mu\sim 1000 \gev$ and then start to
rise if $\mu$ is increased further.
Increasing $\mu$, the splitting between the two sbottoms gets larger (while
the stop masses stay nearly constant), which implies also an increase of the
splitting between stops and sbottoms. The counteracting terms in $\Delta
\rho$ (see the discussion in \refse{sec:mssmlimsec}) lead to the observed behaviour.

\section{Conclusions}
\label{sect:conclusions}

We have presented the currently most accurate prediction for the
$W$~boson mass in the NMSSM, in terms of the $Z$ boson mass, the fine-structure constant, the Fermi constant, and model-parameters entering via higher-order contributions. 
This result includes the
full one-loop determination and all available 
higher-order corrections of SM and SUSY type.  
These improved predictions have been compared to the state--of--the--art predictions in the SM and the MSSM within a coherent framework, and we have presented numerical results illustrating the similarities and the main differences between the predictions of these models. 

Within the SM,
interpreting the signal discovered at the LHC as the SM Higgs boson with
$\MHSM = 125.09 \gev$,
there is no unknown parameter in the $\MW$ prediction anymore. 
We have updated the SM prediction for $\MW$ making use of the most up to
date higher-order contributions. 
For $\MHSM = 125.09 \gev$ this yields
$\MW^{\text{SM}}=80.358 \gev$ (with a theory uncertainty from unknown
higher-order corrections of about $4 \mev$). The comparison with the 
current experimental value of $\MW^{\rm exp} = 80.385 \pm 0.015 \gev$
shows the well-known feature that the SM prediction lies somewhat below the
value that is preferred by the measurements from LEP and the Tevatron (at
the level of about $1.8\,\si$).
The loop contributions
from supersymmetric particles in general give rise to an upward shift in
the prediction for $\MW$ as compared to the SM case, which tend to bring
the prediction into better agreement with the experimental result. 

For the calculation of the $\MW$ prediction, we made use of the 
highly automated programs 
{\tt FeynArts} and {\tt FormCalc}.
Our evaluation is based on a framework which was developed in \citere{Benbrik:2012rm}, consisting in particular of a
NMSSM model file for the program {\tt FeynArts} and a {\tt Fortran} driver for the evaluation of the masses, 
mixing angles, etc.\ needed for the numerical evaluation.
The code {\tt NMSSMTools} is used for
the evaluation of the loop-corrected Higgs boson masses.
The implementation of another result for the NMSSM Higgs masses,
obtained in \citere{Peternmssm}, is in progress.

Our improved prediction for the $W$ boson mass in the NMSSM consists of the
full one-loop result, all available higher-order corrections of SM-type, 
stop-loop and sbottom-loop contributions with gluon and gluino exchange of 
${\cal O}(\alpha \alpha_s)$, relevant reducible higher-order
contributions, as well as an approximate treatment of the MSSM-type
Yukawa-enhanced electroweak two-loop corrections of
${\cal O}(\alpha_t^2)$, ${\cal O}(\alpha_t \alpha_b)$, ${\cal
O}(\alpha_b^2)$. Analytic expressions for all those contributions are
implemented, except for the electroweak two-loop contributions of SM-type, 
for which we make use of the fit formula
given in \citere{Awramik:2006uz}. The latter allows us to properly evaluate $\Delta
r^{\rm SM}$ at an NMSSM value for the $W$ boson mass.

We presented a detailed investigation of the prediction for $M^{\rm NMSSM}_W$,
focussing on the parameter regions which are allowed by Higgs searches
(tested by {\tt HiggsBounds}), SUSY searches and further theoretical
constraints. As a first step we analysed the size of the 
contributions from stops/sbottoms.
Since the sfermion sector is unchanged in the NMSSM with respect to the
MSSM, we have done this study in the MSSM limit, yielding an important check
of our NMSSM implementation.
We have investigated the size of the SUSY two-loop corrections to $\MW$ and 
found that the ${\cal O}(\alpha \alpha_s)$ corrections beyond the pure
gluon exchange contributions, which were incorporated in the previous result
of \citere{Domingo:2011uf}, can give sizeable contributions.
On the other hand, the effect of 
the Yukawa-enhanced electroweak two-loop corrections of
${\cal O}(\alpha_t^2)$, ${\cal O}(\alpha_t \alpha_b)$, ${\cal O}(\alpha_b^2)$
stays numerically relatively small in the allowed parameter region.

Concerning the investigation of genuine NMSSM effects, 
we started our discussion with the Higgs sector contributions to $\MW$.
The tree-level prediction for the lightest $\cp$-even Higgs mass is modified
by an additional term in the NMSSM as compared to the MSSM, 
which (for small $\tan \beta$) leads to an upward shift of the tree-level 
Higgs mass. Therefore, in that region, the radiative corrections needed to push the Higgs mass to about $125 \gev$ can be smaller than in the MSSM, 
which implies that lighter stop masses and a smaller stop mixing are possible.
We investigated a scenario where this additional tree-level term gives rise
to a higher $M_{h_1}$ prediction than in the MSSM limit.  
The impact on the 
$M_W$ prediction is a downward shift (of $\sim 25 \mev$ in the considered
example) as compared to the corresponding prediction in the MSSM. In the
prediction for $\MW$ this contribution from the Higgs sector enters
together with other SUSY loop contribution to $\Delta r$ yielding an upward
shift in $\MW$ compared to the SM. The overall effect is such that also in a scenario of this
kind a very good agreement between the theoretical prediction and the
experimental result can be reached. We have furthermore 
investigated the effect of doublet--singlet mixing. While a 
sizeable doublet--singlet mixing can occur in the region where the two NMSSM 
Higgs states ${h_1}$ and ${h_2}$ are close to each other in mass, we find
that it has only a minor effect on the $\MW$ prediction.

In the NMSSM the Higgs signal seen at the LHC can be interpreted both as the
lightest and the second-lightest $\cp$-even Higgs boson of the
spectrum.
Both interpretations give predictions for the $W$ boson mass in good agreement with the $\MW$ measurement.
In the NMSSM the interpretation of the LHC signal as the 
second-lightest $\cp$-even Higgs $h_2$ is possible 
together with a relatively heavy charged Higgs. 
This is different from the situation in the MSSM, where all Higgs states 
have to be light in this case, so that such a scenario can be probed by
searches for charged Higgs bosons in top-quark decays. As a consequence, the
interpretation of the observed Higgs signal as the 
second-lightest $\cp$-even Higgs boson is much
less constrained in the NMSSM compared to the MSSM.

For completeness, we have nevertheless briefly investigated also the case of
a light charged Higgs boson. We have found that a light charged Higgs boson
(together with a light $\cp$-even Higgs with reduced but non-zero couplings
to gauge bosons) 
can in principle give very significant contributions to $\MW$ (as in the
MSSM).
In that case large deviations from the SM Higgs sector contributions
occur, but as discussed above scenarios of this kind are severely 
constrained by limits from charged Higgs searches at the LHC.
Generally we find that taking all available constrains on the Higgs sector into account, the specific NMSSM effects of the Higgs sector to $\MW$ are relatively small.

On the other hand, the extended neutralino sector of the NMSSM
can lead to a sizeable difference between the $W$ boson mass predictions in the NMSSM and the MSSM.
The chargino/neutra\-lino contributions to $\MW$ can be larger in the NMSSM compared to the MSSM, where in the scenario which we studied 
the difference reaches $\sim 8 \mev$.
Assuming the same values for the soft mass parameters in the MSSM and the NMSSM and choosing $\mu=\mu_{\rm eff}$, the mixing with the singlino
leads to shifts in the neutralino masses as compared to the MSSM case. 
In the considered scenarios the lightest NMSSM states turned out to be lighter than the corresponding MSSM states. 
They also have a relatively small singlino component, which causes the 
resulting contributions to the prediction for $M_W$
to be larger than in the MSSM.
While light wino/bino states typically give positive contributions, light higgsinos can give contributions entering with both signs.

As a final step of our analysis, we compared the $M_W$ prediction
calculated with the full $\Delta r$ to the one where the full result is
approximated by the contribution to $\Delta \rho$.
We found that the difference between the full result and the 
$\Delta \rho$ approximation can be sizeable in the NMSSM,
where the approximation can lead both to an 
over- or an underestimate of the full result.
Light neutralinos with a significant higgsino--singlino mixing tend to give
a positive contribution to $\Delta \rho$, but a negative contribution to the
$\Delta r$ terms beyond 
$\Delta \rho$. It therefore depends on the 
exact patterns of the admixture with the singlino whether the neutralino
sector of the NMSSM leads to an upward or downward shift in the prediction
for $\MW$ in comparison with the MSSM.

We have demonstrated that the prediction for the $W$ boson mass arising
from the relation with the $Z$ boson mass, the Fermi constant and the fine
structure constant 
in comparison with high-precision measurements of those quantities
provides a high sensitivity for discriminating between the SM and possible
extensions of it. With further improvements of the experimental accuracy of
$\MW$, possible improvements in the determination of $\mt$ and further
information on possible mass spectra of supersymmetric particles -- either
via improved limits or the discovery of new states -- the impact of this
important tool can be expected to be even more pronounced in the future.

\subsection*{Acknowledgements}
We are grateful to Florian Domingo for helpful discussions and for providing
numerical values to cross-check our results. 
We also thank
Ayres Freitas,
Thomas Hahn,
Sven Heinemeyer,
Wolfgang Hollik 
and
Pietro Slavich
for helpful discussions.
This work has been supported by 
the Collaborative Research Center SFB676 of the DFG,
``Particles, Strings and the early Universe". O.S.~is supported by the Swedish Research Council (VR) through the Oskar Klein Centre.
This work is part of the D-ITP consortium, a program of the Netherlands Organization for Scientific Research (NWO) that is funded by the Dutch Ministry of Education, Culture and Science (OCW).
This research was supported in part by the European Commission through
the ``HiggsTools'' Initial Training Network PITN-GA-2012-316704. 

\bibliography{ref}
\bibliographystyle{JHEP}


\end{document}